\documentstyle[epsf,12pt]{article}
\begin{document}
\title{\bf Statistical model analysis of multiparticle correlations 
in $e^+e^-$ annihilation}
\author{T.Osada\thanks{e-mail address: osada@nucl.phys.tohoku.ac.jp}
\thanks{Address after June 1998:
{\it Istituto de F\'\hspace*{-1.5mm}{$\imath$}sica, 
Universidade de S\~ao Paulo, C.P.66318, 05389-470 S\~ao Paulo-SP, Brazil}}, 
M.Maruyama, F.Takagi\\
{\small Department of Physics, Tohoku University, Sendai 980-8578, Japan}}
\date{\today}
\maketitle
\begin{abstract}
A new statistical model for multiparticle production in $e^+e^-$ 
annihilation is proposed based on the idea of the longitudinal 
phase space with limited transverse momentum. The longitudinal 
rapidity space is divided into cells of equal size in order to 
take into account the Bose-Einstein correlations(BEC) with a finite 
correlation length $\delta y$. The maximum entropy method is 
used to determine the probability distributions of the final state 
pions (or $\rho$ mesons) for a given mean multiplicity, 
mean transverse momentum and mean total energy. Event simulation 
based on our model is performed in two extreme cases, a 
$\pi$-model and a $\rho$-model. The $\pi$-model assumes 
that only pions are produced directly. 
On the other hand, the $\rho$-model assumes that only 
$\rho$ mesons are produced directly and they decay into pions. 
A good overall fit to experimental data from 
PETRA to LEP energy regions is obtained for $\delta y$ ranging from 
0.6 to 1.2 in the $\pi$-model. 
We found that the BEC plays a very important role to 
reproduce various correlation data.  
In some correlation data, resonance effect and conservation 
laws are also important.
\end{abstract}
\section{Introduction}
With increase in energies of accelerators, accurate data on 
high-multiplicity events at high energies have recently been provided. 
This makes it possible to analyze correlations and fluctuations 
such as the Bose-Einstein correlations (BEC)\cite{BOAL}
and the intermittency \cite{BIALAS-1,BIALAS-2} in detail. 
These correlations and fluctuations of produced particles are 
rather new tools to study multiparticle production. 
By analyzing them, one may be able to extract useful information 
on the space-time size of the particle production region 
and the production mechanism\cite{BOAL}. 
Correlations and fluctuations may also provide a clear signal of quark 
gluon plasma formation expected in high energy nucleus-nucleus collisions. 
On the other hand, multiparticle production is a phenomenon observed 
commonly in various reactions such as $e^+e^-$ annihilations, 
lepton-hadron deep inelastic interactions, hadron-hadron interactions, 
and nucleus-nucleus collisions.
A large amount of correlation data has been accumulated 
on various reactions. It has been found that there are some universal 
characteristics which are independent of the reaction type while some 
other properties depend on it. 
Phenomenological models which can describe some crucial properties 
of many particle correlations will be very useful for systematic 
investigations of those characteristics caused by correlations and 
fluctuations. 

The BEC have lately attracted considerable attention, in particular, 
because they may give information on the space time structure 
of hadronic source and they may cause large fluctuations 
of particle density in the phase space. 
Although many existing phenomenological models have not taken 
into account the BEC completely yet, 
there are some models (or event generators) 
which allow computing the two identical particle correlation 
functions.
In most cases, the BEC are generated more or less artificially. For example, 
they are calculated by using 
(i) the Fourier transformation of the model's source 
function\cite{BOWLER} or (ii) the Wigner function\cite{PRATT}, 
(iii) by weighting every event\cite{ANDESSON}, or 
(iv) by modifying the distribution of momentum difference 
of identical 
particle pairs in each event\cite{KADIJA}. 
However, some of those models cannot describe the BEC in a single event. 
Motivated by those consideration, we would like to propose a new 
statistical model. 
It is constructed by the maximum entropy method \cite{JAYNES,WILK} 
and two versions of the model, the $\pi$-model and 
the $\rho$-model are applied to $e^+e^-$ annihilation in this paper. 
The most characteristic point of our model is that the 
BEC is taken into account in quantum statistical 
level with a characteristic correlation 
length ($\delta y$) defined in the rapidity space. 

This paper is organized as follows. 
In Sec.I$\!$I, our model is explained and the 
distribution functions of the final state hadrons are derived. 
Sec.I$\!$I$\!$I is devoted to the explanation of the simulation method 
including the procedure of parameter determination. 
Comparison of results obtained from simulation 
with the experimental data from PETRA to LEP energies is 
given in Sec.I$\!$V. Concluding remarks and discussions are given 
in Sec.V. 
\section{Statistical model based on Maximum Entropy Method}
We first consider the $\pi$-model, i.e., we assume that only pions 
are produced according to a statistical distribution. 
A system consisting of many pions produced in a single event 
may be decomposed into three subsystems, 
which consist of like sign pions, that is, $\pi^{+}$, 
$\pi^{-}$ and $\pi^{0}$ subsystems. 
For example, consider a $\pi^+$ subsystem 
in the longitudinal phase space with limited 
transverse momentum\cite{LPS}. 
The longitudinal axis may be 
identified with the direction of the initial quark and antiquark 
produced by the virtual photon in center of mass system. 
Then the rapidity space in the longitudinal phase space 
is divided into many cells \cite{KNOX} 
of equal size $\delta y$, and we consider a probability to find 
$\pi^+$ mesons in each cell. 
The probability of finding $n_i$ $\pi^+$s 
in the {\it i}-th cell, ~$P^{(n_i)}_{i}$ is normalized as 
\begin{equation}
\sum_{n_i=0}^{\infty} P^{(n_i)}_{i} =1. \label{NOR}
\end{equation}
The summation index $n_i$ runs from 0 to infinity 
because of the Bose-Einstein statistics. 
The mean $\pi^+$ multiplicity and the mean energy of the 
$\pi^+$ subsystem are given, respectively, by: 
\begin{equation}
\langle n_+ \rangle=
\sum_{i} \sum_{n_i} n_i P^{(n_i)}_{i} \label{MUL}~~
\end{equation} 
and 
\begin{equation}
\langle E_+ \rangle=
\sum_{i} \sum_{n_i} n_i \epsilon_i P^{(n_i)}_{i}.\label {ENE}
\end{equation}
Here $\epsilon_i$ is the energy of a $\pi^+$ in the {\it i}-th cell:
\begin{eqnarray}
&& \epsilon_i = m_{\mbox{\tiny T}} \cosh y_i~, \label{epsi}\\ 
&& y_i = y_{min} + (i-\frac{1}{2}) \delta y~, \\ 
&& y_{min} = -\ln\bigg[~ 
\frac{\sqrt{s}+\sqrt{s-4m_{\mbox{\tiny T}}^2}}{2m_{\mbox{\tiny T}}}~\bigg], 
\end{eqnarray}
where 
$y_i$ is the central rapidity value of the {\it i}-th cell, 
$y_{min}$ is the kinematical minimum value of rapidity, 
$m_{\mbox{\tiny T}}= \sqrt{\langle p_{\mbox{\tiny T}}\rangle^2 +m_{\pi}^2}$ 
is the mean transverse mass, and $\sqrt{s} $ is the total center 
of mass energy. 
The same argument applies to both $\pi^-$ and $\pi^0$ subsystems. 
Therefore the mean multiplicities and the mean energies 
of those subsystems are given as 
\begin{equation}
\langle n_+ \rangle=\langle n_0 \rangle=\langle n_-\rangle
=\langle n_{ch}\rangle /2 ,
\end{equation}
and 
\begin{equation}
\langle E_+ \rangle=\langle E_0 \rangle=\langle E_- \rangle
=\sqrt{s}/3 , 
\end{equation}
where $\langle n_{ch} \rangle$ is the mean charged multiplicity. 
The most probable distribution $P^{(n_i)}_{i}$ for the $\pi^+$ subsystem 
can be determined by the maximum entropy method which is well-known 
in information theory\cite{JAYNES}. 
According to a statistical picture, particles will distribute in the 
phase space in such a way that every possible state is realized with the 
same probability. 
This method has been applied in many fields in science including 
multiparticle production phenomenology.
First, we define the missing information entropy \cite{WILK}:
\begin{equation}
S=-\sum_{i} \sum_{n_i} P^{(n_i)}_{i} \ln P^{(n_i)}_{i}. 
\end{equation}
This $S$ has the maximum value when $P^{(n_i)}_{i}$ is a constant, 
so-called equal {\it a priori}~~probabilities, which does not 
depend on $i$ and $n_i$. 
It is noted here that the probability $P^{(n_i)}_{i}$ does not 
depend on $P^{(n_{j})}_{j}$ when ($i\ne j$). 
Therefore a  probability of finding 
$n_i$ $\pi^+$s in the $i$-th cell and 
$n_j$ $\pi^+$s in the $j$-th cell simultaneously, 
$P^{(n_{i},n_{j})}_{i,j}$, is given by 
\begin{eqnarray}
P^{(n_{i},n_{j})}_{i,j} = P^{(n_i)}_{i} \times P^{(n_j)}_{j}. 
\end{eqnarray}

The entropy $S$ should be maximized with the constraints which 
correspond to the normalization eq.(\ref{NOR}), the mean multiplicity 
eq.(\ref{MUL}) and the mean energy eq.(\ref{ENE}) of the subsystems. 
For that purpose we consider $F$ defined below instead of $S$ 
by introducing Lagrange multipliers $\lambda_1^i$, $\lambda_2$ and 
$\lambda_3$; 
\begin{eqnarray}
F=&-&\sum_{i} \sum_{n_i} P^{(n_i)}_{i} \ln P^{(n_i)}_{i} \nonumber \\
&+&\sum_i \lambda^i_1 (\sum_{n_i} P^{(n_i)}_{i} -1) \nonumber \\
&+&\lambda_2 (\sum_i \sum_{n_i} n_i P^{(n_i)}_{i}
- \langle n_+\rangle) \nonumber \\
&+&\lambda_3 (\sum_i \sum_{n_i} n_i\epsilon_i P^{(n_i)}_{i}
- \langle E_+ \rangle ) ~~. 
\end{eqnarray}
By requiring that the variation of $F$ is vanishing, one obtains
\begin{equation}
P^{(n_i)}_i = \frac{e^{ n_i\left(\mu-\epsilon_i \right)/T}}{Z_i}~. 
\end{equation}
Here, the Lagrange multipliers $\lambda_1^i,
~\lambda_2$ and~$ \lambda_3 $ are rewritten in terms of the
partition function $Z_i$, the ``partition'' temperature $T$ \cite{CHOU-YANG}
and the ``chemical'' potential $\mu$: 
\begin{equation}
e^{-1+\lambda^i_1} = 1-e^{(\mu-\epsilon_i)/T} \equiv 1 / Z_i ~~,
\end{equation}
\begin{equation}
\lambda_2= \mu / T ~~,
\end{equation}
\begin{equation}
\lambda_3 = -1 / T . 
\end{equation}
The parameters $T$ and $\mu$ are determined uniquely for given 
$\langle E_+ \rangle$ and $\langle n_+ \rangle$ and they play a 
dominant role to reproduce the gross features of data on 
single particle spectra. 
The probability distribution functions 
for $\pi^-$ and $\pi^0$ subsystems are 
the same as $P^{(n_i)}_{i}$ for $\pi^+$ subsystem.
The parameter $\delta y$ is determined by fitting to data 
on the second order fluctuations of multiplicity.

In order to study the resonance effect, 
we also consider another extreme case called 
$\rho$-model where it is assumed that only $\rho$ mesons are 
produced in the same way as pions in the $\pi$-model. 
In the $\rho$-model, one must take the spin degree of freedom into account. 
A $\rho$ meson system is decomposed into nine subsystems:
$\rho^+\!\!\uparrow$, $\rho^+\!\!\to$, $\rho^+\!\!\downarrow$,~
$\rho^0\!\!\uparrow$, $\rho^0\!\!\to$, $\rho^0\!\!\downarrow$,~
$\rho^-\!\!\uparrow$, $\rho^-\!\!\to$, and $\rho^-\!\!\downarrow$, 
where $\uparrow,\to$, and  $\downarrow$ 
denote the three eigenstates of the spin $S_{\mbox{\scriptsize z}}=$
+1, 0 and -1, respectively.
The partition temperature $T_{\rho}$, 
the ``chemical potential'' $\mu_{\rho}$ 
and the cell size $\delta y_{\rho}$ in the $\rho$-model 
are determined so as to reproduce the mean charged 
multiplicity, the total energy and the second order fluctuations 
of multiplicity. (The decay process of $\rho$ 
meson will be discussed in Sec.I$\!$I$\!$I B)~

\section{Method of simulation}
The method of generating events based on the 
$\pi$-model and the $\rho$-model is explained in this section. 
\subsection{$\pi$-model}
Events generating procedures are divided into three steps. 
The first step is determination of the parameters $\mu$ and $T$. 
The second step is $\delta y$ search with trial events. By using 
these determined parameters, the final simulations are 
executed in the third step. 
\begin{center}
{\bf 
Step 1: Determination of the parameters $\mu$ and $T$} 
\end{center} 
\noindent
To determine the values of $T$ and $\mu$, we solve the following 
simultaneous equations for $\mu$ and $T$ with a 
fixed trial value of $\delta y$: 
\begin{eqnarray}
\langle n_{ch} \rangle/2 &=& \sum_{i} \sum_{n_i} n_i \frac
{e^{ n_i(\mu-\epsilon_i)/T}}{Z_i} \nonumber \\
&=& \sum_i\frac{e^{(\mu-\epsilon_i)/T}}{1-e^{(\mu-\epsilon_i)/T}}, 
\end{eqnarray}
\begin{eqnarray}\sqrt{s}/3 &=& \sum_{i} \sum_{n_i} n_i \epsilon_i 
\frac{e^{ n_i(\mu-\epsilon_i)/T}}{Z_i} \nonumber \\ 
&=& \sum_i\frac{\epsilon_i e^{(\mu-\epsilon_i)/T}}{1-e^{(\mu-\epsilon_i)/T}},
\end{eqnarray}
where observed value is used for the mean charged multiplicity 
$\langle n_{ch} \rangle$. 
It should be noted here that $\mu$ and $T$ are 
determined as functions of $\delta y$, i.e. 
$\mu=\mu(\delta y)$, $T=T(\delta y)$~.\\

\begin{center} 
{\bf 
Step 2: Generation of trial events and determination of
the parameter $\delta y$} 
\end{center} 

\noindent 
\begin{enumerate}
\item
In order to avoid a possible artifact due to a particular 
cell location, we now take the central value $y_i$ of the $i$-th 
cell as $y_i=y_{min}+(i-\frac{1}{2}+r)\delta y$, where $r$ is 
a homogeneous random number between 0 and 1.

\item 
The $\pi^+$, $\pi^0$ and $\pi^-$ subsystems 
are generated in all cells in the rapidity space 
according to the probability distribution $P^{(n_i)}_{i}$ 
with a trial $\delta y$, $\mu(\delta y)$ and $T(\delta y)$. 
When a pion is produced in a certain cell, 
its rapidity is assigned by using a random number 
so that pions distribute homogeneously in that cell. 
This smearing of pion rapidity causes slight changes 
of $\langle n_{ch} \rangle$ and $\sqrt{s}$ 
in comparison with the result of the step 1. 
We readjust the values of $\mu $ and $T$ 
to reproduce correctly the mean multiplicity and the 
``mean'' energy of the system. 
\item 
Charge conservation is demanded event by event. 
Events are discarded if the charge conservation is violated. 
\item 
The transverse momenta are generated according to the following 
distribution: 
\begin{eqnarray}
\frac{d\sigma}{dp_{\scriptscriptstyle T}} \propto 
p_{\scriptscriptstyle T}~e^{-2p_{\scriptscriptstyle T}/
\langle p_{\scriptscriptstyle T} \rangle}\label{pt}~~.
\end{eqnarray} 
Here $\langle p_{\scriptscriptstyle T}{\scriptstyle} \rangle$ 
denotes the observed mean transverse momentum at each $\sqrt{s}$.

The azimuthal angle $\varphi$ of the transverse momentum 
is distributed at random between 0 and 2$\pi$. 
The transverse momenta generated in this way 
are assigned to all pions independent of their rapidities. 
A four momentum of a pion is then given as  
\begin{eqnarray}
{\it p}^{\mu}&=& (\sqrt{m_{\pi}^2 +p_{\scriptscriptstyle T}
^{~2}}\cosh y,
~p_{\scriptscriptstyle T}\cos \varphi,~  \nonumber \\ 
&&~~~~~~~~~~~~~p_{\scriptscriptstyle T}\sin \varphi,~ 
\sqrt{m_{\pi}^2+p_{\scriptscriptstyle T}^{~2}} \sinh y ~) . 
\end{eqnarray}

\item 
Approximate energy-momentum conservation is also required. 
For energy conservation, we set an energy window with width 
$\pm \delta E$ around $\sqrt{s}$. For momentum conservation, 
we also set the longitudinal and the transverse momentum windows 
with widths $\pm\delta P_{\scriptscriptstyle L}$and $\delta P_
{\scriptscriptstyle T}$, respectively, around $0$ total momentum. 
The total energy $E_{tot}$, the longitudinal component of the 
total momentum $P_{L}$ and the transverse component $P_{T}$ are 
evaluated in every event.  
An event is adopted only when it satisfies the inequalities 
\begin{equation}
\sqrt{s}-\delta E \leq ~~E_{tot}=
\sum_{all~\pi} \epsilon~~\leq \sqrt{s}+\delta E, 
\end{equation}
and
\begin{equation}
-\delta P_{\scriptscriptstyle L} \leq ~~P_{L}= \sum_{all~\pi} 
p_{\scriptscriptstyle L}
~~\leq \delta P_{\scriptscriptstyle L},
\end{equation}
\begin{equation}
{\vec P}_{T}^2= \mid \sum_{all~\pi} {\vec p}_{\scriptscriptstyle T} \mid^{2}
\leq \left( \delta P_{\scriptscriptstyle T} \right)^{2}  ~~. 
\end{equation} 
As an example, the distribution of $E_{tot}$ and the applied energy 
window for $\sqrt{s}$=34.5~GeV are shown in Fig.1. 
One has to take the window size as small as possible to keep the 
quality of conservation law while one would like to take it as 
large as possible to save the computation time. 
Thus one has to determine an optimum value of window size.
We decided to take $\delta E=0.2\sqrt{s}$ and 
$\delta P_{\mbox{\scriptsize L}}=0.2p_{max}$, where $p_{max}$ is the 
maximum value of the c.m. momentum carried by a pion. By the way, 
we found that some physical quantities, e.g., the dispersion of the 
multiplicity distributions are rather sensitive to $\delta E$ when 
$\delta E$ is large. We have confirmed that this strong 
$\delta E$-dependence disappears when $\delta E=0.2\sqrt{s}$, 
which implies that the physical observables are insensitive to 
$\delta E$ if $\delta E \leq 0.2\sqrt{s}$.
For $p_{\scriptscriptstyle T}$-conservation, we use a multiplicity 
dependent window size
$\delta P_{\scriptscriptstyle T}=\sqrt{n_++n_0+n_-}
~\langle p_{\scriptscriptstyle T}\rangle$~. 

\item 
The jet axis, thrust or sphericity, is determined event by event.
New four components of a momentum vector of a pion are calculated by 
referring to this jet axis. 

\item 
The value of the parameter $\delta y$ is determined by fitting 
to appropriate experimental data which are sensitive to $\delta y$. 
We have used the rapidity interval dependence of 
$D_{ch}^2/\langle n_{ch}\rangle^2$ for the fitting. 
Here, $D_{ch}=\sqrt{\langle n_{ch}^2 \rangle - \langle n_{ch} \rangle^2}$. 
\end{enumerate}
\begin{center}
------------------------------------------\\
{\bf Figure 1.}\\ 
------------------------------------------\\
\end{center}

\begin{center} 
{\bf Step 3: Final event generation }
\end{center} 

\noindent
A sufficiently large number of events 
are generated using the probability distribution 
$P_i^{(n_i)}$ with $\mu$, $T$ and $\delta y$ determined 
in the preceding steps. 
\vspace*{0.5cm}

Final results of those parameters are shown 
in Table~I~and the results of the fitting are shown 
in Fig.2(a)-(c), where $y_{cut}$ is the half size of the rapidity 
interval $[-y_{cut},y_{cut}]$. 
Experimental data are taken from refs.\cite{TASSO-1,DELPHI-1}.
~As shown in Table I, 0.6, 0.8 and 1.2 are 
chosen as the values of the $\delta y$ at $\sqrt{s}=$ 
14.0, 34.5 and 91.2 GeV, respectively. 
In the determination of the best fit value $\delta y$, 
we have put a special weight on the data point for 
$y_{cut}$ = $y_{max}$ in order to reproduce 
the multiplicity distribution for full phase space 
as precisely as possible. 
It should be noted that $y_{cut}$ dependence of the 
$D_{ch}^2/\langle n_{ch} \rangle^2$, a kind of 2nd order fluctuations 
of the particle density in the rapidity space, is reproduced well.
\begin{center}
------------------------------------------\\
{\bf Figure 2.~(a)-(c)}\\
{\bf Table I}\\
------------------------------------------\\
\end{center} 

\subsection{$\rho$-model.} 
The parameter $T_{\rho}$, $\mu_{\rho}$ and $\delta y_{\rho}$ 
in the $\rho$-model are determined in a way similar to the $\pi$- model. 
The results are given in Table I$\!$I.
\begin{center} 
------------------------------------------\\
{\bf Table I$\!$I}\\
------------------------------------------\\
\end{center}

Now we discuss the decay of $\rho$ mesons e.g. $\rho^+ \to \pi^+\pi^0$. 
Rapidities have to be assigned to each pion produced 
from one $\rho$ meson of some rapidity $y_{\rho}$. 
In the rest frame of the $\rho$ meson, 
the maximum and the minimum rapidities of the two 
pions produced by the $\rho$ meson are given by $\pm y_d$ where
\begin{eqnarray} 
y_{d}\!&=&\!\frac{1}{2}\ln(\frac{E_{d}+p_{d}}{E_{d}-p_{d}})~~,\\ 
&& E_{d}=\sqrt{p_{d}^2+m_{\pi}^2}\quad ,\qquad 
p_{d}=\sqrt{M_{\rho}^2/4-m_{\pi}^2}\quad, \nonumber 
\end{eqnarray} 
and $M_{\rho}$ and $m_{\pi}$ are the masses 
of the $\rho$ and $\pi$ mesons. 
We use the following Breit-Wiger form $f(M_{\rho})$ as the 
invariant mass distribution of $\rho$ meson:
\begin{eqnarray}
f(M_{\rho}) \propto  \frac{1}{(M_{\rho}-m_{\rho})^2+\Gamma^2/4},
\end{eqnarray} 
where $m_{\rho}$=770 MeV and $\Gamma$=150 MeV.

To impose the Bose-Einstein correlations between 
like charge pions, we use an ``enhancement factor'' method. 
When a parent $\rho$ meson has a rapidity $y_{\rho}$, 
we divide the longitudinal rapidity 
interval $[y_{\rho}-y_d,y_{\rho}+y_d]$ into $K$ cells of equal size 
$\delta y_{\pi}$. 
Here we take $\delta y_{\pi}=\delta y$. 
In order to determine the cell to which a pion, say $\pi^+$, belongs, 
we introduce the enhancement factor: 
\begin{eqnarray} 
\frac{n^+_i+1}{K+N^{+}} \qquad \mbox{for $i$-th cell},\label{ENHA}
\end{eqnarray} 
where $n^+_i$ and $N^{+}$ are the number of $\pi^+$s 
already produced in the $i$-th cell and that in the interval 
$[y_{\rho}-y_d,y_{\rho}+y_d]$, respectively. 
Transverse momenta $\vec{p}_{\scriptscriptstyle T1}$ and 
$\vec{p}_{\scriptscriptstyle T2}$ are assigned to the two pions,
in the same way as in the $\pi$- model. 
To reproduce the invariant mass of the parent $\rho$ meson $M_{\rho}$, 
the azimuthal angle difference, 
${\scriptstyle \Delta}\phi=\phi_1-\phi_2$, needs to satisfy 
the following equation:
\begin{eqnarray} 
&&M_{\rho}^2=2m_{\pi}^2+2( \epsilon_1\epsilon_2
-p_{\scriptscriptstyle L1}p_{\scriptscriptstyle L2}
-|\vec{p}_{\scriptscriptstyle T1}||\vec{p}_{\scriptscriptstyle T2}|~
\cos{\scriptstyle \Delta} \phi),\\ 
&&\epsilon_{i}= 
\sqrt{m_{\pi}^2 + |\vec{p}_{\scriptscriptstyle Ti}^2}|\cosh y_{i},\\ 
&& p_{\scriptscriptstyle Li}= 
\sqrt{m_{\pi}^2 + |\vec{p}_{\scriptscriptstyle Ti}^2}|\sinh y_{i} 
\qquad \mbox{for $i=1$ or $2$.}
\end{eqnarray} 

\section{Comparison of results with experimental data}
In this chapter we compare the results of event simulation with 
experimental data. 
We investigate effects of charge and energy-momentum 
conservations and also the resonance effect by 
comparing the results from $\pi$- and $\rho$- models. 
20000 events of multiparticle production in $e^+e^-$ 
annihilation have been generated for both the $\pi$-model and 
the $\rho$-model at each energy $\sqrt{s}=14.0, 34.5$ and $91.2$ GeV. 
We use the experimental data reported by the TASSO 
Collaboration\cite{TASSO-1,TASSO-2,TASSO-intermittency,TASSO-BEC}~
for $\sqrt{s}=14.0,34.5$ GeV, and by the DELPHI 
\cite{DELPHI-1,DELPHI-intermittency,DELPHI-BEC,DELPHI-FBC}~
and ALEPH Collaborations\cite{ALEPH-intermittency,ALEPH-BEC}~for 
$91.2$ GeV. We present all results obtained from the 
$\pi$-model. When significant differences are found between the two models, 
we present also the result of the $\rho$- model.  
\subsection{Multiplicity distributions} 
Charged multiplicity distributions $P(n_{ch})$ 
are shown in Figs.3 (a)-(c).
The experimental data\cite{TASSO-1,DELPHI-1} 
are well reproduced by our $\pi$-model. 
This is reasonable because the mean multiplicity and dispersion 
of the multiplicity distribution are used in the determination of 
the model parameters. 
It should be noted, however, that our model is able 
to predict the detailed shape of the multiplicity distribution at each energy. 
We also investigate the multiplicity distribution in terms of KNO variable. 
Our model gives good scaling behavior for both the 
full phase space(see Fig.4(a)) and small rapidity windows(not shown). 
The $y_{cut}$-dependence of the multiplicity distribution is 
reproduced correctly as shown in Fig.4(b). Note that our results are
approximately consistent with the negative binomial distributions. 
\begin{center}
------------------------------------------\\
{\bf Figure 3.~(a)-(c)}\\
{\bf Figure 4.~(a) and (b)}\\
------------------------------------------\\
\end{center}

Dependence of multiplicity distributions on $\delta y$ has already 
been demonstrated in Figs.2(a)-(c). 
Decreasing $\delta y$ reduces the dispersion of the 
multiplicity distribution. Conservation 
laws also affect the multiplicity distribution. 
In addition to ``full conservation'' event for which 
both energy-momentum and charge conservation are imposed, we 
have also generated events with ``only charge conservation'', 
``only energy-momentum conservation'' or ``no conservation'' 
by imposing only charge conservation or only energy-momentum 
conservation or none of them in order to investigate the effects of 
conservation laws\footnote{
It should be noted here that both energy-momentum and charge 
conservations hold on the average even in events with no 
conservation.}. 
It is found that the charge conservation causes the 
broadening of the multiplicity distribution, 
while the effect of energy conservation is opposite 
as shown in Fig.5.
The effect of energy-momentum conservation is stronger than 
that of charge conservation. 

We have investigated the charge correlations due to charge conservation. 
For this purpose, we consider the rapidity window 
dependence of the quantity $D_{ch}^2/2D_{+}^2$, where 
$D_{+}^2$ is the dispersion of the $\pi^+$ 
multiplicity distribution,  
$\sqrt{\langle n_+^2 \rangle - \langle n_+ \rangle^2}$. 
The quantity $D_{ch}^2/2D_{+}^2$ can  be written as
\begin{eqnarray*}
\frac{D_{ch}^2}{2D_{+}^2}&=&\frac{\langle (n_+ + n_- )^2 \rangle 
-\langle n_+ + n_- \rangle^2}{2D_{+}^2}\\
&=& 1+\frac{\langle n_+n_-\rangle-\langle n_+ \rangle^2}{D_{+}^2}, 
\end{eqnarray*}
where use was made of the equality 
$\langle n_+ \rangle =\langle n_- \rangle$. 
If there is the maximum correlation 
$n_+ = n_-$ due to charge conservation, 
one has $D_{ch}^2/2D_{+}^2=2$ because 
$\langle n_+n_- \rangle = \langle n_+^2 \rangle$. 
On the other hand, if there is no correlation 
between $n_+$ and $n_-$, $\langle n_+n_-\rangle$ 
is reduced to $\langle n_+ \rangle^2$ and hence $D_{ch}^2/2D_{+}^2=1$. 
Therefore, it is obvious that $D_{ch}^2/2D_{+}^2=2$ for $y_{cut}=y_{max}$. 
On the other hand, one can expect that $D_{ch}^2/2D_{+}^2=1$ for very 
small $y_{cut}$ because the correlation due to charge conservation will 
be maximally weakened when $y_{cut}$ is small. 
Those expectations are indeed realized by our simulation 
as shown in Fig.6. 
The correlation measure $D_{ch}^2/2D_{+}^2$ increases linearly from 
about unity as $y_{cut}$ increases and then tends to saturate at 
the value 2 for $y_{cut}\geq$3.5.
The slope of the linear rise may be regarded as a measure of 
local charge conservation. 
Unfortunately, it appears that corresponding experimental data 
are not available at present. We therefore urge experimentalists 
to provide such data. 
\begin{center}
------------------------------------------\\
{\bf Figure 5}\\
{\bf Figure 6 (a) and (b)}\\
------------------------------------------\\
\end{center}

\subsection {Rapidity distribution}
The single particle rapidity distributions 
are calculated and the results are shown in Fig.7. 
Our model reproduces the experimental data\cite{TASSO-2} 
except for the observed dip structure at $y\cong 0$\footnote{
The result of our model is systematically larger than TASSO data 
at $\sqrt{s}$=14.0 and 34.5 GeV, because integration of 
the $dn/dy$ data reported by TASSO is smaller than the reported 
mean multiplicity.}.
In particular, the energy dependence of the central height 
is reproduced correctly by our model. 
Although our model produces a dip at $y$=0 in 
qualitative agreement with experimental data, it is 
too deep and too narrow. Furthermore it tends to diminish as 
$\sqrt{s}$ increases. This tendency appears opposite to what is 
observed experimentally. 
This situation does not change even 
if one uses the $\rho$- model. The discrepancy may be due to the fact that 
three jet events are not included in our model\footnote{
This dip structure 
had been investigated by TASSO collaboration\cite{TASSO-2}. 
They report their analyses using the Lund-string model\cite{LUND-STRING} 
and the independent jet model\cite{INDEP-JET}.}. 
By the way, we found that the dip is not reproduced if one uses 
the sphericity instead of the thrust to define the jet axis.
We have also confirmed 
that the rapidity distribution is insensitive to both 
conservation laws and the value of $\delta y$. 
\begin{center}
------------------------------------------\\
{\bf Figure 7.}\\
------------------------------------------\\
\end{center}

\subsection{Rapidity dependence of the mean transverse momentum} 
There are interesting experimental data on the rapidity dependence 
of the mean transverse momentum, $\langle p_T(y)\rangle$\cite{TASSO-2}. 
The observed $\langle p_T(y)\rangle$ is almost constant in the central 
region while it decreases as $y$ approaches the kinematical limit as 
shown in Fig.8(a). On the other hand, transverse momentum of a pion is 
generated according to the exponential formula (18) with 
$y$-independent mean transverse momentum 
$\langle p_{\mbox{\tiny T}} \rangle$ in our model. 
Therefore, it is very interesting 
to see if our model can reproduce the observed $y$-dependence of the 
mean transverse momentum. The result of our model is shown and 
compared with data in Fig.8(a).
Our model reproduces correctly the observed $y$-dependence of the 
mean transverse momentum. Therefore, one can conclude that the observed 
$y$-dependence is totally due to kinematical reasons. To confirm this 
conclusion, we have calculated $\langle p_T(y)\rangle$ for events 
generated without energy-momentum conservation and/or without 
assigning the thrust axis.
The result for 34.5 GeV is shown in Fig. 8(b). 
It is clearly seen that the energy-momentum conservation is most 
responsible for reproducing the correct $y$-dependence but the use 
of thrust axis has also some effect.
\begin{center}
------------------------------------------\\
{\bf Figure 8.~(a) and (b)}\\ 
------------------------------------------\\
\end{center}

\subsection{Bin size dependence of the scaled factorial moments} 
We study the scaled factorial moments in order to 
investigate the fluctuations of particle density in 
the rapidity space. 
The following two kinds of scaled factorial moment 
are used for comparison with experimental data. 

The factorial moments are evaluated for the multiplicity of 
particles produced in the rapidity window $y_0 \leq y \leq y_0+\Delta Y$. 
The rapidity window is divided into M bins of equal size $\Delta Y/M$. 
When $N$ particles are produced in the window while 
$k_m$ particles are found in the $m$-th bin, 
the `exclusive' scaled factorial moment\cite{BIALAS-1} is defined as
\begin{equation}
F^{ex}_i = M^{i-1} \bigg\langle \sum_{m=1}^M 
\frac{k_m(k_m-1) \cdots (k_m-i+1)}
{N(N-1) \cdots (N-i+1)} \bigg\rangle, 
\end{equation}
where $\langle ~~\rangle$ means an event 
average with fixed $N$. 
The `inclusive' scaled factorial moment used 
in many literatures\cite{BIALAS-2} is given by 
\begin{equation}
F^{in}_i = M^{i-1} \bigg\langle \sum_{m=1}^M
\frac{k_m(k_m-1) \cdots (k_m-i+1)}
{\langle N \rangle^i } \bigg \rangle, 
\end{equation}
where $\langle ~~\rangle$ now means an event average 
for distributed $N$. 
Both $F^{ex}_i$ and $F^{in}_i$ are equal to unity if there are only 
statistical fluctuations( a binomial distribution for $F^{ex}_i$ or a 
Poisson distribution for $F^{in}_i$). 
On the other hand, the Bose-Einstein distribution (eq.(12)) gives 
a typical example of non-statistical fluctuations. 
It is the multiplicity distribution in a single cell of size 
$\delta y$ in our model, corresponding to $M\approx \Delta Y/\delta y$. 
In this case the inclusive factorial moments of the second order for 
like charged and all charged particles are given, respectively, by 
\begin{eqnarray}
F^{in}_2 &=& \frac{\langle n_+(n_+-1)\rangle}{\langle n_+ \rangle ^2 } 
\nonumber \\ 
&=&  2 ~~\mbox{ (for like sign particles) },
\end{eqnarray} 
and 
\begin{eqnarray} 
F^{in}_2 &=& \frac{\langle (n_+ + n_- )^2 -(n_+ +n_-) \rangle }
{\langle (n_+ + n_-)\rangle^2} \nonumber \\
&=& \frac{\langle n_{+}(n_{+}-1) \rangle + \langle n_{+} \rangle^2}
{2 \langle n_{+} \rangle^2 } \nonumber \\
&=& 1.5~~~\mbox{ (for charged particles) }. \label{FAC-ES}
\end{eqnarray}
For $\sqrt{s}$=34.5GeV, 
one can investigate the fluctuations in a single cell, 
i.e. fluctuations 
induced by pure Bose-Einstein distributions(cf. eq~(\ref{FAC-ES})~) 
at $M$=5 since $\delta y$=0.8 and $\Delta Y$=4.0. 
Our simulation gives $F^{in}_2(M=5) \approx$ 1.27, and ~$F^{in}_2(M>16) 
\approx$ 1.38. The reduction from 1.5 can be attributed to the 
smearing due to varying cell location(cf. the item 1 of the step 2 
of sec.I$\!$I$\!$I~A ) and also to the global charge conservation. 
In fact, we obtain ~$F^{in}_2(M=5) \approx $1.32, and $F^{in}_2(M >16) 
\approx$ 1.46 when all the conservation laws are not imposed. 

Now the results are compared with experimental 
data\cite{TASSO-intermittency,ALEPH-intermittency} for both 
exclusive and inclusive factorial moments in Figs. 9 and 10. 
Our results approximately reproduce the experimental data. 
In particular, the agreement is very good for $\sqrt{s}=34.5$ GeV. 
In our model, the parameter $\delta y$ plays a crucial role in 
reproducing the bin size dependence of the factorial moments. 
As $M$ increases from unity, the factorial moments increase due to 
the BEC and then start to saturate for 
$M>\Delta Y/\delta y$\footnote{
TASSO Collaboration compared their data\cite{TASSO-intermittency} 
(exclusive factorial moments at $\sqrt{s}=$ 34.5 GeV ) with 
results of various event simulators of multiparticle production model: 
Webber, Hoyer and Lund model version 
6.2 and 6.3 (see Refs in \cite{TASSO-intermittency}). 
Those models can also roughly reproduce the experimental data. 
The BEC are not taken into account while cascade processes are 
included in them.}. 
In this sense, the parameter $\delta y$ in our model may 
be interpreted as an effective correlation length in the 
rapidity space due to the BEC. 
Our results show that the BEC considerably contribute to the 
increase of the scaled factorial 
moments(the so-called intermittency).
However, at $\sqrt{s}=91.2$ GeV our results of both exclusive and 
inclusive factorial moment systematically underestimate 
compared with experimental data for all orders. 
This underestimation becomes remarkable as 
the order of the moment increases. 
It may be due to three jets events at high energy not 
taken into account in our calculation. 
At LEP energy, events with three jets take place with 
a significant probability. 
\begin{center}
----------------------------------------------\\
{\bf Figure 9.~(a)-(c)} \\ 
{\bf Figure 10.~(a)-(c)} \\ 
-----------------------------------------------\\
\end{center}

\subsection{Bose-Einstein correlations}

Here the results for the Bose-Einstein correlation functions 
are compared with data at $\sqrt{s}=34.5$\cite{TASSO-BEC}, 
and $91.2$ GeV \cite{DELPHI-BEC,ALEPH-BEC}. 
The following correlation functions $C(Q)$ (or $C(Q^2)$) are used 
for this analysis: 
\begin{eqnarray}
C(Q)=N^{\pm\pm}(Q)/N^{ref}(Q)
\qquad \mbox{or} \qquad 
C(Q^2)=N^{\pm\pm}(Q^2)/N^{ref}(Q^2), 
\end{eqnarray}
where $N^{\pm\pm}(Q)~$(or$~N^{\pm\pm}(Q^2)$) is the number of 
like sign pairs with 
four momentum difference:~$Q^2=(p_1-p_2)^2$ or $Q=\sqrt{Q^2}$. 
The so-called reference 
sample, the denominator $N^{ref}$ is the number of pairs without 
Bose-Einstein correlations. TASSO and ALEPH collaboration use the 
number of unlike sign pairs as $N^{ref}$. 
`Mixed pair reference' is also used by ALEPH. 
In this case, a pair is taken from different events. 
Our results of both the $\pi$- and the 
$\rho$- models are shown in 
Figs. 11 and 12, respectively. 
We also present energy dependence of the BEC correlation function 
$C(Q)$ in Figs.13(a) for the $\pi$-model and (b) for the 
$\rho$-model.
\begin{center}
-----------------------------------------------\\
{\bf Figure 11.~(a),(b)} \\ 
{\bf Figure 12.} \\ 
-----------------------------------------------\\
\end{center}

Our model produces a characteristic enhancement of $C(Q)~(C(Q^2))$ 
in small $Q~(Q^2)$ region. 
The result of the $\rho$-model is in better agreement 
with experimental data than the $\pi$-model for the whole region 
of $Q^2$. 
We found that the enhancement of the correlation 
function $C(Q)~(C(Q^2))$ strongly depends on the size of $\delta y$. 
When $\delta y$ becomes large, the slope of the 
correlation function becomes small. If $\delta y$ is larger 
than the rapidity interval for the full phase space, i.e. 
$\delta y \ge  2y_{\mbox{\scriptsize max}}$, the correlation function 
has no $Q$ dependence, $C(Q) \equiv 2$. 
On the contrary when $\delta y$ approaches 
to zero, the slope of the correlation function becomes 
large and $C(Q)$ approaches to unity for $Q>0$. 

Although our model has no explicit information on the 
space-time structure of particle 
emission points, we can extract the size and the lifetime 
of the particle sources by fitting a theoretical 
formula to the result with mixed pair reference from our 
simulation. Here we use the following fitting formula for $Q>$5 MeV/c: 
\begin{eqnarray} 
&&C(Q)=~c[1+\lambda \exp(-RQ)],\\
&&C(Q^2)=~c[1+\lambda \exp(-R^2Q^2)], 
\end {eqnarray} 
where $c$ is the normalization constant, $\lambda$ is the so-called 
chaoticity parameter, and $R$ is the size parameter. 
We don't use data for $Q<$5 MeV/c in the  fitting because the 
statistical errors are very large there. 
\begin{center}
------------------------------------------------------\\
{\bf Figure 13.~(a) and (b)}. \\ 
------------------------------------------------------\\
\end{center}

In Table I$\!$I$\!$I, the results of the fitting 
at various energies are summarized for both the 
$\pi$-model and the $\rho$-model.
With increasing energy of the system, 
$R$ decreases while $\lambda$ tends to increase. 
We find that $R$ in the $\rho$-model is larger than that 
in the $\pi$-model. 
We also try to extract the longitudinal size 
$R_{{\mbox \scriptsize L}}$, transverse size 
$R_{{\mbox \scriptsize T}}$, and 
lifetime $\tau$ by using the following formulae in our fit:. 
\begin{eqnarray}
C(Q_{{\mbox \scriptsize L}})\!\!&=&\!\!~
c_{{\mbox \scriptsize L}}[1+\lambda_{{\mbox \scriptsize L}} 
\exp(-R_{{\mbox \scriptsize L}}^2Q_{{\mbox \scriptsize L}}^2)],\quad 
C(Q_{{\mbox \scriptsize T}})=
c_{{\mbox \scriptsize T}}[1+\lambda_{{\mbox \scriptsize T}}
\exp(-R_{{\mbox \scriptsize T}}^2Q_{{\mbox \scriptsize T}}^2)],\nonumber \\  
&& C(\Delta E)=c_{\tau}[1+\lambda_{\tau} \exp(-\tau^2\Delta E^2)], 
\end{eqnarray} 
where $Q_{{\mbox \scriptsize L}}^2=
|p_{{\mbox \scriptsize L1}}-p_{{\mbox \scriptsize L2}}|^2$, 
$Q_{{\mbox \scriptsize T}}^2=
|\vec{p}_{{\mbox \scriptsize T1}}-\vec{p}_{{\mbox \scriptsize T2}}|^2$ 
and $\Delta E=|\epsilon_1-\epsilon_2|$. 
The results are presented in Table I$\!$V. 
The source size decreases with increasing $\sqrt{s}$, 
while the life time is almost independent of $\sqrt{s}$. 
It is reasonable that $C(Q_T)$ has no strong 
$Q_{{\mbox \scriptsize T}}$-dependence because the BEC is not 
taken into account in the transverse momentum space in our model. 
\begin{center} 
------------------------------------------ \\
{\bf Table I$\!$I$\!$I and I$\!$V} \\
------------------------------------------ \\
\end{center} 

\vspace*{5mm} 

The value of the normalization constant $c$ may be understood as follows. 
The number of unlike charged pairs and that of (++) pairs 
for large $Q$ may be proportional to (statistical) combinatorial numbers 
$\langle n_+n_-\rangle$ and 
$\langle n_+(n_+ -1) \rangle$, respectively. 
Therefore the asymptotic value of the correlation function 
may be estimated as 
\begin{eqnarray}
      \lim_{Q\to\infty}C(Q)&\equiv& ~~c \\ 
    &\cong& 
      \frac{\langle n_+ ( n_+-1)\rangle}
           {\langle n_+ n_-\rangle } \nonumber \\
  &\cong& 1- \frac{\langle n_{+} \rangle}
                  {\langle n_+^2 \rangle}~
\label{BEC-NOR}
\end{eqnarray} 
where we have used charge conservation in deriving the 
second line. 
Eq.~(\ref{BEC-NOR}) with the approximation 
$\langle  n_+^2 \rangle \sim \langle  n_+ \rangle^2$ and 
the experimental values of 
$\langle n_{ch}\rangle (=2\langle  n_+ \rangle)$ yields 
$c$=0.785, 0.855 and 0.901 for 
$\sqrt{s}$=14.0, 34.5 and 91.2 GeV,respectively 
in qualitative agreement with the result given in Table I$\!$I$\!$I.

We also investigate the influence of the charge and 
energy-momentum conservation on the Bose-Einstein correlations. 
Both energy-momentum conservation and charge conservation reduce the 
values of the correlation functions in the whole 
$Q$ (or $Q^2$) range as shown in Fig.14. 
The effect of charge conservation is stronger 
than that of energy-momentum conservation. 
If ``no conservation'' events are used, 
$C(Q)$ approaches to a larger constant as $Q$ increases. 
\begin{center}
------------------------------------------\\
{\bf Figure 14.} \\ 
------------------------------------------\\
\end{center}
\noindent 

As mentioned in Section I$\!$I$\!$I B, 
the BEC in the $\rho$-model is imposed for like-charged pions 
by using the enhancement factor. 
By changing this enhancement factor, we can control the strength of 
the BEC. 
For example, one can reduce the BEC substantially by 
putting $n_i^{+}= N^{+}=0$ in eq.(\ref{ENHA}) as shown in Fig.15.
\begin{center}
------------------------------------------\\
{\bf Figure 15.} \\ 
------------------------------------------\\
\end{center}

\subsection{2-particle rapidity correlations} 
We investigate the 2-particle correlations in terms of rapidity 
variables in detail. The experimental data reported 
by TASSO Collaboration\cite{TASSO-BEC} at $\sqrt{s}$=34.5 GeV~ 
is used in comparison with our calculations. 
When the 2-particles of type ``$a$'' and ``$b$''
~( for example, a,b~=~$\pi^+$, $\pi^-$) have rapidities 
$y_1$ and $y_2$, respectively, the 
two particle rapidity correlation function 
$R^{(a,b)}_2(y_1,y_2)$ is defined as follows:
\begin{eqnarray}
R^{(a,b)}_2(y_1,y_2) \equiv \frac{\rho^{(a,b)}(y_1,y_2)}
{f~\rho^{(a)}(y_1)~\rho^{(b)}(y_2)} -1, 
\end{eqnarray} 
where $\rho^{(a,b)}(y_1,y_2)$ ~and~$\rho^{(a)}(y_1)$~
is the two particle density and the one particle density, respectively:
\begin{eqnarray} 
\rho^{(a,b)}(y_1,y_2)&=&\frac{1}{\sigma}\frac{d^2\sigma}{dy_1dy_2},\\
\rho^{(a)}(y_1)&=& \frac{1}{\sigma}\frac{d\sigma}{dy}~~. 
\end{eqnarray} 
Here $\sigma$ is the total cross section and 
$f$ is the normalization constant given by
\begin{eqnarray}
f=\frac{<n_a(n_b-\delta_{ab})>}{<n_a>~<n_b>}, 
\end{eqnarray}
where 
\begin{equation}
\delta_{ab} =\left\{
\begin{array}{ll}
1&\mbox{~~~~~~~~~~for a$= $b}\\
0&\mbox{~~~~~~~~~~for a$\ne$b}\\
0&\mbox{~~~~~~~~~~when particle type is not distinguished,}
\end{array}\right.
\end{equation} and $n_a$ and $n_b$ are the multiplicities of 
``$a$'' and ``$b$'' ,respectively. 

Instead of using $R^{(a,b)}_2(y_1,y_2)$,
in practical measurements by TASSO Collaboration, they use 
\begin{eqnarray}
R^{(a,b)}_2(y_1,y_{trigg}) \equiv 
\frac{1}{y_{\beta}-y_{\alpha}}~\int_{y_{\alpha}}^{y_{\beta}}\,\,dy_{t}
~~R^{(a,b)}_2(y_1,y_t), 
\end{eqnarray}
where $y_{trigg}$ refers to 
a rapidity interval ($y_{\alpha}$, $y_{\beta}$). 
These trigger rapidity intervals are~~
trigger I~( -5.50$\leq$ y $\leq$ -2.50),~
trigger I$\!$I~( -2.50$\leq$ y $\leq$ -1.50),~
trigger I$\!$I$\!$I~( -1.50$\leq$ y $\leq$ -0.75) and 
trigger I$\!$V~( -0.75$\leq$ y $\leq$ -0.00). 
First we investigate a case where sign of charge is not 
distinguished and hence $\delta_{ab}$=0. 
The results are compared with experimental data in Fig.16(a)-(d). 
The experimental data are approximately reproduced by the simulation for 
every trigger regions. 
Although our model predicts a negative correlation 
for large $y$ ($y\ge$2 ) in the trigger~I, 
a positive correlation is reported by TASSO. 
For the trigger regions I$\!$I, I$\!$I$\!$I and I$\!$V, calculated 
correlation functions show positive bumps around the trigger 
rapidity interval in agreement with experiment. 
The effect is apparently caused by the BEC. 
\begin{center}
------------------------------------------\\
{\bf Figure 16(a)-(d)}\\ 
------------------------------------------\\
\end{center}

For confirmation, we calculate the 
2-particle rapidity correlation 
$R^{++}(y,y_{trigg})$ and $R^{+-}(y,y_{trigg})$. 
See Fig.17(a)-(d).
The clear positive correlations are seen near the trigger rapidities 
in the correlation function $R^{(++)}(y_1,y_2)$. 
On the other hand, such positive correlations are absent in 
$R^{(+-)}(y_1,y_2)$. 
From these investigations, one can conclude that 
the BEC (identical particle effect) plays a crucial role in 
an enhancement of the $R^{++}(y,y_{trigg})$ 
near the trigger rapidity $y_{trigg}$. 
The correlation functions are also affected by 
the effects of $\rho$ meson production and their decay.
Results of the $\rho$-model are shown in Fig.18. 
It is found that the $\rho$-model gives better agreement with 
experimental data than the $\pi$-model for all 
trigger rapidity intervals.
\begin{center}
------------------------------------------\\
{\bf Figure 17~(a)-(d)}\\ 
------------------------------------------\\
\end{center}

There are some discrepancies between the model result for 
$R(y,y_{trigg})$ and the data as most evidently seen in Fig.16(a). 
The situation is qualitatively the same in both the 
$\pi$- and the $\rho$- models.
This is an indication that there are unknown long range as well as 
short range correlations besides BEC, e.g. those due to higher resonances 
and/or three jets.
\begin{center}
------------------------------------------\\
{\bf Figure 18.~(a)-(d)} \\ 
------------------------------------------\\
\end{center}

In analogy to $C(Q)$, another 2-particle rapidity correlation 
function can be defined by using the rapidity deference 
$\Delta y=|y_1-y_2|$ in place of $Q$: 
\begin{eqnarray} 
     {\tilde C}(\Delta y) \equiv \frac{N^{++}(\Delta y)}{N^{ref}(\Delta y)}. 
\end{eqnarray} 
This function allows us to measure the size of the model 
parameter $\delta y$. We expect that slope of ${\tilde C}(\Delta y)$ 
will change at $\Delta y \approx \delta y $. 
In this paper, we assume that $\delta y$ is independent of $y$. 
This assumption can be checked by measuring ${\tilde C}(\Delta y)$ 
in various rapidity regions. 
We strongly urge experimentalists to measure ${\tilde C}(\Delta y)$. 
Our prediction for this correlation function is shown in Fig.19(a)-(c). 
\begin{center}
------------------------------------------\\
{\bf Figure 19.}\\
------------------------------------------\\
\end{center}
\subsection{Forward-backward correlations} 
Finally we compare the results of the forward-backward correlation 
with experimental data\cite{TASSO-1,DELPHI-FBC}. 
It is known experimentally that there are weak (positive) 
correlations between 
the multiplicity $n_B$ in the backward hemisphere and the mean 
multiplicity $\langle n_F\rangle$ in the forward hemisphere. 
The direction of the ``forward'' or ``backward'' 
is assigned at random in the simulation. 
The comparison of our results with experimental data is shown in Fig.20. 
It is found that our model reproduces well experimental data 
for every $\sqrt{s}$. 
This correlation is usually parameterized by a linear form: 
\begin{eqnarray}
\langle n_F \rangle =a+b~n_B .  
\end{eqnarray} 
The values of the parameters $a$ and $b$ extracted from 
the $\pi$-model and the corresponding experimental values 
are given in Table~V. 
\begin{center}
------------------------------------------\\
{\bf Figure~20.} \\
{\bf Table V}\\
------------------------------------------\\
\end{center}

The correlation strength $b$ depends on the size of $\delta y$. 
When $\delta y$ becomes large, 
the correlation strength $b$ also becomes large. 
When the multiplicity distribution is extremely narrow, 
the correlation strength $b$ is negative. For example, 
if the multiplicity distribution is delta function type, 
$P(n_{ch})= \delta(n_{ch}-\langle n_{ch}\rangle)$, 
the correlation strength $b$ is $-1$. Note that 
the dispersion of 
the multiplicity distribution is also sensitive 
to the size of $\delta y$. 
Increase of $\delta y$ enhances the dispersion of the multiplicity 
distribution. This enhancement of the dispersion also increases the 
correlation strength $b$. We found that the forward-backward correlation 
is not significantly 
affected by the energy-momentum conservation.

\section{Conclusions and discussions}
We have investigated various correlations and fluctuations observed 
in multiparticle production in $e^+e^-$ annihilation by using a new 
statistical model constructed on the basis of the
maximum entropy method. 
This model allows us to investigate the crucial roles of the 
Bose-Einstein Correlations (BEC) in multiparticle production phenomena. 
The model has three parameters $T$, $\mu$ and $\delta y$. 
The ``partition'' temperature $T$ and the ``chemical'' potential $\mu$ 
characterize the single particle spectra. The BEC is characterized 
by the third parameter $\delta y$. 
Based on this model we have constructed two kinds of event 
generators which correspond to two extreme cases. 
One case is the $\pi$-model which assumes that all pions are produced 
directly, i.e., $e^+e^- \to \pi~\pi~\cdots~$. 
The other is the $\rho$-model where it is assumed that all pions are 
produced via the decay of $\rho$ mesons, i.e., 
$e^+e^- \to \rho~\rho~\cdots\to~\pi~\pi~\cdots$. 
In the course of the event generation, energy-momentum 
and charge conservations are imposed by event selection. 
Thus we can study the effects on single particle spectra 
and correlations caused by these conservation laws.
We found that some observables are significantly affected by them.\\  

Various kinds of the experimental data on single particle spectra 
and many particle correlations are well reproduced by our model. 
We have compared the results obtained from our simulations 
with experimental 
data on multiplicity distributions, rapidity distributions and 
rapidity dependence of the mean transverse momentum. 
For many particle correlations, 
we compare our calculations with experimental data 
on factorial moments, Bose-Einstein correlations, 
2-particle rapidity correlations and forward-backward correlations. 
We have also calculated the charge correlations in the rapidity 
space. We hope experimentalists will measure this observable.\\ 

Our results for all observables on correlations, 
including the dispersion of the multiplicity distributions, exhibit a strong 
$\delta y$ dependence. We found that 
$\delta y$ plays an essential role in explaining the behavior of 
some 2-particle correlation functions, scaled factorial moments and so on. 
Once the value of the parameter $\delta y$ is determined 
by fitting our result on the rapidity dependence of the dispersion 
of the multiplicity distribution to that of the experimental data, 
other correlation and fluctuation data are systematically 
reproduced by using the same $\delta y$ value. 
The parameter $\delta y$ may correspond to an effective correlation 
length (in the rapidity space) of the BEC. 
Constancy of the size of the $\delta y$ in the whole rapidity space is 
the most characteristic point in our model. This simple assumption may 
be checked by measuring the correlation ${\tilde C}(\Delta y)$ 
at various rapidity regions.
There are possibilities that it depends on the rapidity $y$ 
and/or the multiplicity.\\

In Bose-Einstein correlations and 2-particle rapidity correlations, 
in particular, we observe a clear difference between 
the prediction from the $\pi$-model and that from the $\rho$-model. 
We found that the 
$\rho$-model gives better agreement with those experimental data 
than the $\pi$-model. 
This means that two-particle correlations are 
sensitive to production and decay processes of resonances and 
experimental data on those observables suggest that 
there are significant contribution from resonance decay. 
No significant difference between the $\pi$-model 
and the $\rho$-model is observed in the single particle 
spectra and other correlation data.  

Finally, we would like to point out two important aspects of our model. 
First, we would like to emphasize that the BEC are incorporated on 
event-by-event basis in our model. It thus provides a useful theoretical 
tool for event-by-event analysis of the BEC. This is an important feature 
of our model not shared by any other event generators.
Second, information on the space-time structure of multiparticle 
production is apparently not contained in our statistical model. 
Nevertheless, it gives the BEC from which one can extract the 
information on the ``apparent'' source size and the chaoticity. 
It thus appears that the information on the space-time structure is 
``hidden'' in our model. The fundamental parameter $\delta y$ may have 
some relevance to the question.

\noindent 
\vspace*{1cm}

{\large \bf Table captions}
\begin{description}
\item {\bf Table I:~}  
Best fit values of $\mu$, $T$ and $\delta y$ for various $\sqrt{s}$ with 
the observed values of $\langle n_{ch}\rangle$ 
and $\langle p_{\mbox{\tiny T}}\rangle$ used in the $\pi$-model. 
\item {\bf Table I$\!$I:~}  
Values of the parameters in the $\rho$-model. 
\item {\bf Table I$\!$I$\!$I:~}
The results of $R$, $\lambda$ and $c$ in the $\pi$- and the $\rho$- models.
\item {\bf Table I$\!$V:~}
The results of $R_L$, $R_T$ and $\tau$ in the $\pi$-model. 
Corresponding results of the chaoticity parameters 
$\lambda_L$, $\lambda_T$ and $\lambda_{\tau}$ are also presented. 
\item {\bf Table V:~}
The results of the parameters which characterize the 
forward-backward correlation. The experimental values are given in the 
parentheses.
\end{description}
\vspace*{1cm}

\noindent 
{\large \bf Figure captions}
\begin{description}
\item {\bf Fig.1:~} 
Total energy of events generated for the mean total energy 
$\langle E_{tot}\rangle=\sqrt{s}=34.5$ GeV and the energy window 
of the full width 2$\delta E$=0.4$\sqrt{s}$. 
\item {\bf Fig.2:~}
Comparison of the results of the $\pi$-model with the experimental 
data on the $y_{cut}$ dependence of $D_{ch}^2/\langle n_{ch} \rangle$ 
for (a) $\sqrt{s}$=14.0 GeV, (b)~34.5GeV and (c)~91.2 GeV. 
\item {\bf Fig.3:~}
Comparison of multiplicity distributions in the $\pi$-model for 
(a) $\sqrt{s}$=14.0 GeV, (b)~34.5GeV and (c)~91.2 GeV. 
\item {\bf Fig.4~} 
(a): KNO scaling of the multiplicity distributions in the $\pi$- model 
at energies $\sqrt{s}$=14.0, 34.5 and 91.2 GeV. 
(b): Comparison of the $\pi$- model(solid line) 
with the experimental data for energy $\sqrt{s}$=91.2 GeV with 
$y_{cut}\sim$ 6.5(full phase space), 2.0 and 1.0. 
\item {\bf Fig.5:~} 
Effects of conservation laws on $y_{cut}$ dependence of 
$D_{ch}^2/\langle n_{ch} \rangle^2$. 
\item {\bf Fig.6~} 
(a):~Prediction of charge correlation $D_{ch}^2/2D_+^2$ for 
$\sqrt{s}=$ 14.0, 34.5 and 91.2 GeV. 
(b):~Effects of conservation laws on charge correlation $D_{ch}^2/2D_+^2$. 
\item {\bf Fig.7:~}
Rapidity distribution $dN/dy$ for $\sqrt{s}=$ 14.0, 34.5 and 91.2 GeV
in the $\pi$-model(open squares). Experimental data for 
 $\sqrt{s}=$ 14.0 and 34.5 GeV are shown by solid circles.
\item {\bf Fig.8~}
(a):~Rapidity dependence of the mean transverse momentum 
$\langle p_{\mbox{\tiny T}} \rangle$ for $\sqrt{s}=$ 14.0, 34.5 and 91.2 GeV. 
(b):~Effect of the thrust axis and/or the energy-momentum 
conservation for energy $\sqrt{s}=$ 34.5 GeV. 
\item {\bf Fig.9:~} 
The exclusive factorial moment for (a) $\sqrt{s}=$ 14.0, (b)~34.5 GeV 
and (c)~91.2 GeV in the $\pi$-model(solid line). Experimental data 
are shown by open circles. 
\item {\bf Fig.10:~} 
The same as in Fig.9(a)-(c) for the inclusive factorial moments 
in the $\pi$-model. 
\item {\bf Fig.11:~} 
Comparisons of the results for $C(Q^2)$ from the 
$\pi$- and $\rho$- model with the experimental data 
for (a)~$\sqrt{s}=$ 34.5 GeV and (b) $\sqrt{s}=$ 91.2 GeV. 
The number of unlike pairs is used as the reference. 
\item {\bf Fig.12:~}
The same as in Fig.11 for $\sqrt{s}=$ 91.2 GeV with the mixed pairs 
as the reference sample. 
\item {\bf Fig.13:~} 
Energy dependence of $C(Q)$ with mixed pair reference in the (a) 
$\pi$- and (b) $\rho$- models. 
\item {\bf Fig.14:~} 
Effect of conservation laws on the BEC correlation function $C(Q)$. 
\item {\bf Fig.15:~} 
The role of the enhancement factor in the $\rho$- model. 
\item {\bf Fig.16~} 
(a):~Comparison of the result from the $\pi$-model 
with experimental data on the 2-particle rapidity correlation function 
$R(y,y_{trigg})$ at $\sqrt{s}$= 34.5 GeV 
for (a) the trigger I~[-5.50,-2.50],(b)~the trigger I$\!$I~[-2.50,-1.50], 
(c)~the trigger I$\!$I$\!$I~[-1.50,-0.75] 
and (d) the trigger I$\!$V~[-0.75,0.00].  
\item {\bf Fig.17:~}
The 2-particle rapidity correlation functions $R^{++}(y,y_{trigg})$ and 
$R^{+-}(y,y_{trigg})$ in the $\pi$-model for $\sqrt{s}$= 34.5 GeV with 
(a) the trigger I~[-5.50,-2.50],(b)~the trigger I$\!$I~[-2.50,-1.50], 
(c)~the trigger I$\!$I$\!$I~[-1.50,-0.75] 
and (d) the trigger I$\!$V~[-0.75,0.00].  
\item {\bf Fig.18:~}
Comparison of the results of $\rho$- model with experimental data on the 
2-particle rapidity correlation function $R(y,y_{trigg})$ at 
$\sqrt{s}$= 34.5 GeV 
for (a) the trigger I~[-5.50,-2.50],(b)~the trigger I$\!$I~[-2.50,-1.50], 
(c)~the trigger I$\!$I$\!$I~[-1.50,-0.75] 
and (d) the trigger I$\!$V~[-0.75,0.00].
\item {\bf Fig.19:~} 
Predictions for the 2-particle rapidity correlation function 
$\tilde{C}(\Delta y)$
for $\sqrt{s}$=14.0, 34.5 and 
91.2 GeV. 
\item {\bf Fig.20:~} 
Comparison of the forward-backward correlation 
in the $\pi$-model with experimental data for $\sqrt{s}$=14.0, 34,5 
and 91.2 GeV. 
\end{description}
\onecolumn 
\begin{table}
\begin{center}
\begin{tabular}{|c|c|c|c|c|c|}
\hline
$\sqrt{s}$\,\,$\left[ GeV \right]$&
\,\,\,$\delta y $\,\,\,&
$\mu$ \,\,$\left[ GeV \right]$&
T\,\,$\left[ GeV \right]$ 
&$\langle n_{ch}\rangle$
& $\langle p_{\scriptscriptstyle T}\rangle$~[GeV]
\\ \hline
14.0&0.60&-1.71&\,2.47 &9.30&0.334 \\ \hline
22.0&0.70&-2.43&\, 4.26&11.30&0.377\\ \hline
34.5&0.80&-3.08&\,5.75 &13.59&0.422\\ \hline
43.6&0.85&-3.77&\,7.83&15.08&0.446 \\ \hline
91.2&1.20&-4.73&15.4&20.80&0.521   \\ \hline 
\end{tabular}

{\vspace*{5mm} {\bf Table I} } 
\end{center}
\end{table}
\begin{table}
\begin{center}
\begin{tabular}{|c|c|c|c|c|}
\hline
$\sqrt{s}$\,\,$\left[ GeV \right]$&
\,\,\,$\delta y_{\rho} $\,\,\,&
\,\,\,$\delta y_{\pi}$\,\,\,&
$\mu_{\rho}$ \,\,$\left[ GeV \right]$&
$T_{\rho}$ \,\,$\left[ GeV \right]$ \\ \hline
14.0&1.20&0.60&-3.61&\,3.05  \\ \hline 
34.5&1.60&0.80&-10.64&\,9.00 \\ \hline 
91.2&2.40&1.20&-17.34&\,20.27 \\ \hline
\end{tabular}

{\vspace*{5mm} {\bf Table I$\!$I} } 
\end{center}
\end{table}
\begin{table}
\begin{center}
\newcommand{\lw}[1]{\smash{\lower1.5ex\hbox{#1}}}
\begin{tabular}{|c|cc|l|c|c|c|}
\hline 
$\sqrt{s}$~[GeV]
&&model &function&$R$[fm]&$\lambda$&c\\ \hline\hline
14.0&\lw{$\pi$}&\lw{~($\delta y$=0.6)}&Gaussian&
0.56$\pm$0.02&0.46$\pm$0.02&0.84$\pm$0.00\\ 
&&&Exponential&
0.72$\pm$0.04&0.66$\pm$0.02&0.83$\pm$0.00\\ \cline{2-7}
&\lw{$\rho$}&~($\delta y_{\rho}$=1.2)&Gaussian
&0.77$\pm$.04&0.28$\pm$.01&0.85$\pm$.00\\ 
&&($\delta y_{\pi}$=0.6)&Exponential&1.17$\pm$0.09&0.43$\pm$0.03&0.85$\pm$.00
\\ \hline\hline
34.5&\lw{$\pi$}&\lw{($\delta y$=0.8)}&Gaussian&
0.50$\pm$0.02&0.58$\pm$0.02&0.90$\pm$0.00\\ 
&&&Exponential&
0.62$\pm$0.02&0.83$\pm$0.02&0.88$\pm$0.00\\ \cline{2-7}
&\lw{$\rho$}&~($\delta y_{\rho}$=1.6)
&Gaussian&0.61$\pm$0.02& 0.39$\pm$0.01&0.91$\pm$0.00\\ 
&&($\delta y_{\pi}$=0.8)&Exponential
&0.80$\pm$0.04&0.54$\pm$0.02&0.91$\pm$0.00
\\ \hline\hline
91.2&\lw{$\pi$}&\lw{($\delta y$=1.2)}&Gaussian&
0.36$\pm$0.01&0.62$\pm$0.02&0.96$\pm$0.01\\ 
&&&Exponential&
0.42$\pm$0.01&0.92$\pm$0.01&0.91$\pm$0.00\\ \cline{2-7}
&\lw{$\rho$}&~($\delta y_{\rho}$=2.4)
&Gaussian&0.48$\pm$0.01&0.34$\pm$0.01&0.96$\pm$0.00\\ 
&&($\delta y_{\pi}$=1.2)
&Exponential&0.63$\pm$0.02&0.49$\pm$0.01&0.95$\pm$0.00\\
\hline 
\end{tabular}

{\vspace*{5mm} {\bf Table I$\!$I$\!$I}} 
\end{center}
\end{table}
\clearpage 
\begin{table}
\newcommand{\lw}[1]{\smash{\lower1.5ex\hbox{#1}}}
\begin{center}
\begin{tabular}{|c|c|c||c|c||c|c||}
\hline 
$\sqrt{s}$~[GeV]
&$R_{{\mbox \scriptsize L}}$[fm]
&$\lambda_{{\mbox \scriptsize L}}$[fm]
&$R_{{\mbox \scriptsize T}}$[fm]
&$\lambda_{{\mbox \scriptsize T}}$[fm]
&$\tau$[fm]
&$\lambda_{\tau}$\\ \hline 
14.0& 0.55$\pm$0.03& 0.32$\pm$0.02& 0.29$\pm$0.16& 0.07$\pm$0.03& 
0.20$\pm$0.00& 0.20$\pm$0.01\\  
91.2& 0.34$\pm$0.01& 0.34$\pm$0.01& 0.13$\pm$0.05& 0.04$\pm$0.01& 
0.20$\pm$0.01& 0.13$\pm$0.00\\  \hline 
\end{tabular}

{\vspace*{5mm} {\bf Table I$\!$V }} 
\end{center}
\end{table}

\begin{table}
\begin{center} 
\begin{tabular}{|c|c|c|}
\hline
~~~$\sqrt{s}$~[GeV]~~~~~& ~~~$a$~~~~& ~~~$b$~~~ \\ \hline
~~14.0~~~($\delta$y=0.60)&4.36&0.054 ~~(0.085$\pm$ 0.014)\\ \hline
~~34.5~~~($\delta$y=0.80)&6.18&0.074 ~~(0.089$\pm$ 0.003)\\ \hline
~~91.2~~~($\delta$y=1.20)&9.03&0.137 ~~(0.118$\pm$ 0.009)\\ \hline
\end{tabular}

{\vspace*{5mm} {\bf Table V}} 
\end{center}
\end{table}
\clearpage 

\begin{figure}
\begin{center}\leavevmode
\epsfysize=10.0cm
\epsfbox{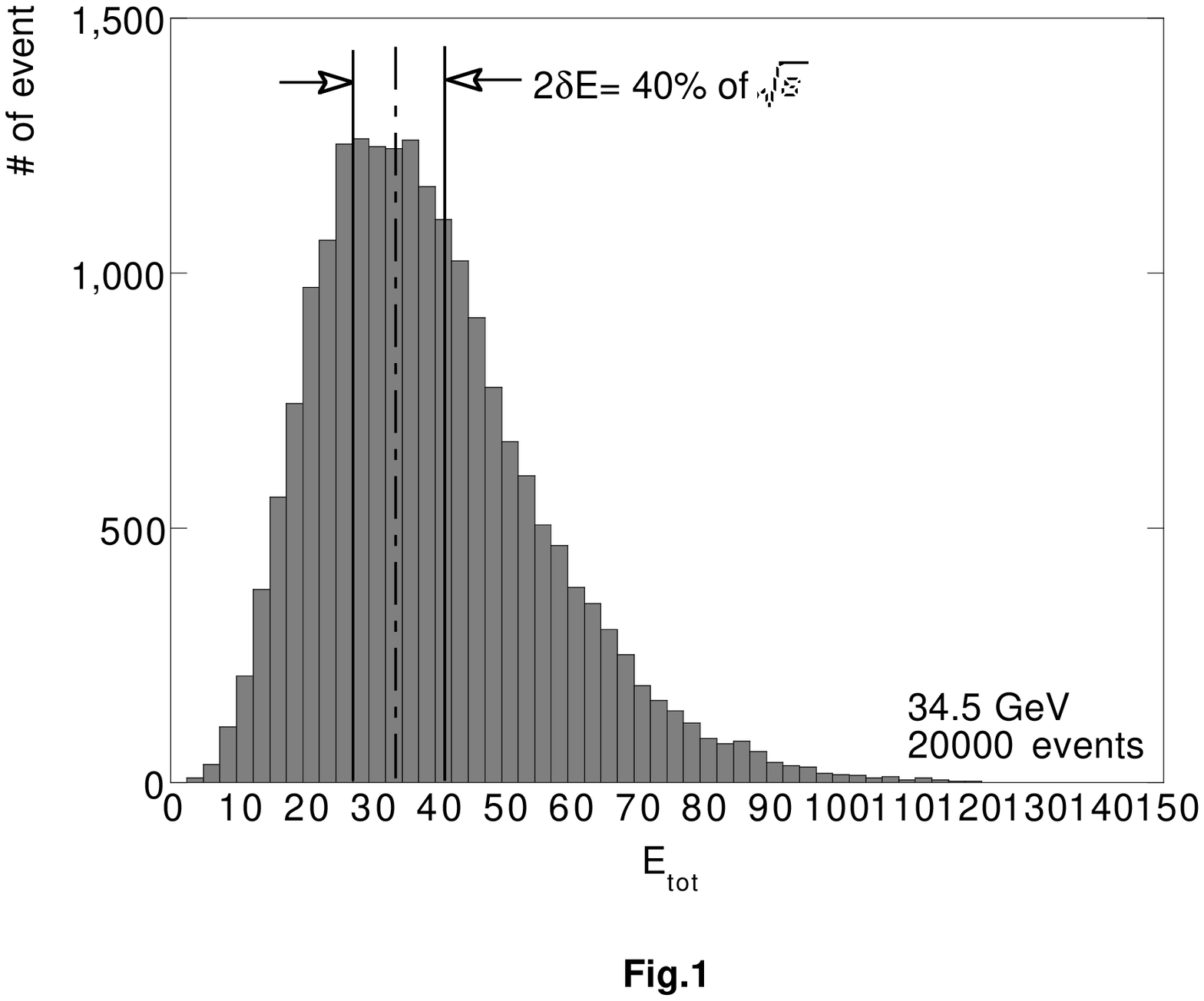}
\end{center}
\end{figure}
\begin{figure}
\begin{center}\leavevmode
\epsfysize=10.0cm
\epsfbox{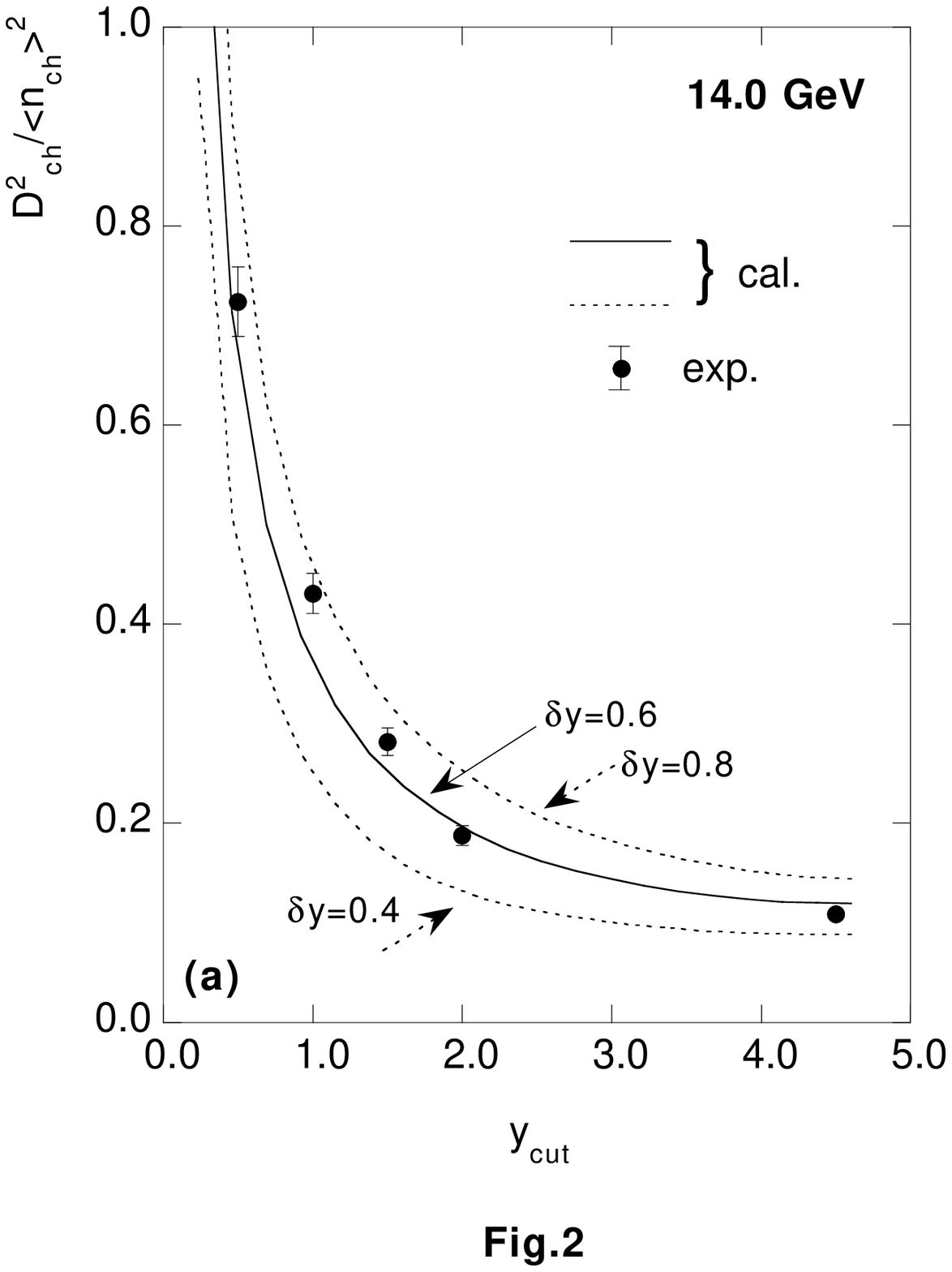}
\end{center}
\end{figure}
\begin{figure}
\begin{center}\leavevmode
\epsfysize=10.0cm
\epsfbox{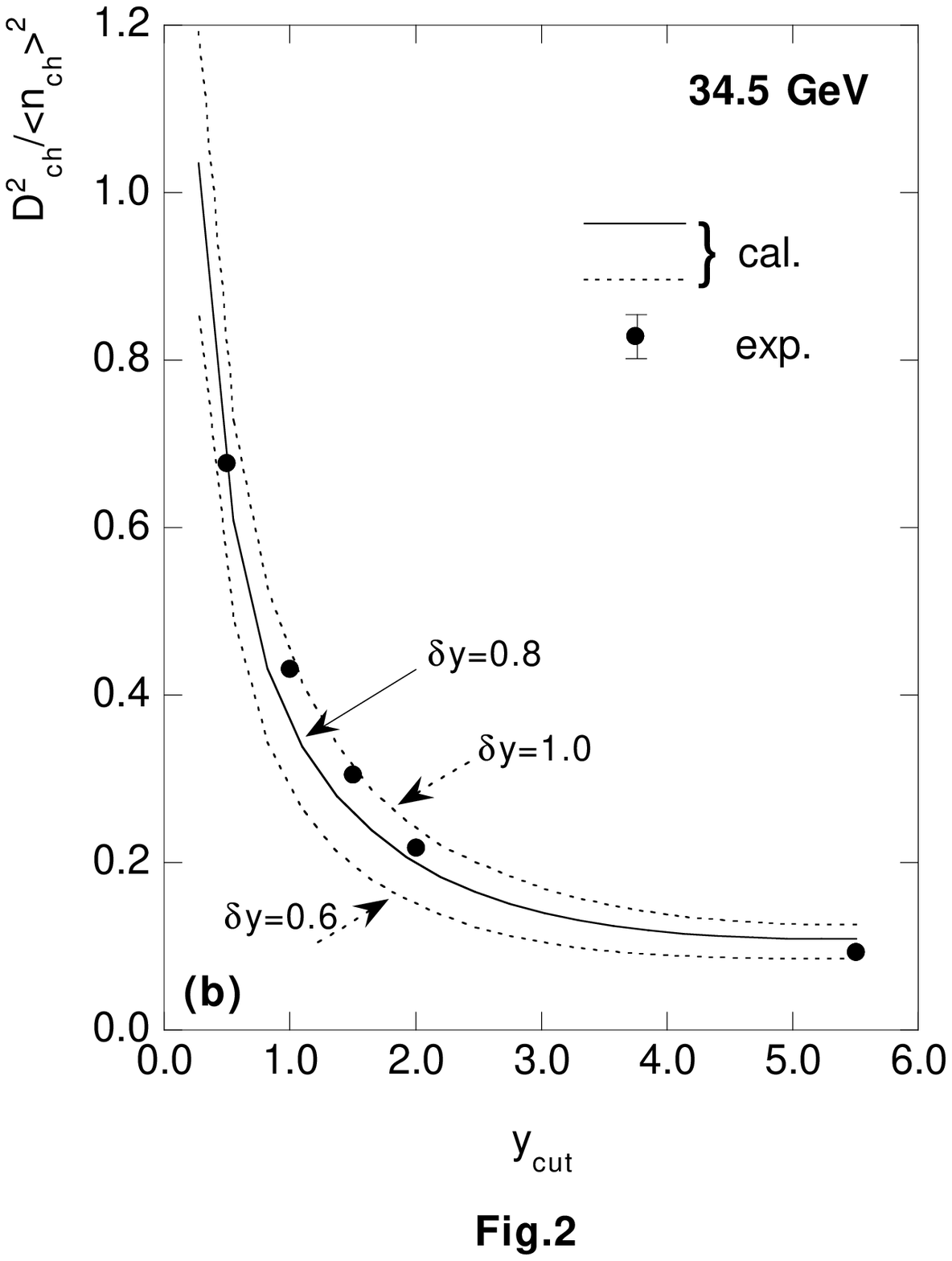}
\end{center}
\end{figure}
\begin{figure}
\begin{center}\leavevmode
\epsfysize=10.0cm
\epsfbox{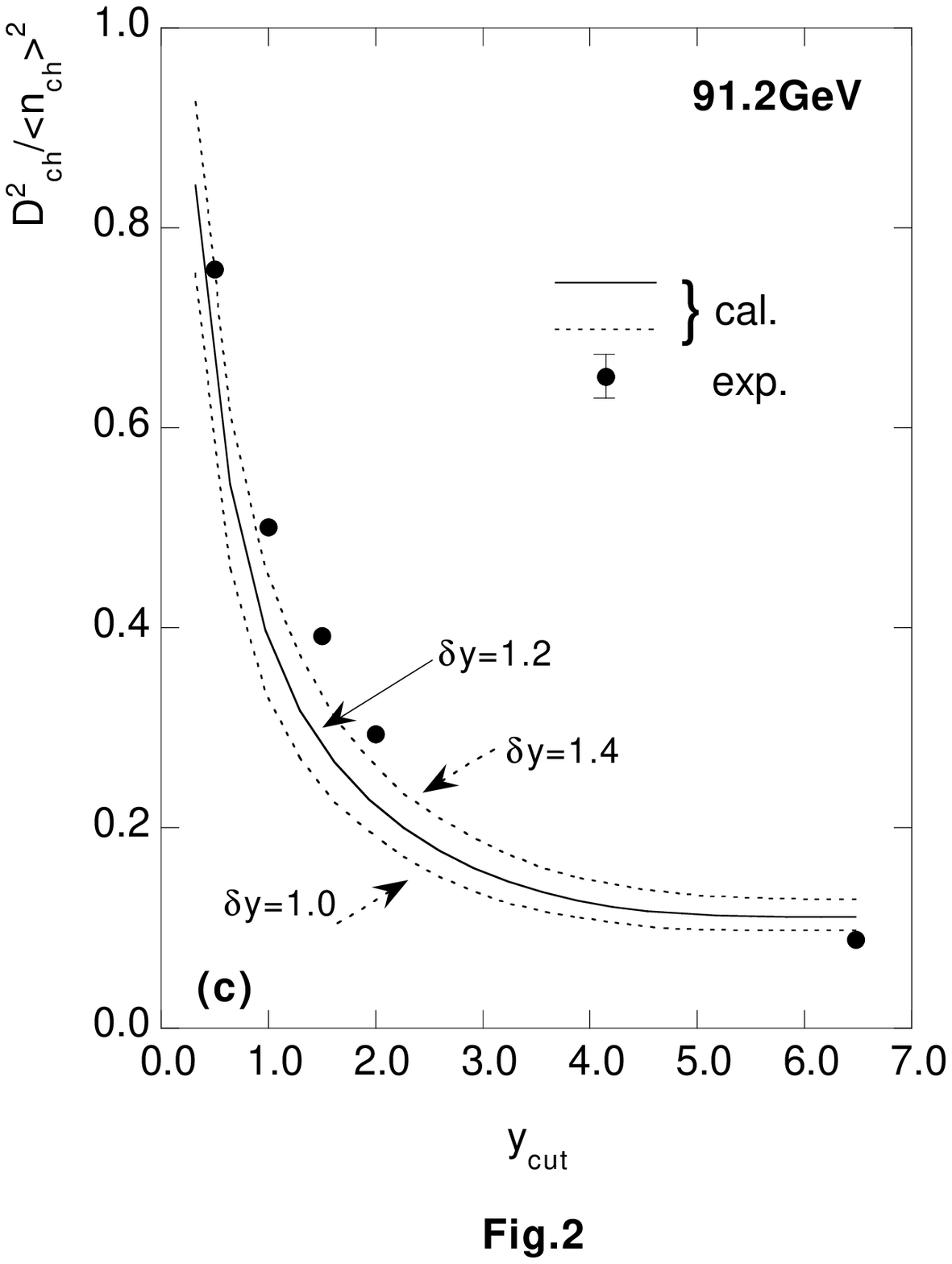}
\end{center}
\end{figure}
\begin{figure}
\begin{center}\leavevmode
\epsfysize=10.0cm
\epsfbox{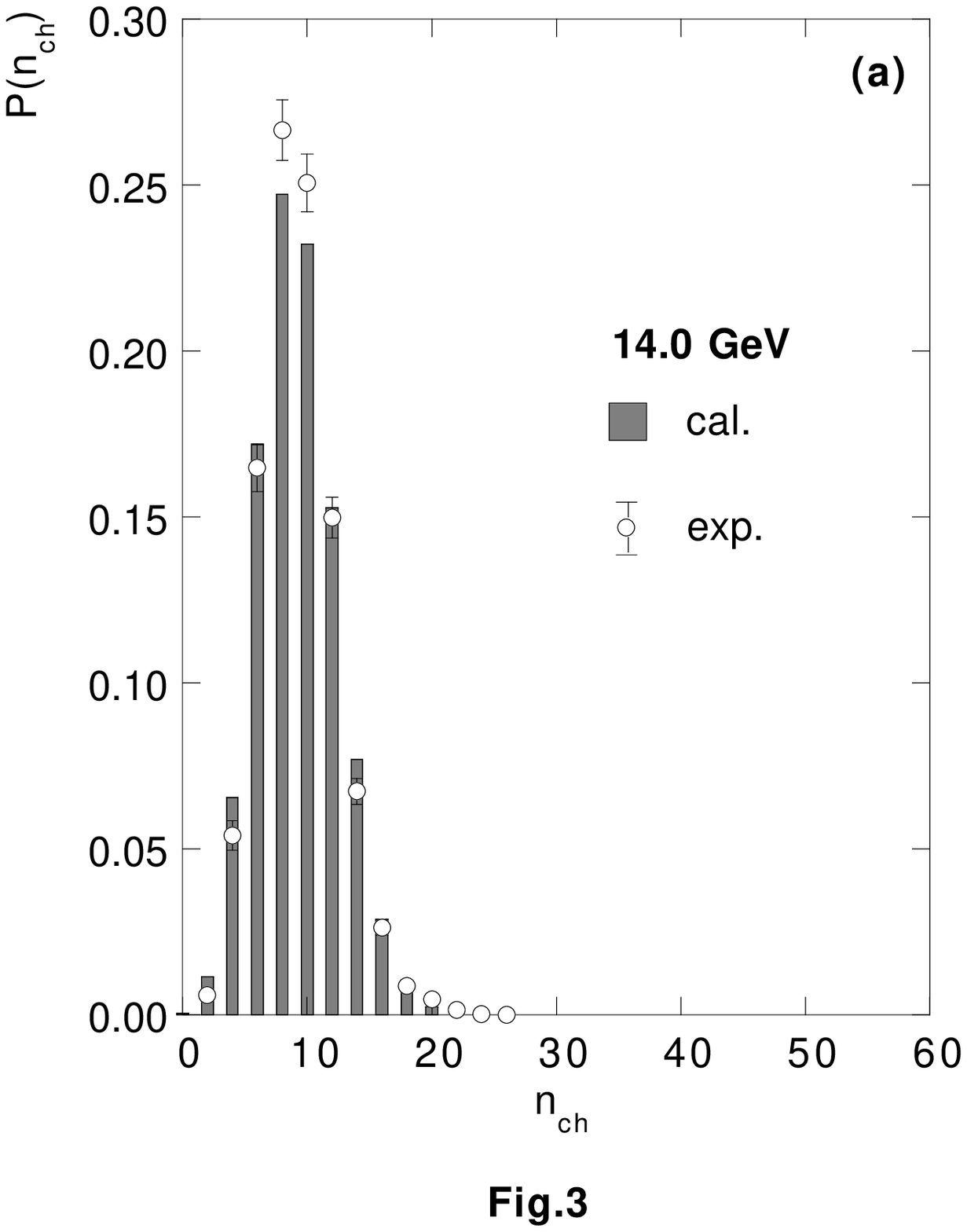}
\end{center}
\end{figure}
\begin{figure}
\begin{center}\leavevmode
\epsfysize=10.0cm
\epsfbox{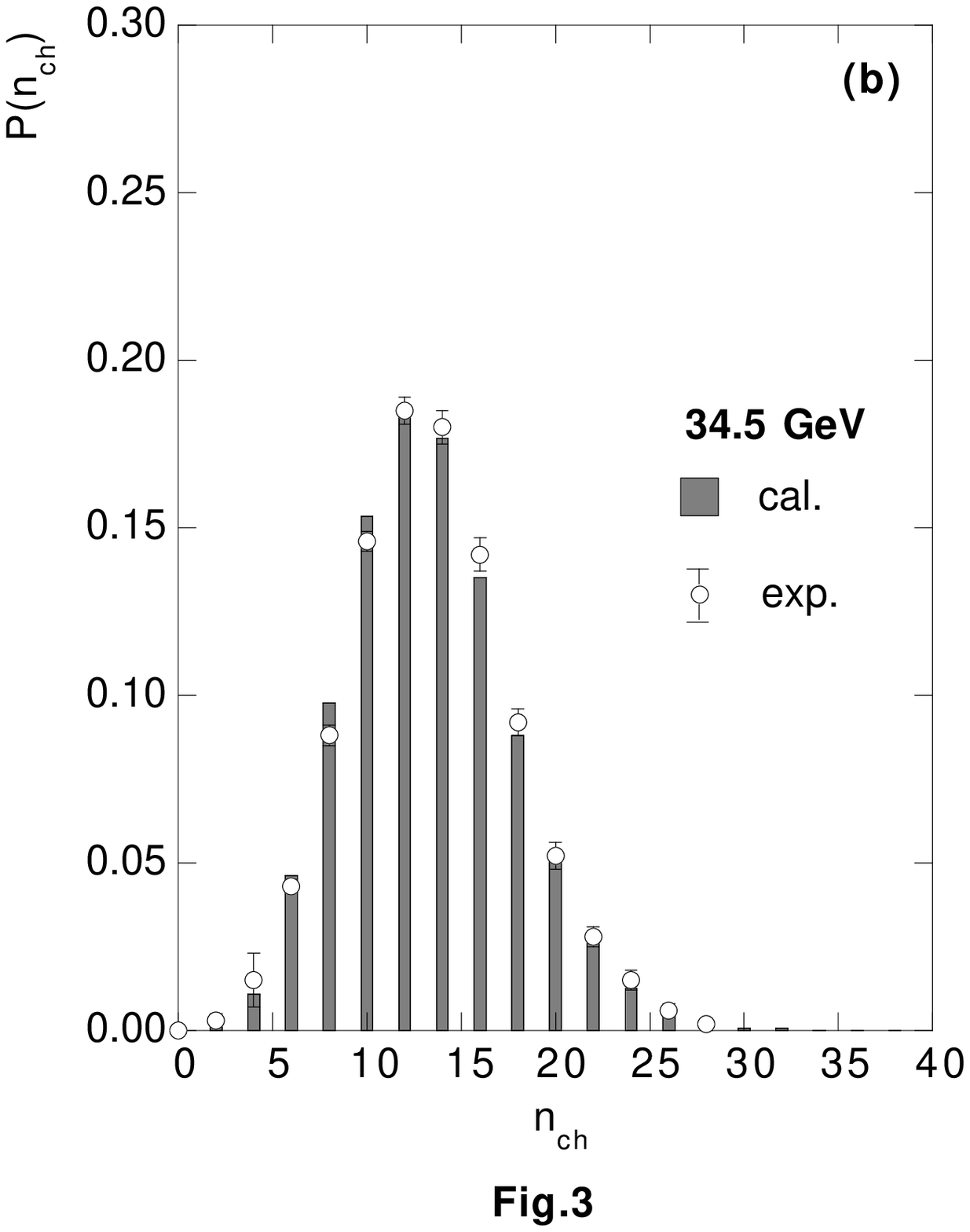}
\end{center}
\end{figure}
\begin{figure}
\begin{center}\leavevmode
\epsfysize=10.0cm
\epsfbox{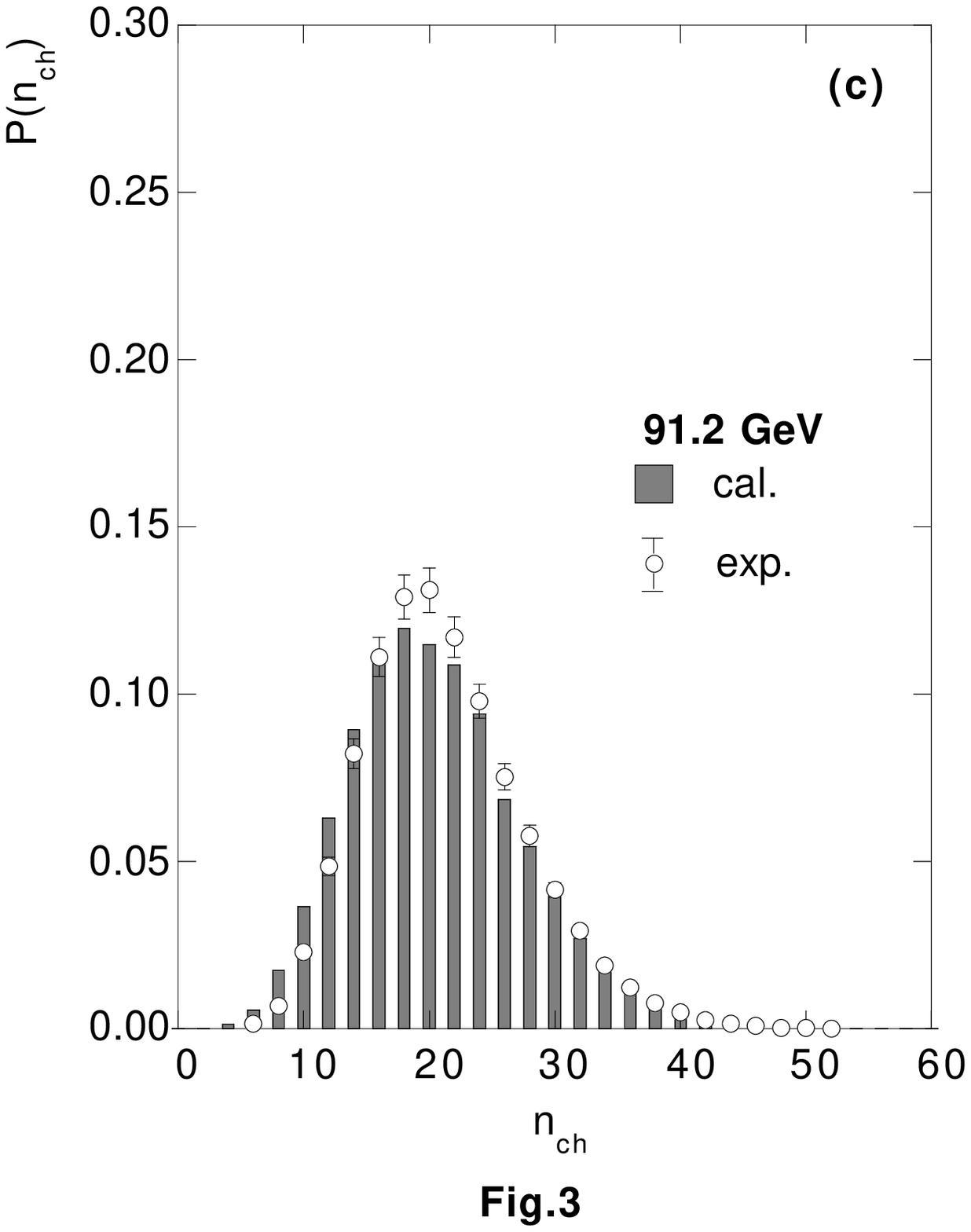}
\end{center}
\end{figure}
\begin{figure}
\begin{center}\leavevmode
\epsfysize=10.0cm
\epsfbox{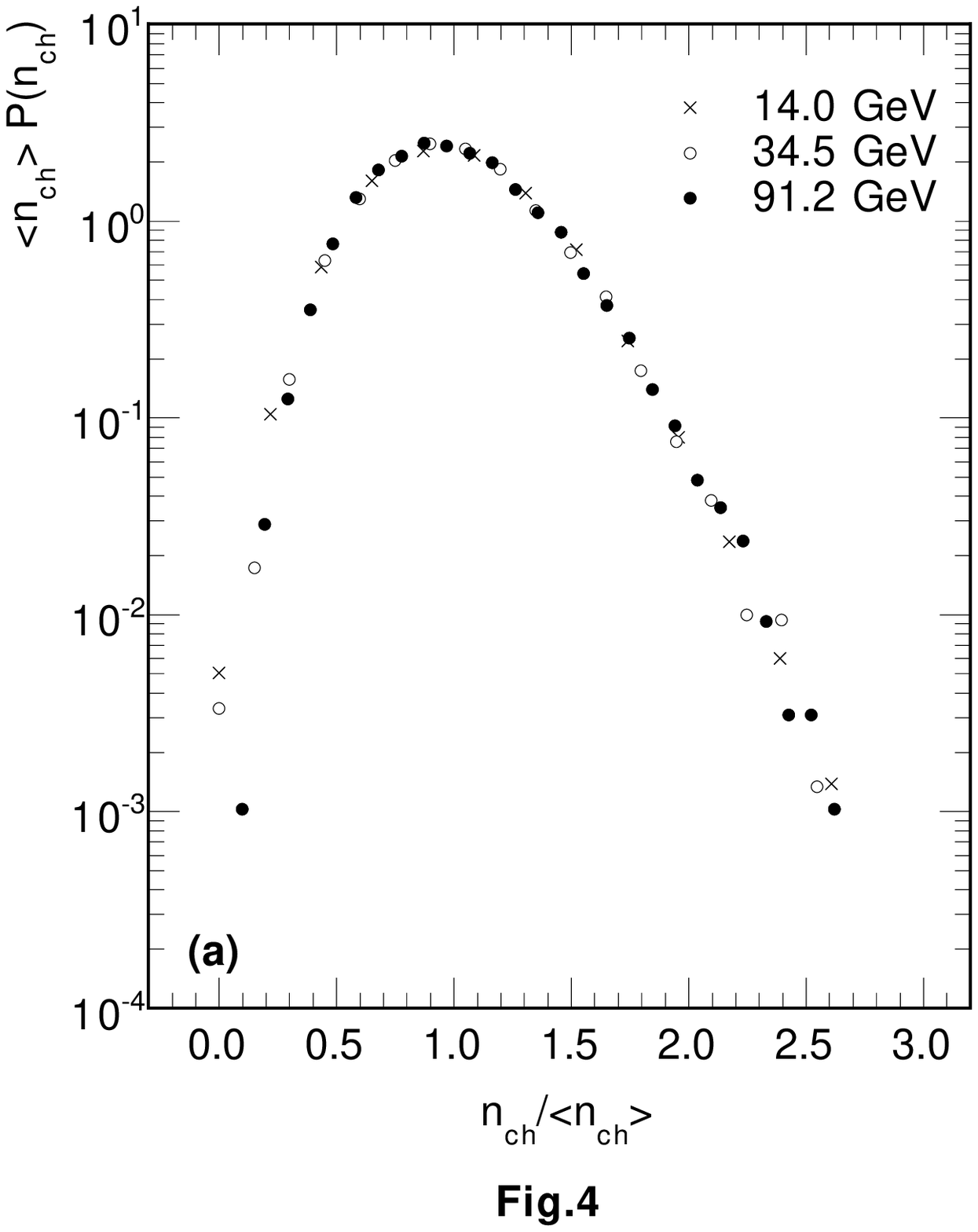}
\end{center}
\end{figure}
\begin{figure}
\begin{center}\leavevmode
\epsfysize=10.0cm
\epsfbox{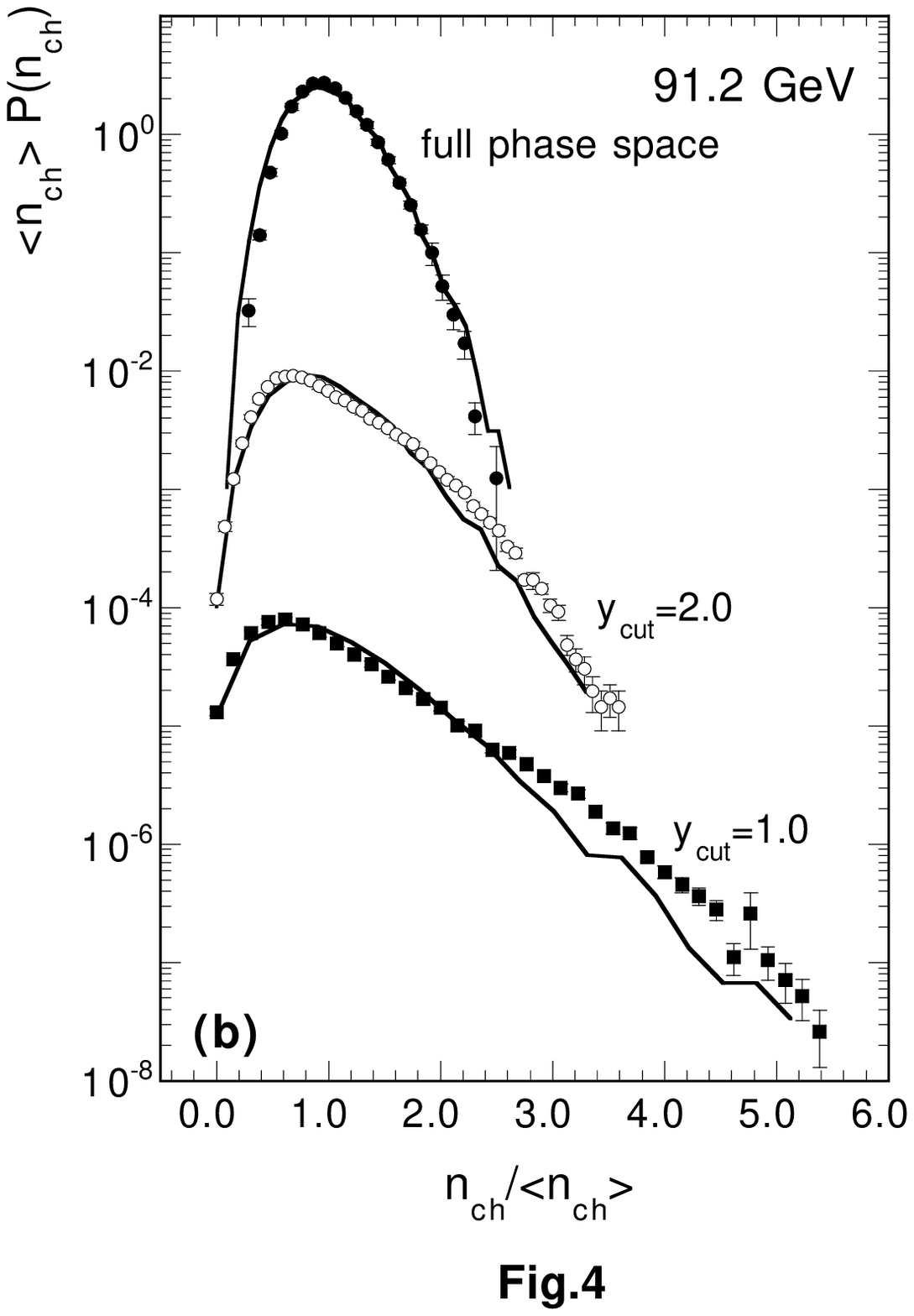}
\end{center}
\end{figure}
\begin{figure}
\begin{center}\leavevmode
\epsfysize=10.0cm
\epsfbox{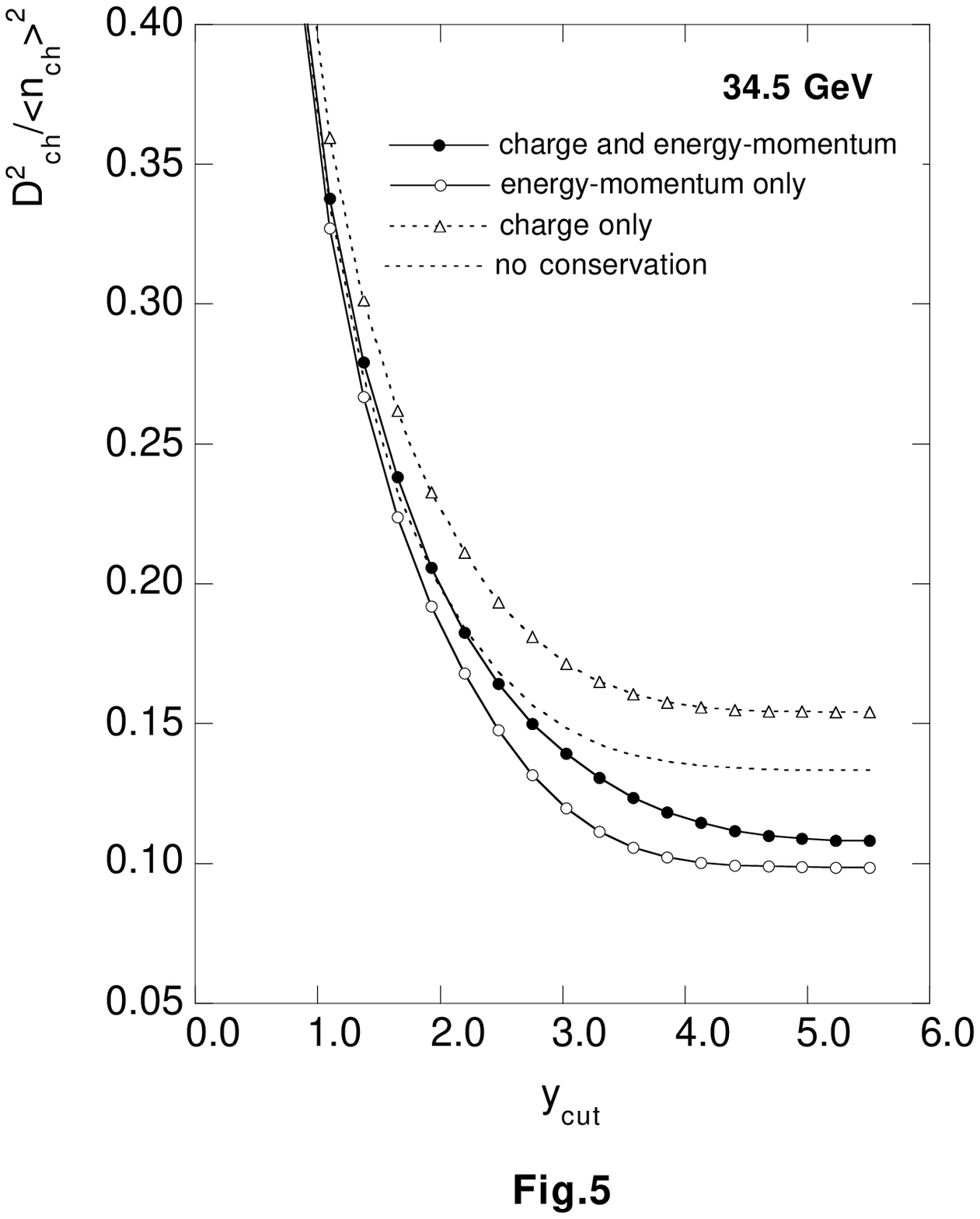}
\end{center}
\end{figure}
\begin{figure}
\begin{center}\leavevmode
\epsfysize=10.0cm
\epsfbox{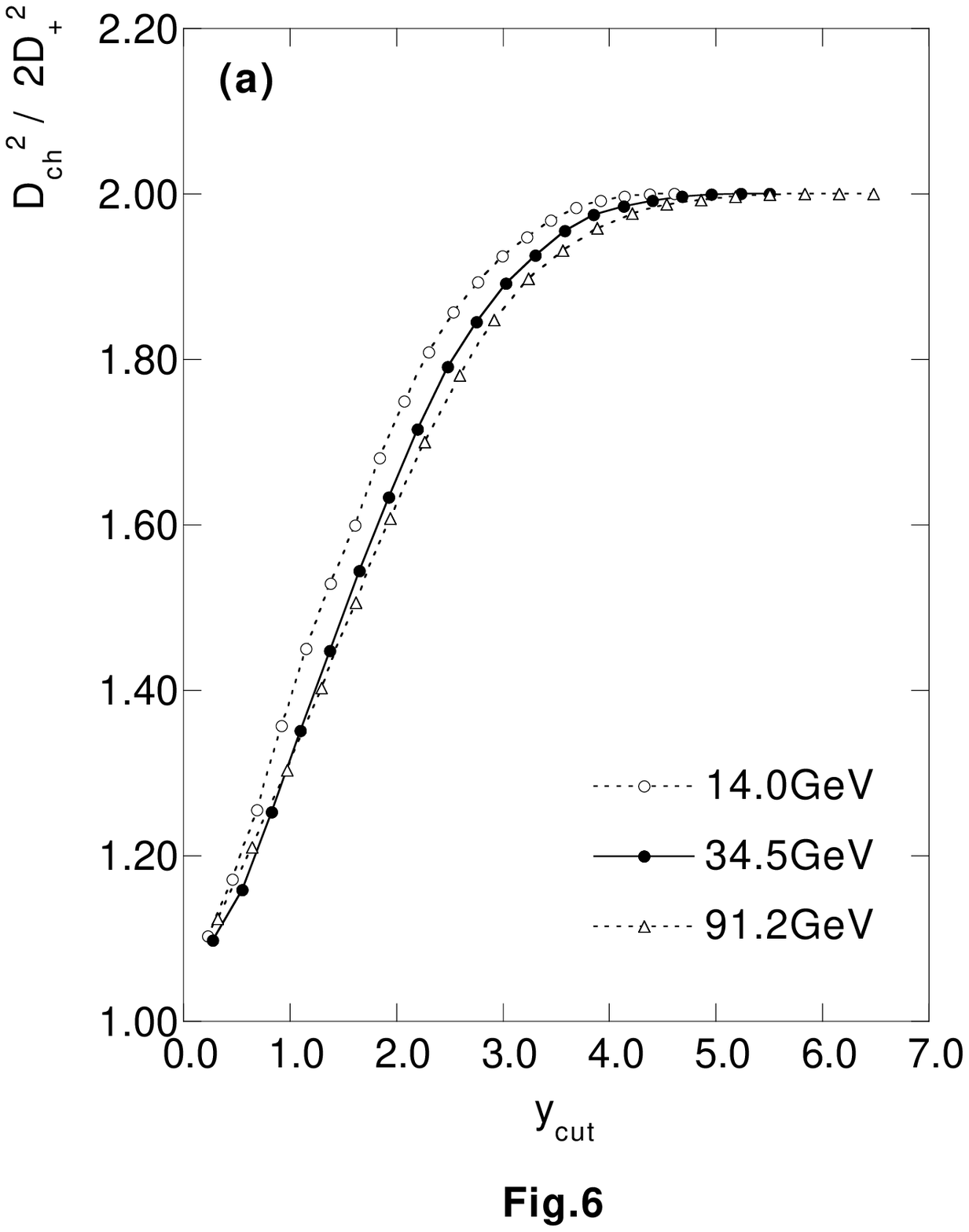}
\end{center}
\end{figure}
\begin{figure}
\begin{center}\leavevmode
\epsfysize=10.0cm
\epsfbox{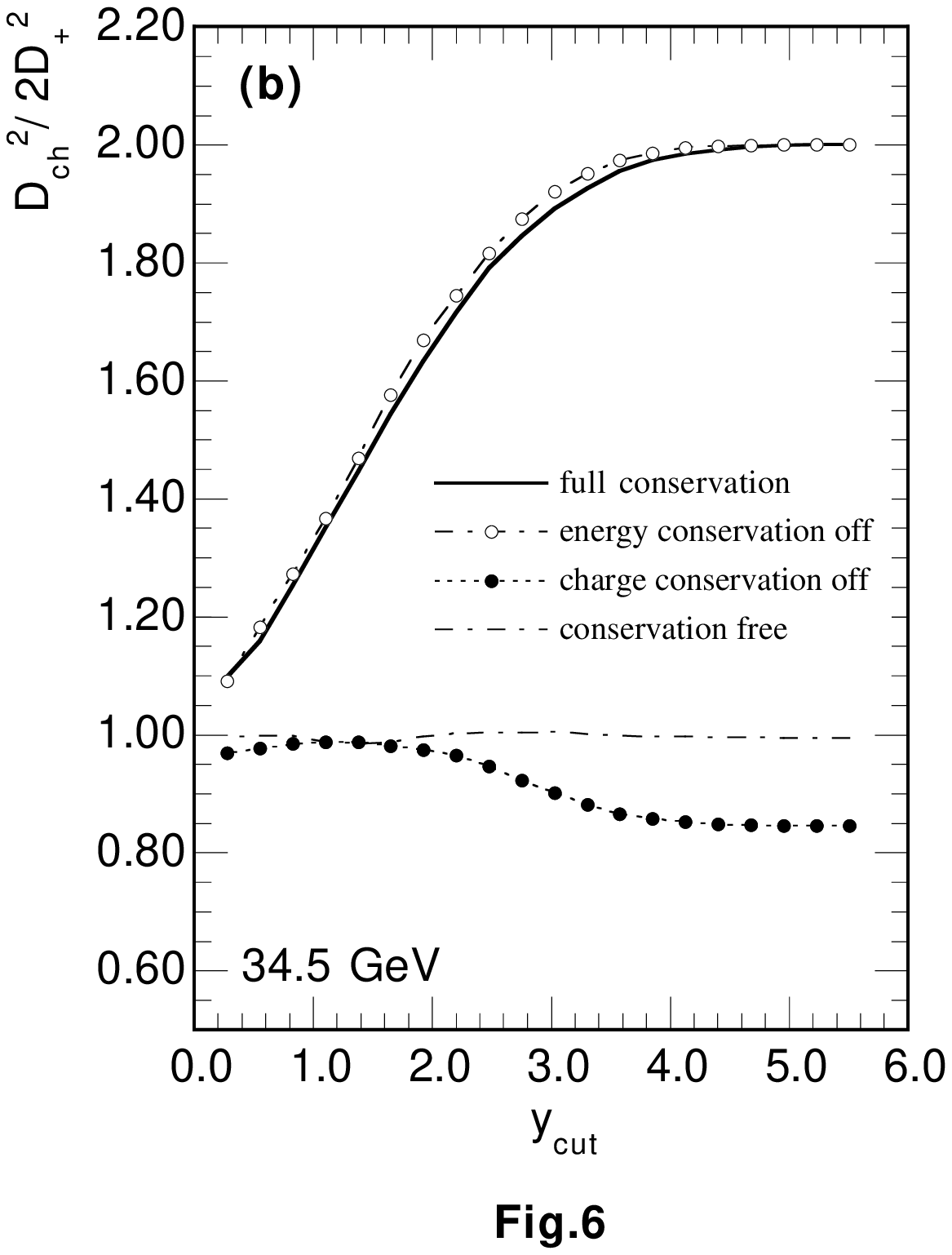}
\end{center}
\end{figure}
\begin{figure}
\begin{center}\leavevmode
\epsfysize=10.0cm
\epsfbox{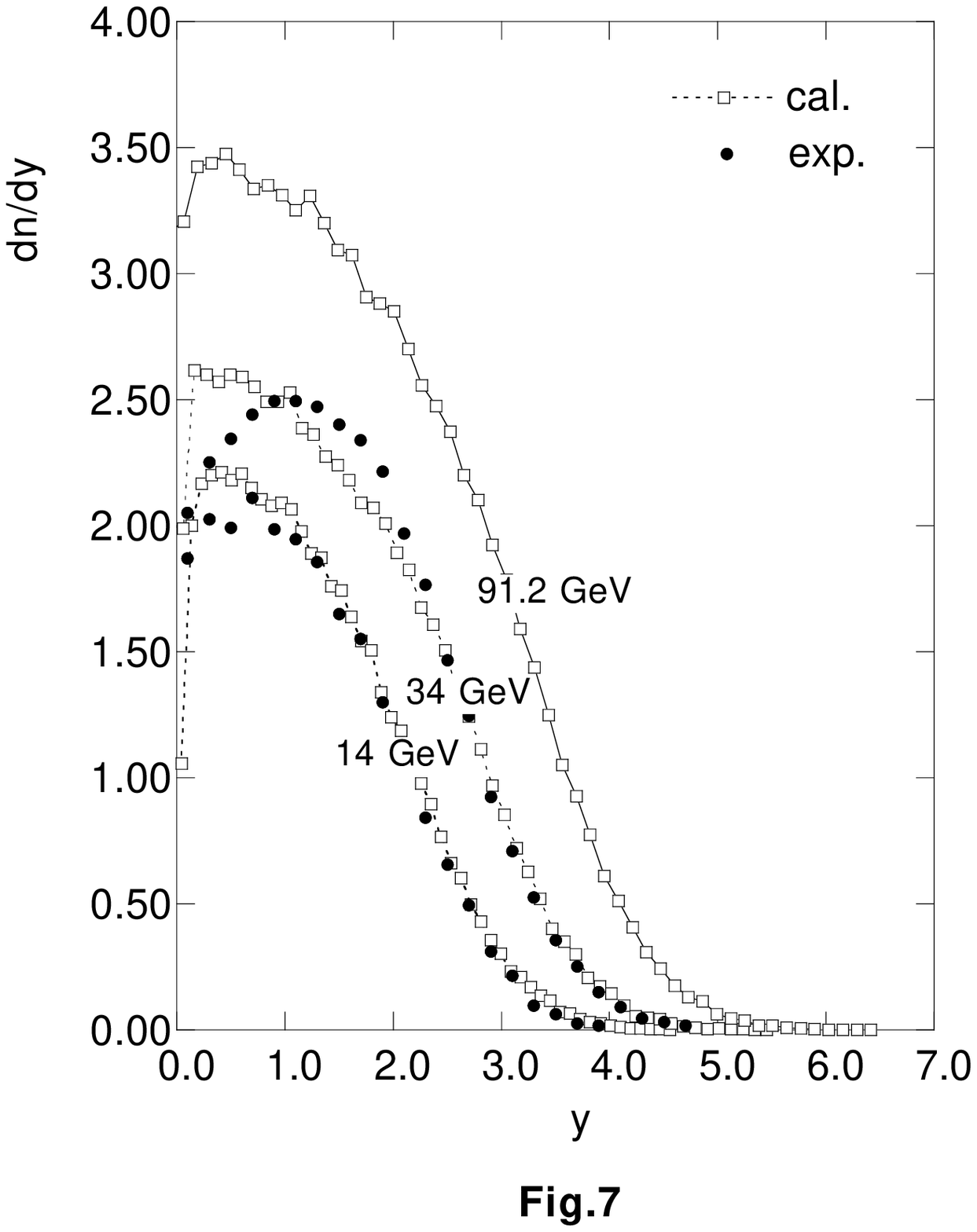}
\end{center}
\end{figure}
\begin{figure}
\begin{center}\leavevmode
\epsfysize=10.0cm
\epsfbox{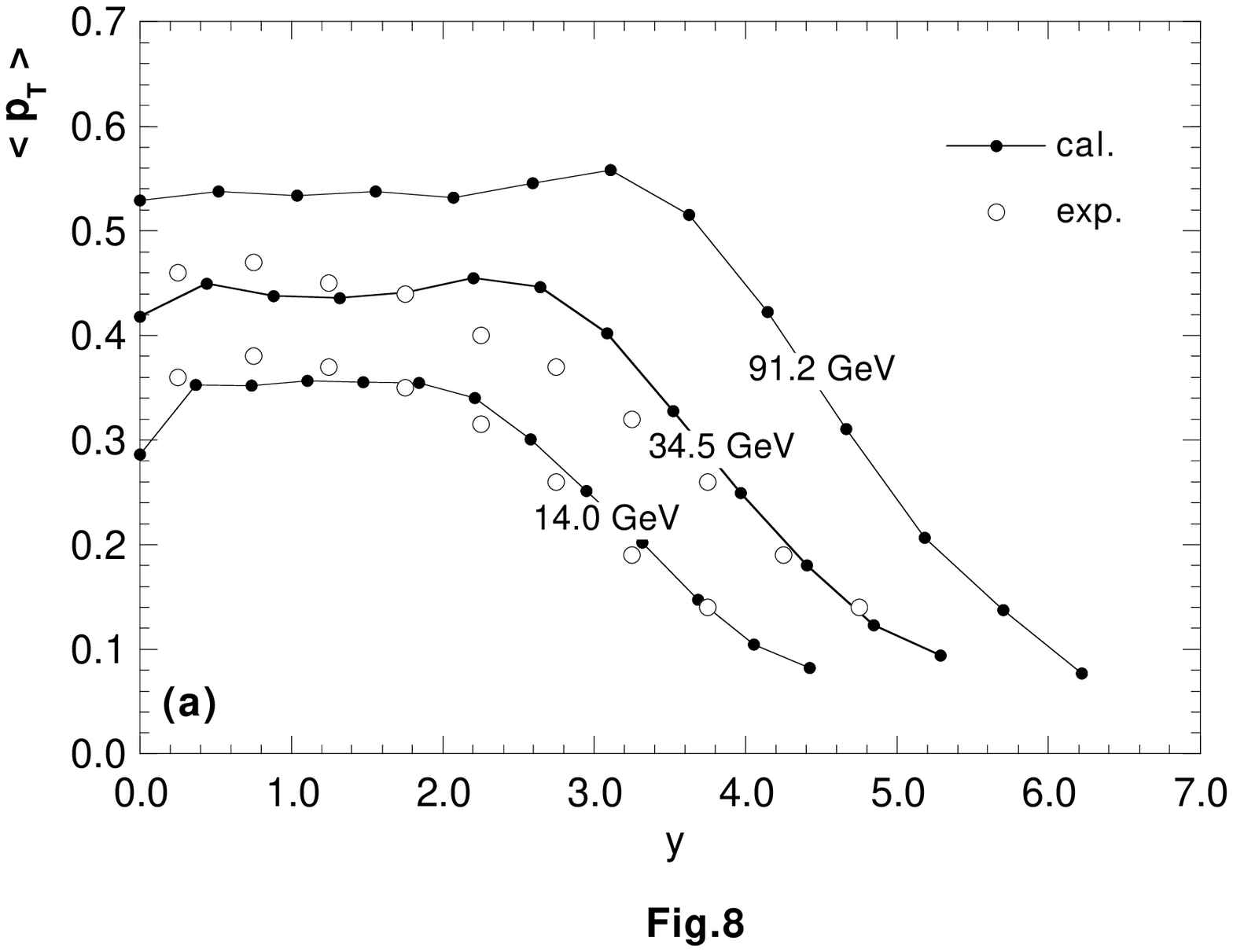}
\end{center}
\end{figure}
\begin{figure}
\begin{center}\leavevmode
\epsfysize=10.0cm
\epsfbox{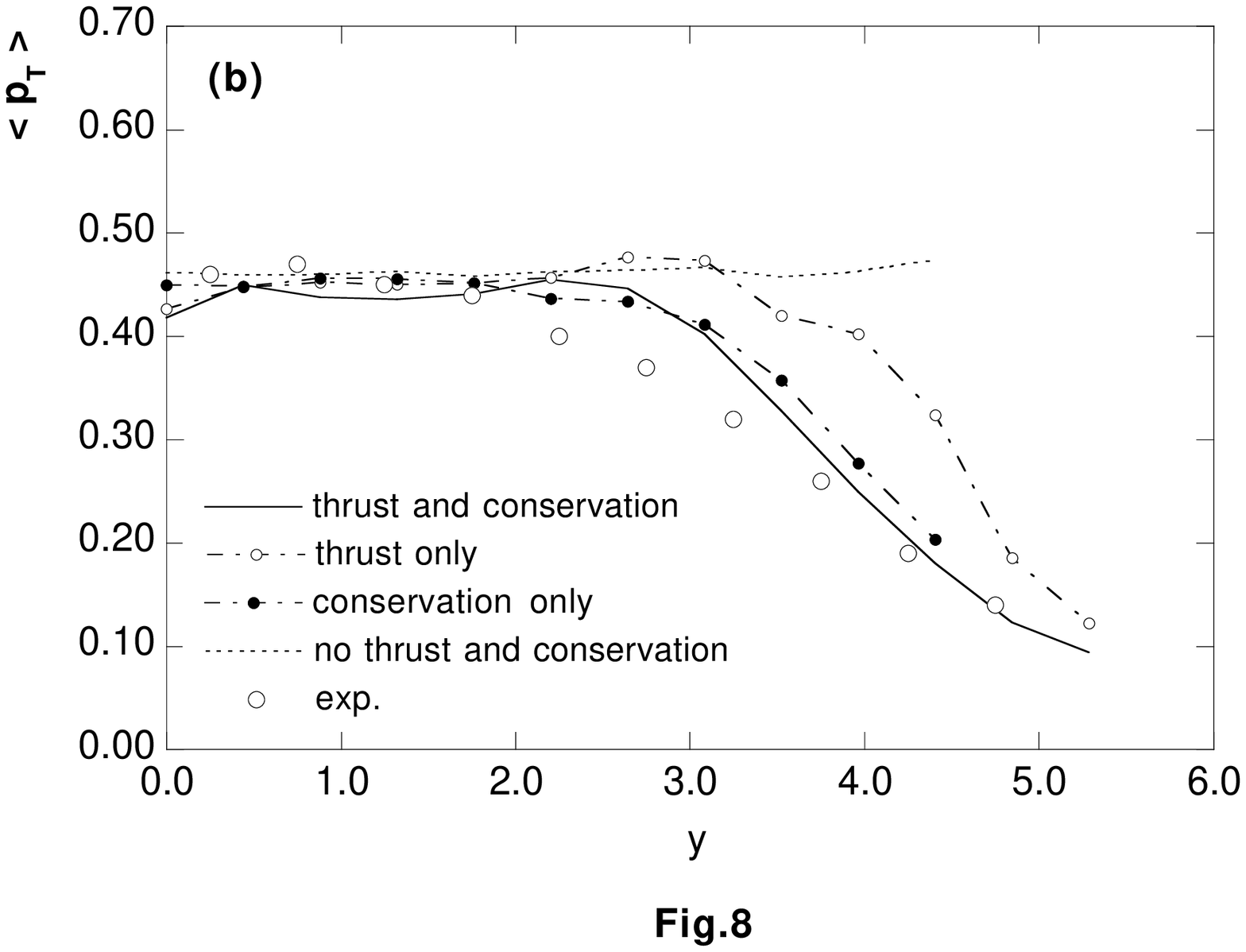}
\end{center}
\end{figure}
\begin{figure}
\begin{center}\leavevmode
\epsfysize=10.0cm
\epsfbox{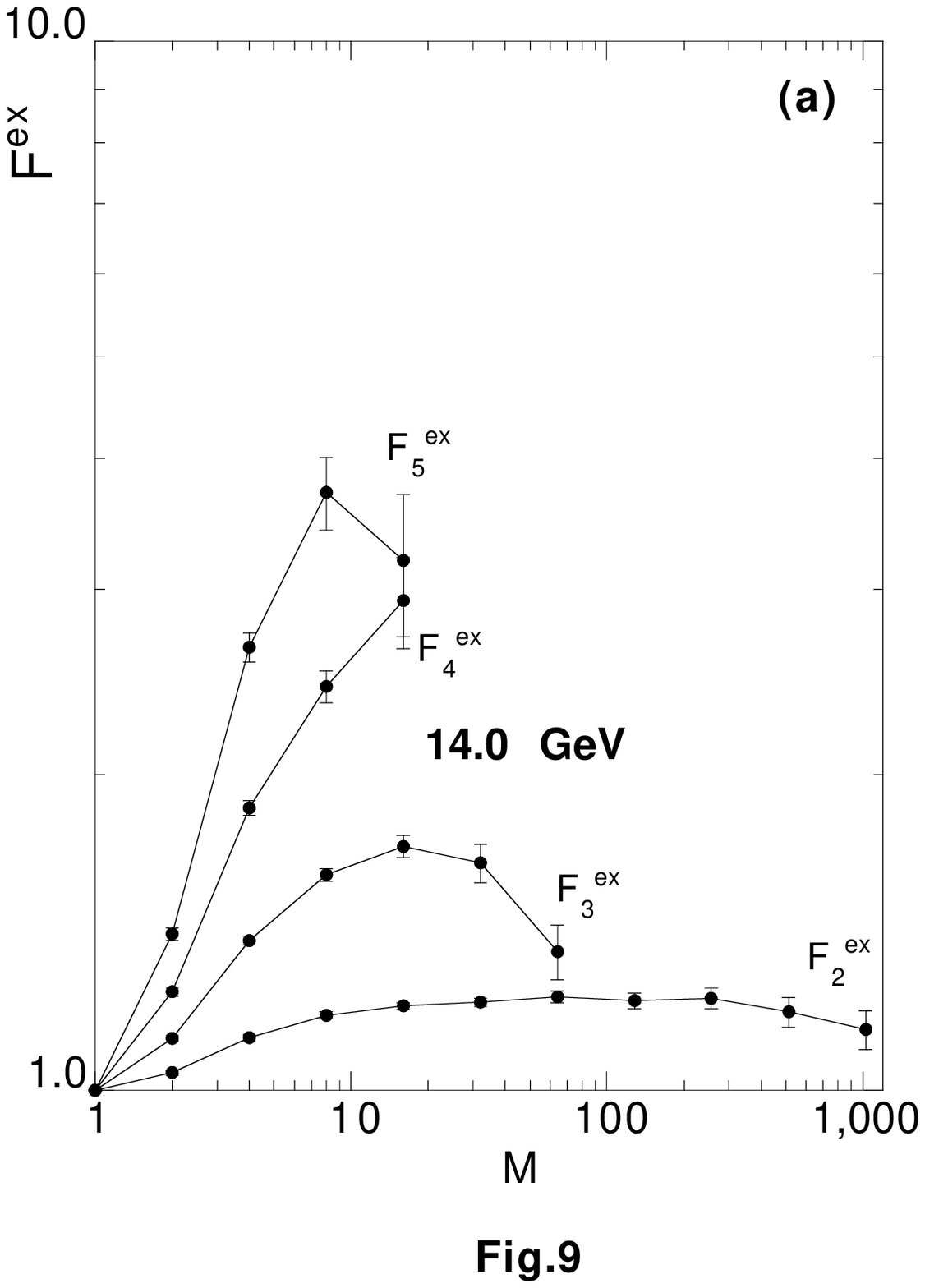}
\end{center}
\end{figure}
\begin{figure}
\begin{center}\leavevmode
\epsfysize=10.0cm
\epsfbox{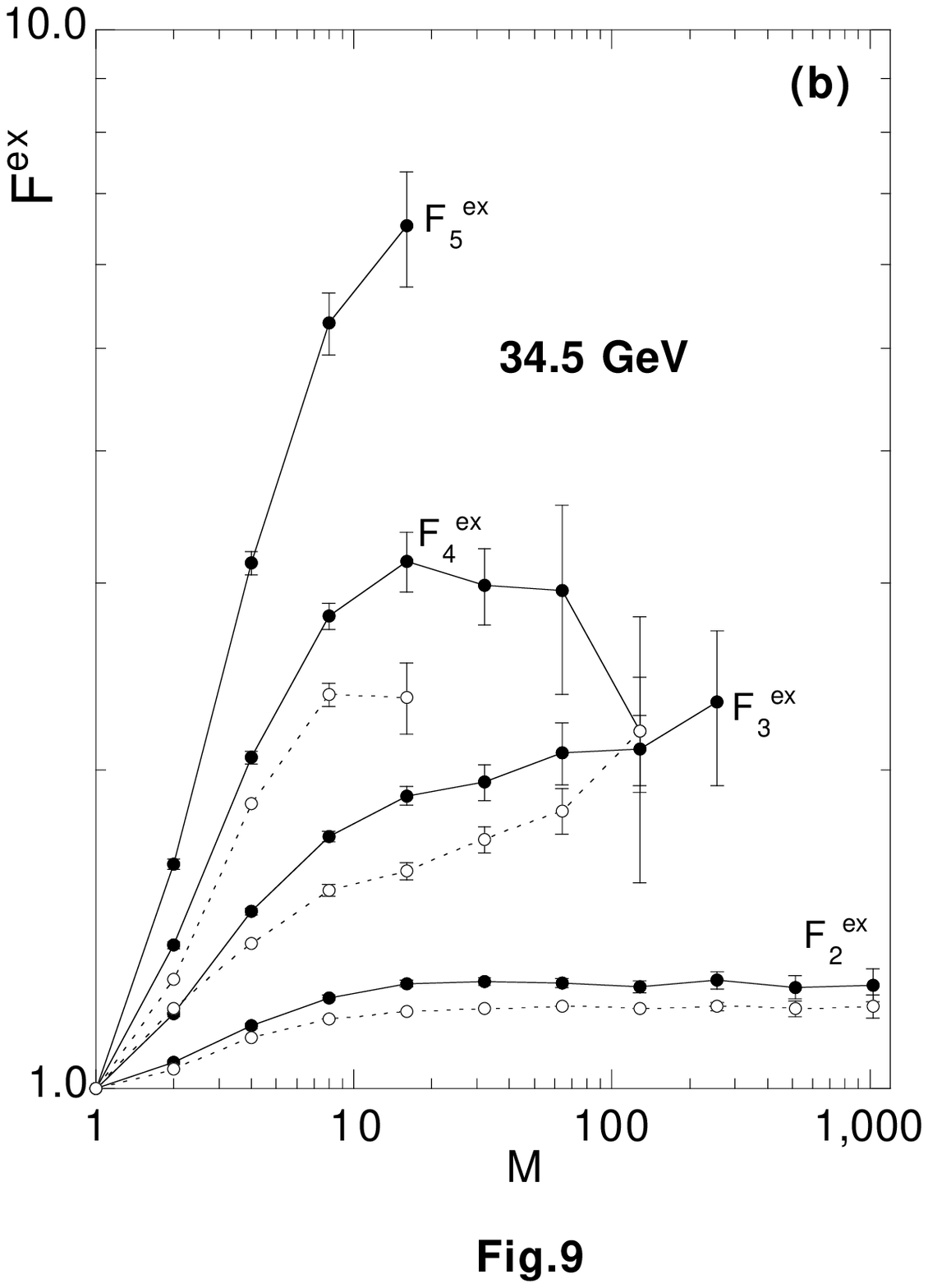}
\end{center}
\end{figure}
\begin{figure}
\begin{center}\leavevmode
\epsfysize=10.0cm
\epsfbox{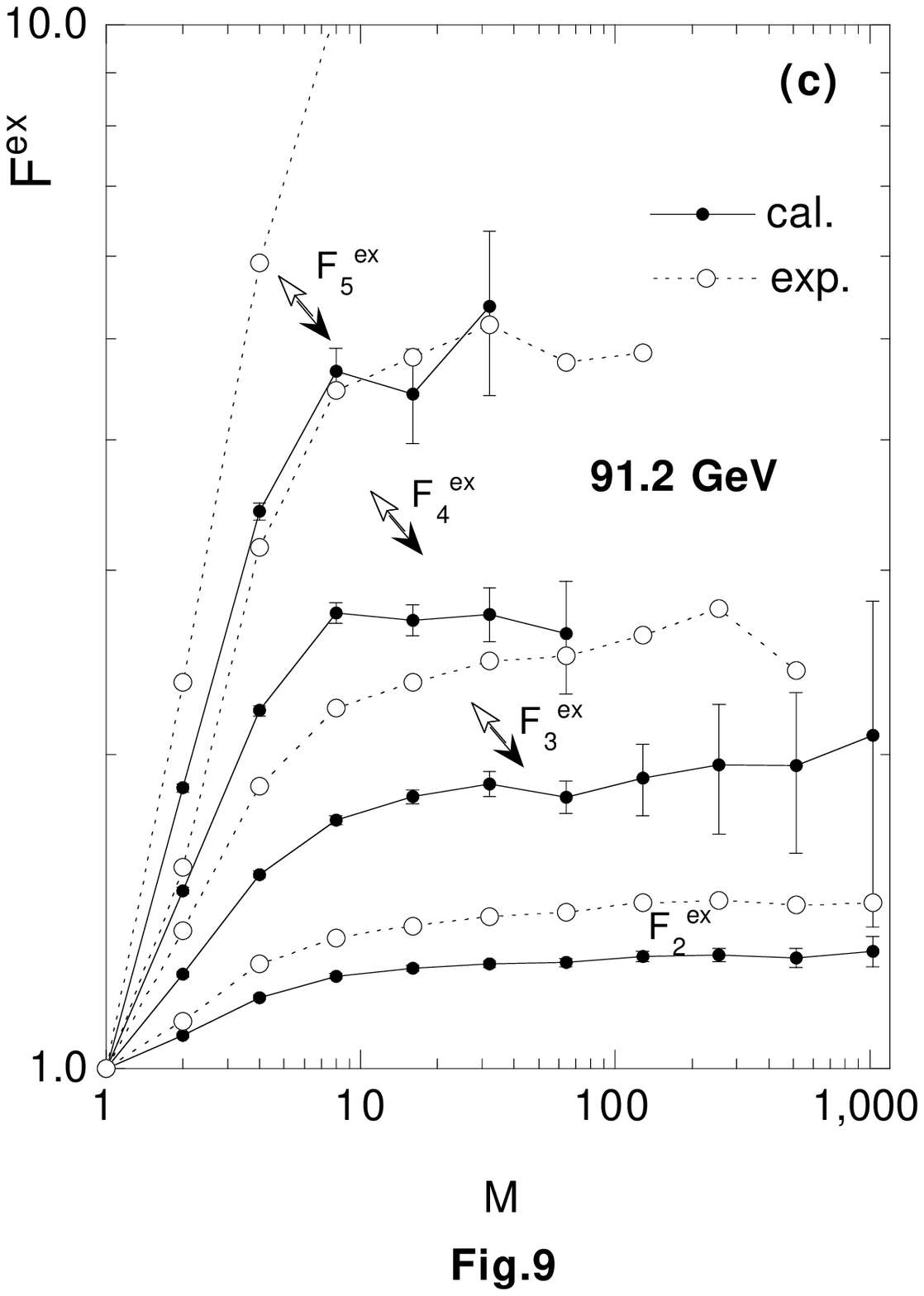}
\end{center}
\end{figure}
\clearpage 
\begin{figure}
\begin{center}\leavevmode
\epsfysize=10.0cm
\epsfbox{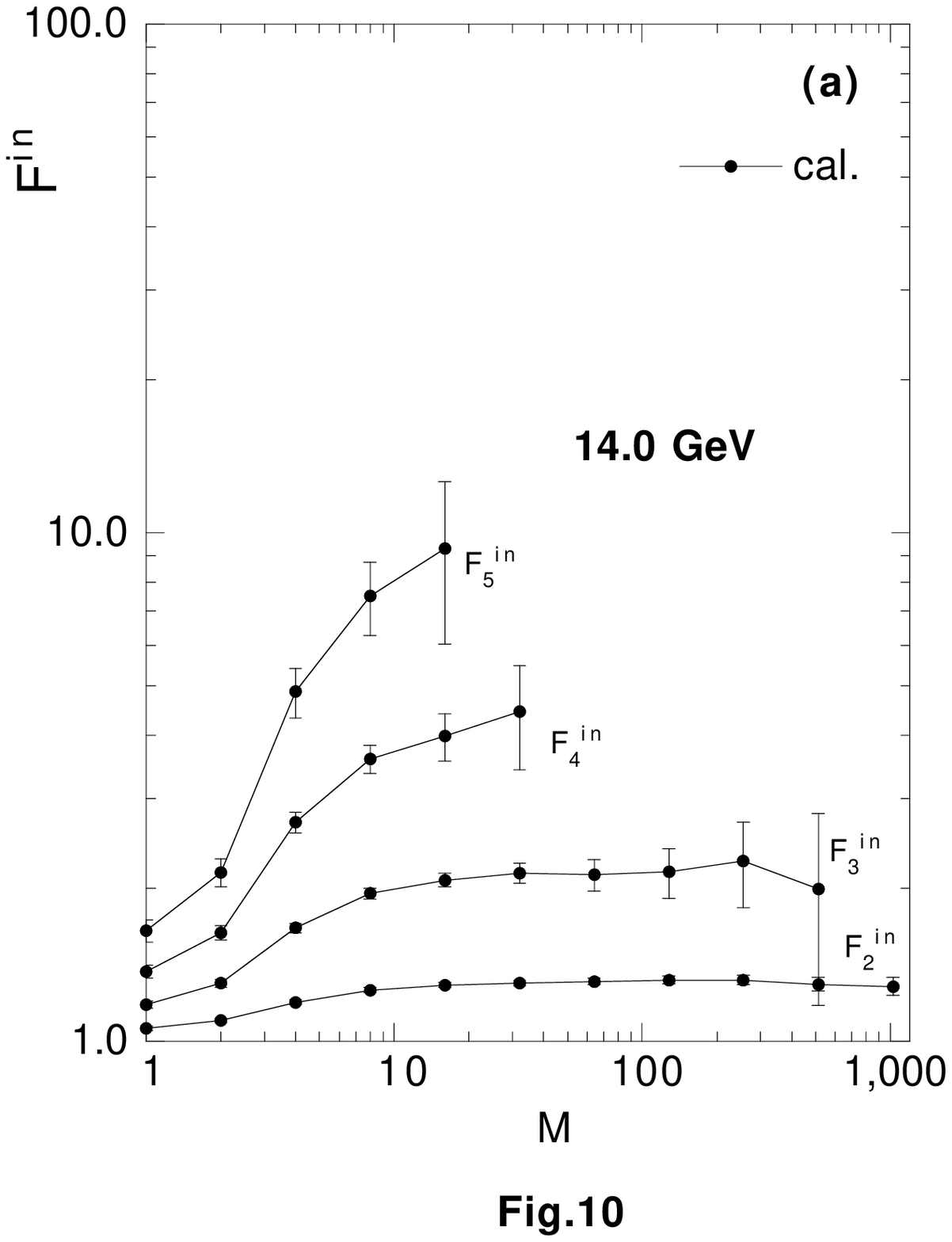}
\end{center}
\end{figure}
\begin{figure}
\begin{center}\leavevmode
\epsfysize=10.0cm
\epsfbox{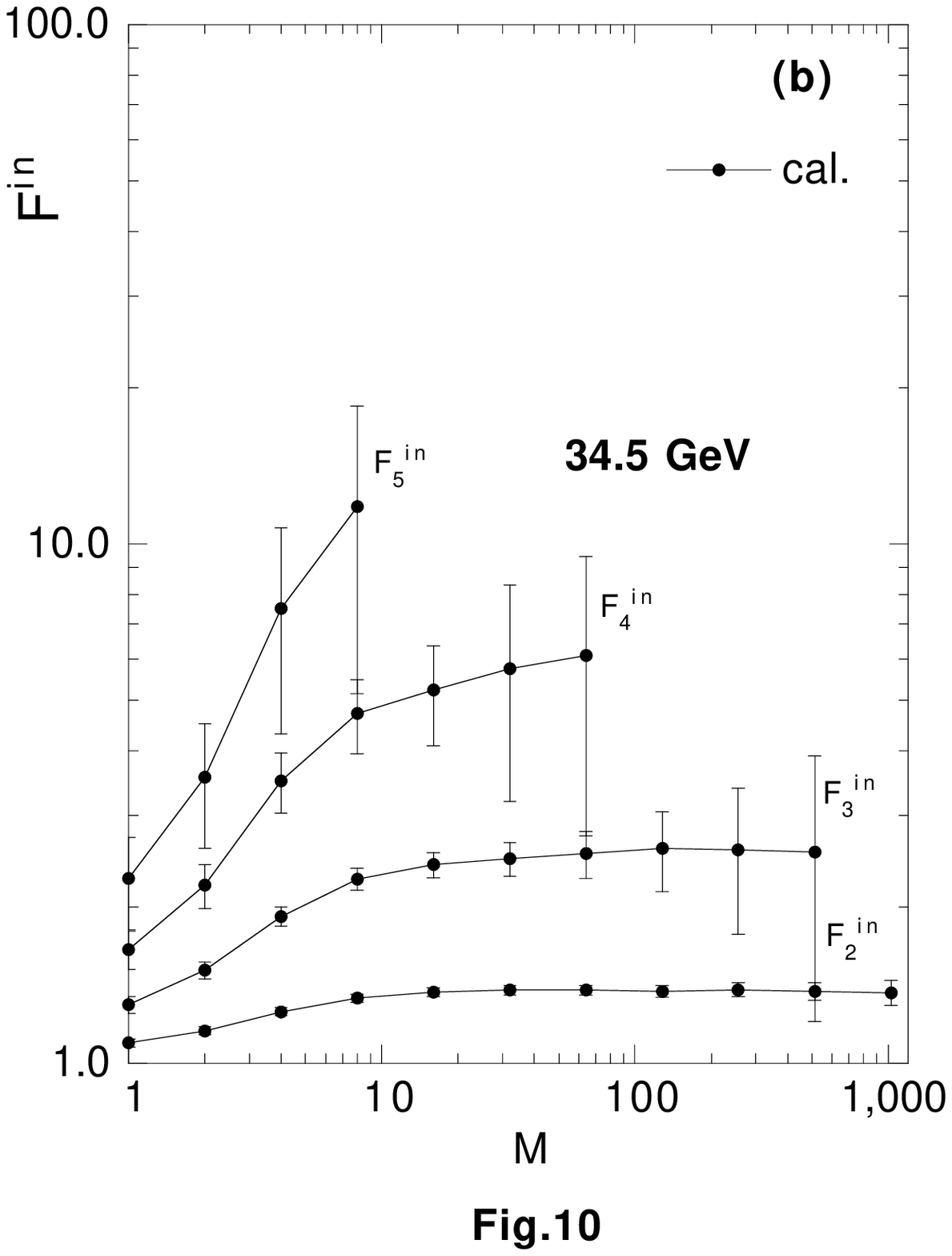}
\end{center}
\end{figure}
\begin{figure}
\begin{center}\leavevmode
\epsfysize=10.0cm
\epsfbox{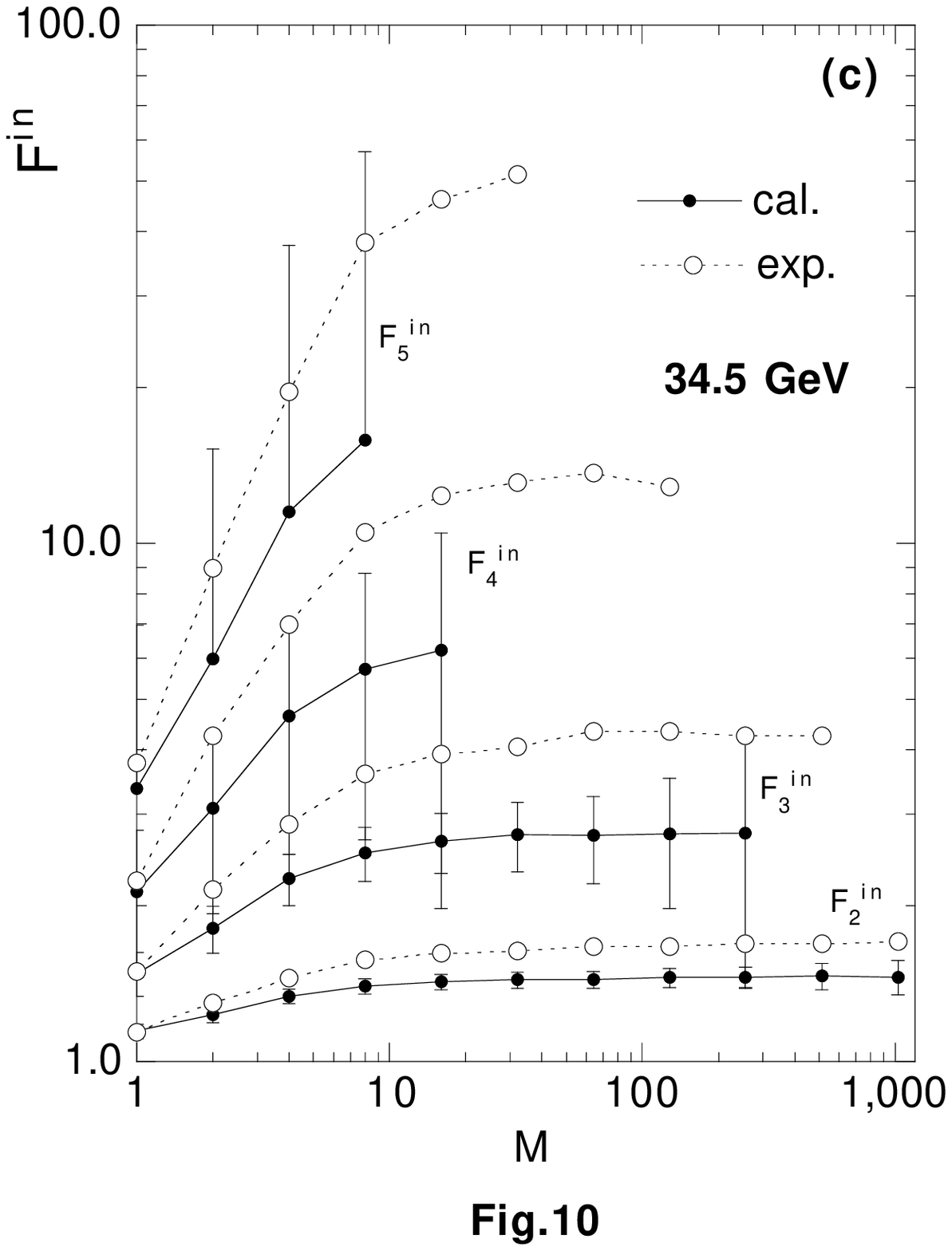}
\end{center}
\end{figure}
\begin{figure}
\begin{center}\leavevmode
\epsfysize=10.0cm
\epsfbox{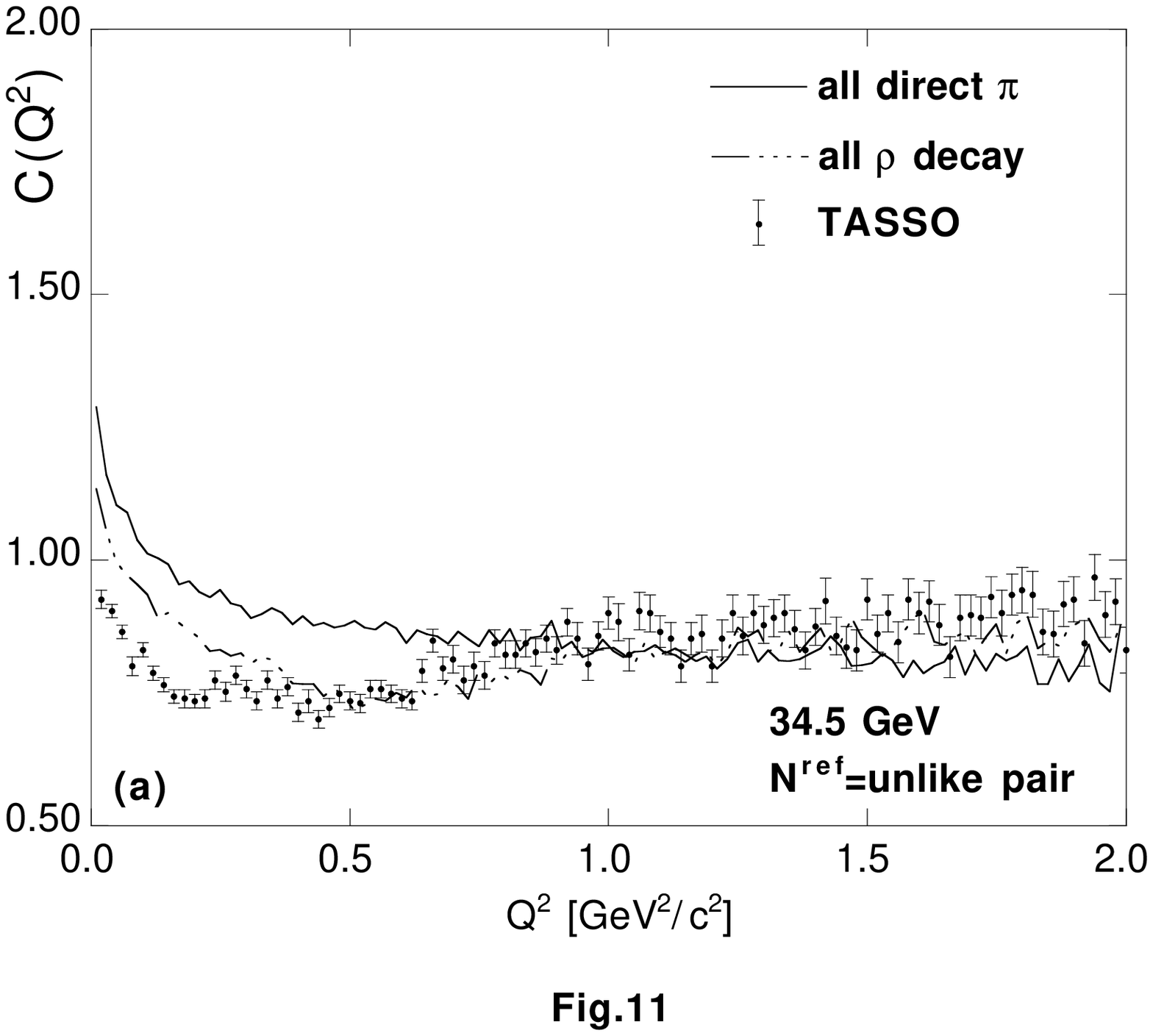}
\end{center}
\end{figure}
\begin{figure}
\begin{center}\leavevmode
\epsfysize=10.0cm
\epsfbox{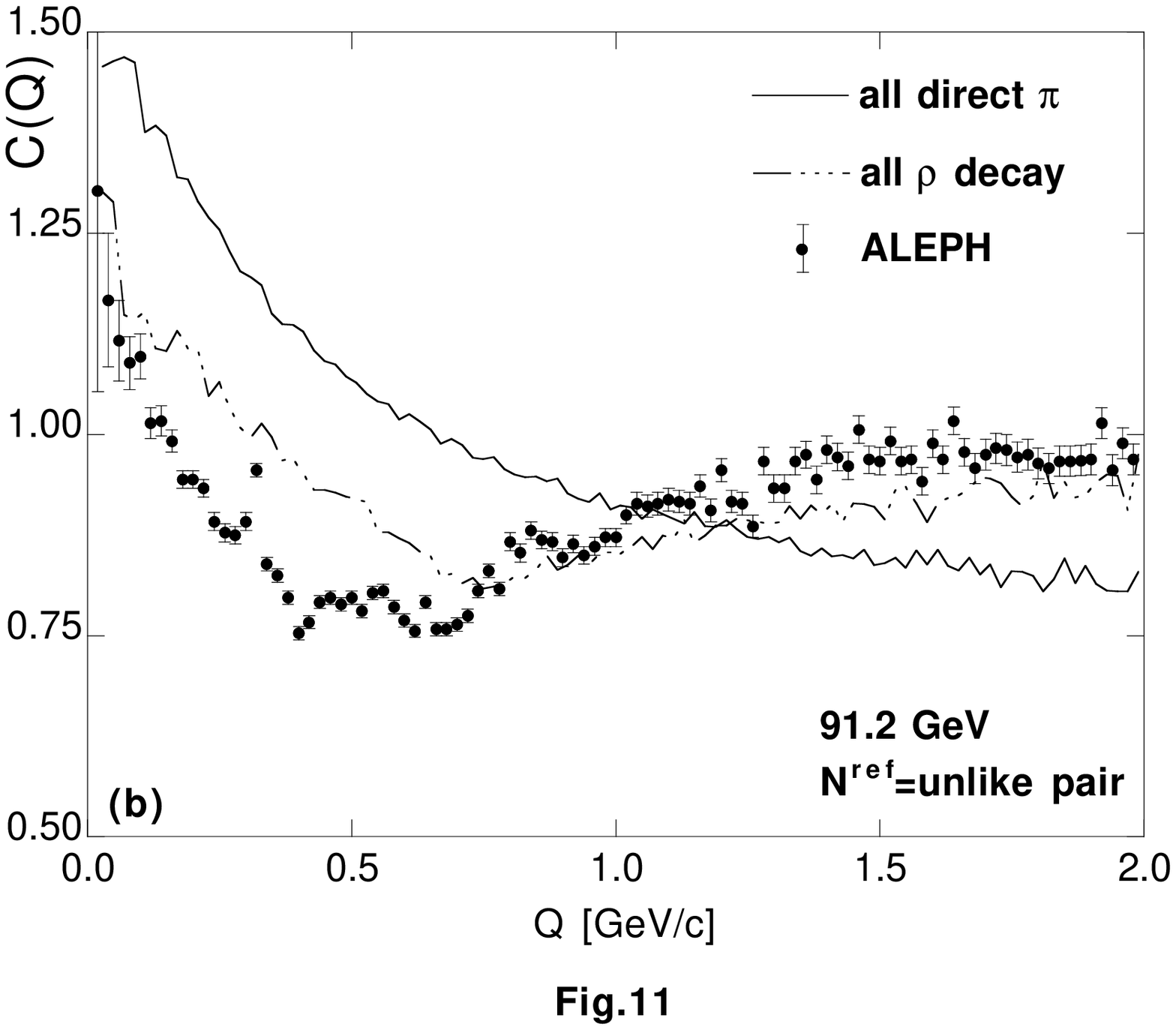}
\end{center}
\end{figure}
\begin{figure}
\begin{center}\leavevmode
\epsfysize=10.0cm
\epsfbox{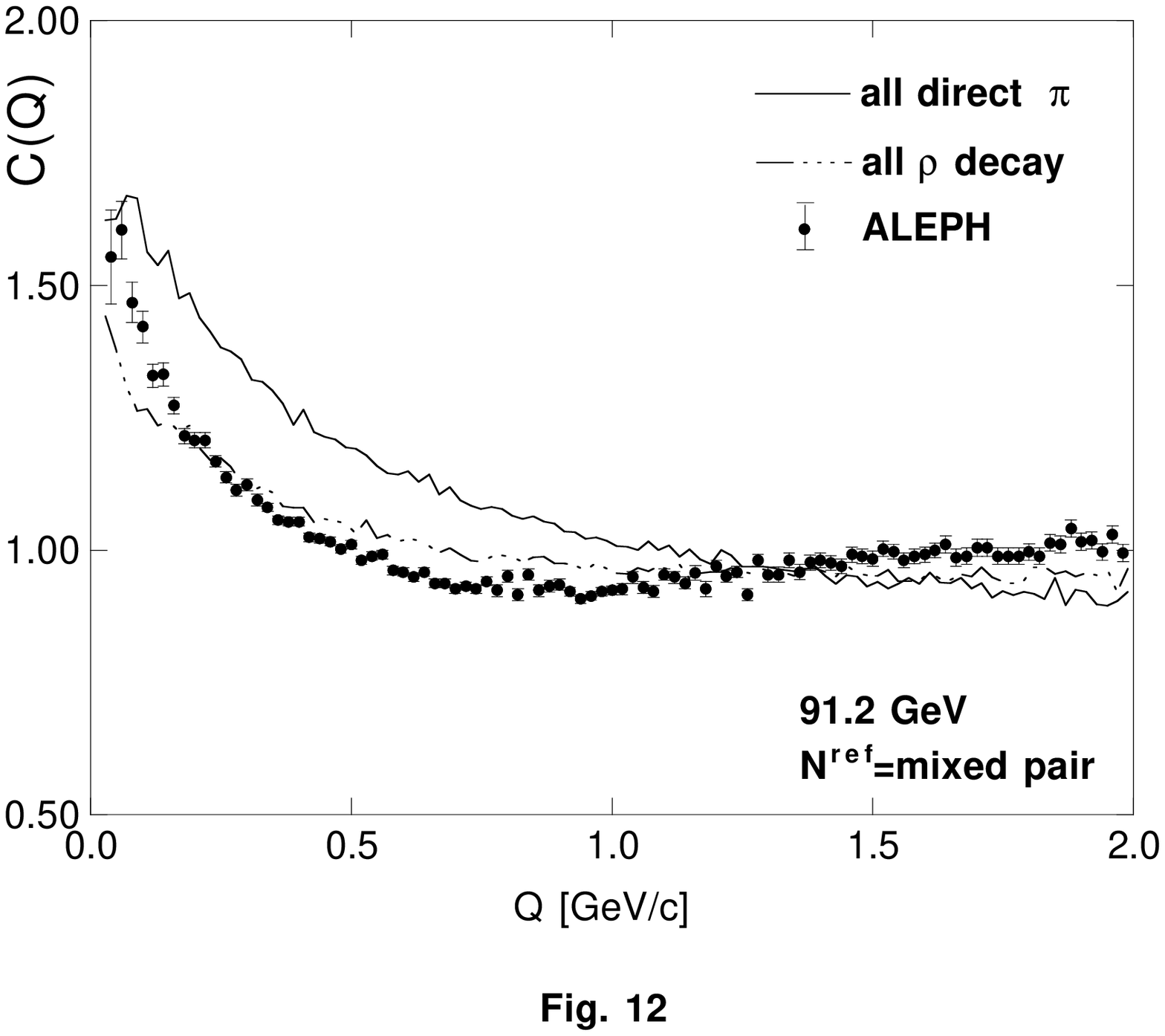}
\end{center}
\end{figure}
\begin{figure}
\begin{center}\leavevmode
\epsfysize=10.0cm
\epsfbox{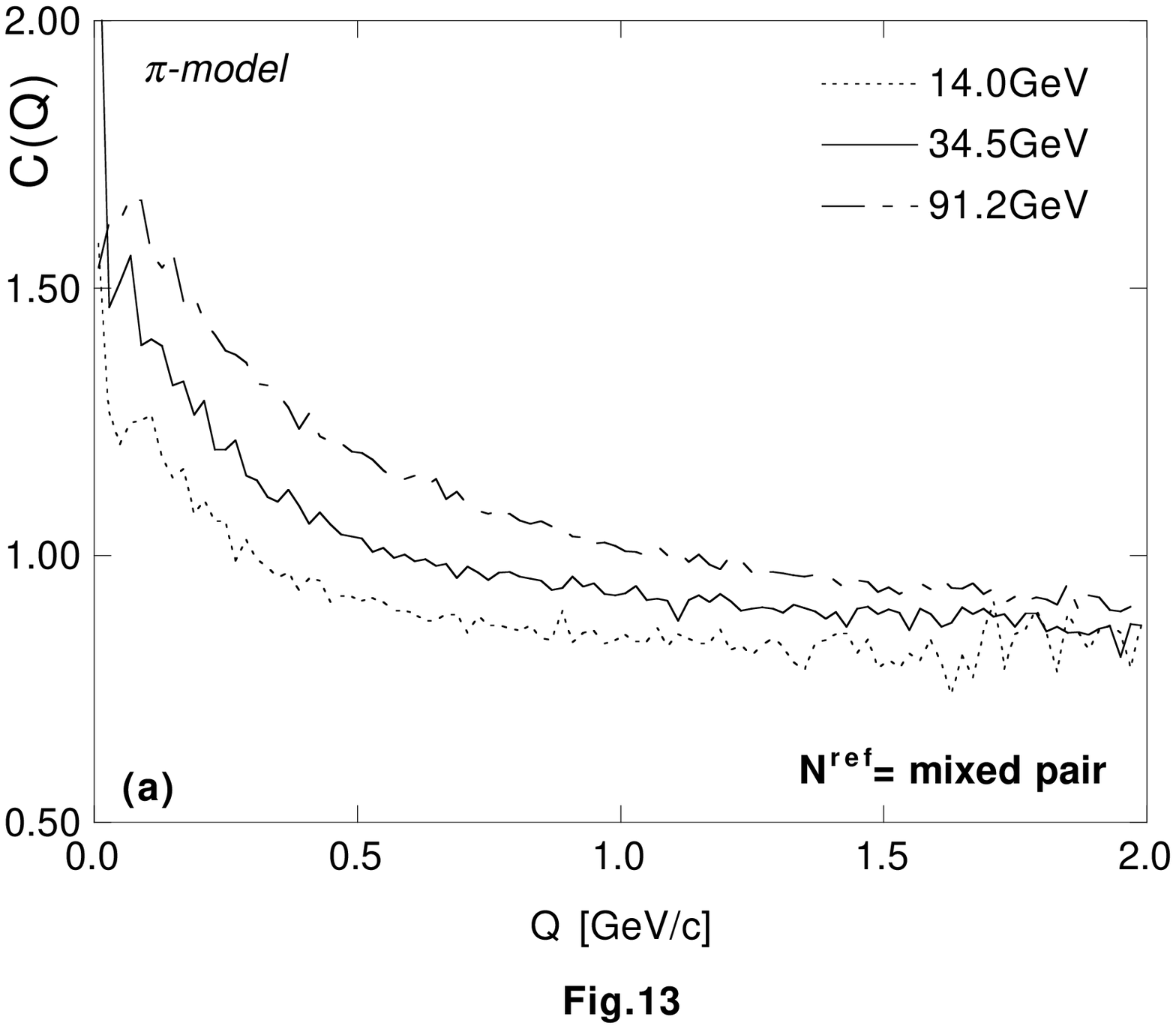}
\end{center}
\end{figure}
\begin{figure}
\begin{center}\leavevmode
\epsfysize=10.0cm
\epsfbox{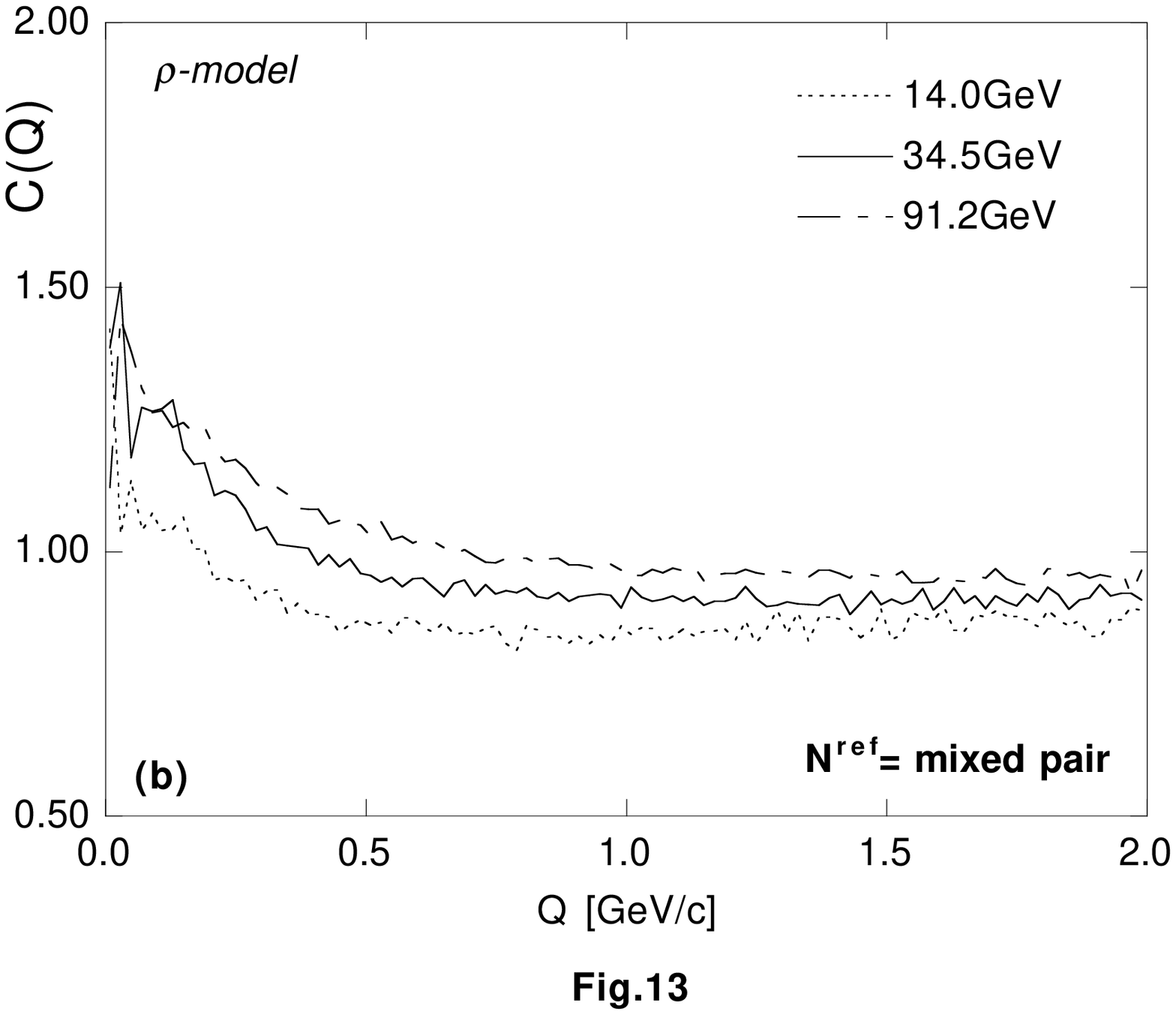}
\end{center}
\end{figure}
\begin{figure}
\begin{center}\leavevmode
\epsfysize=10.0cm
\epsfbox{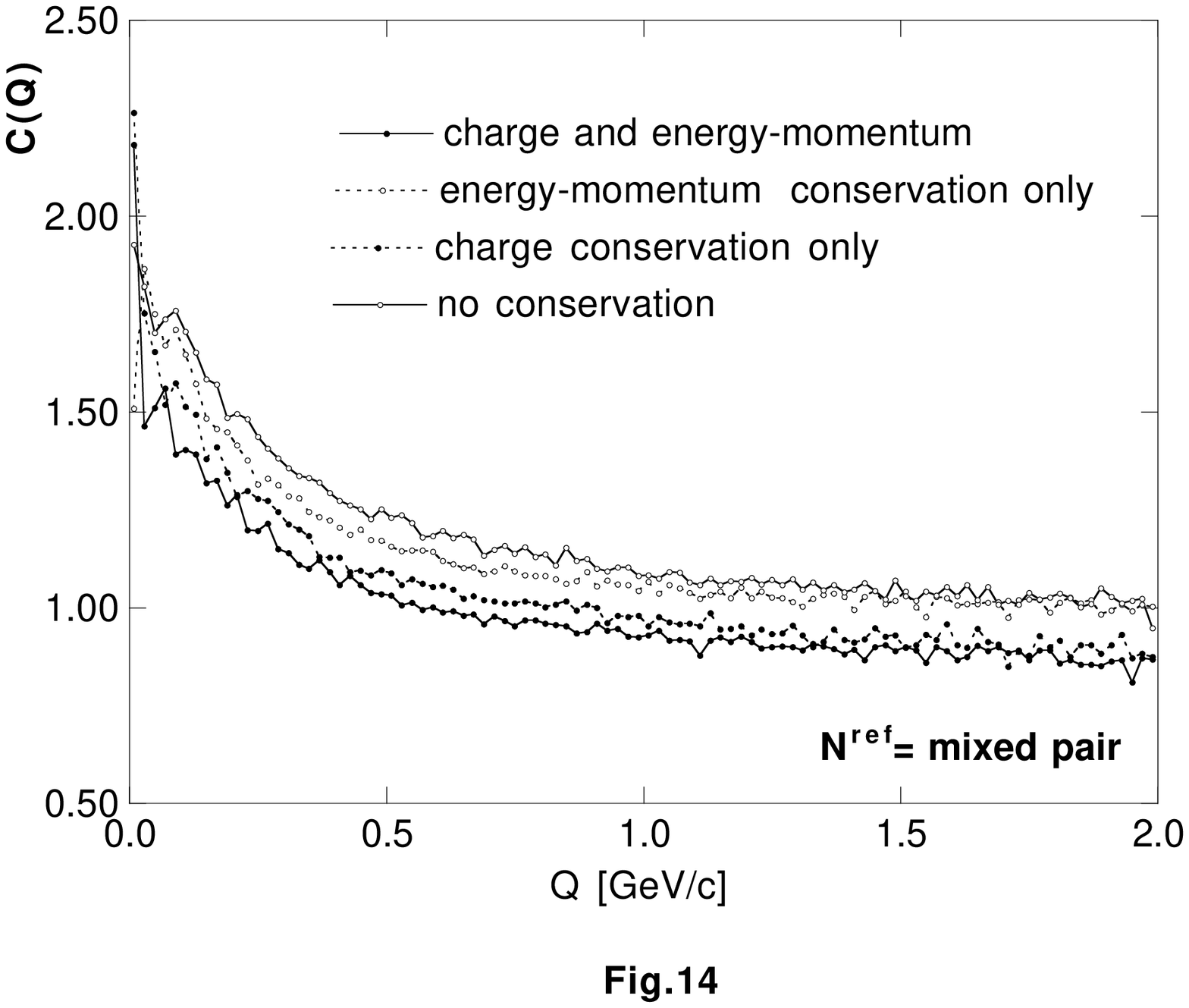}
\end{center}
\end{figure}
\begin{figure}
\begin{center}\leavevmode
\epsfysize=10.0cm
\epsfbox{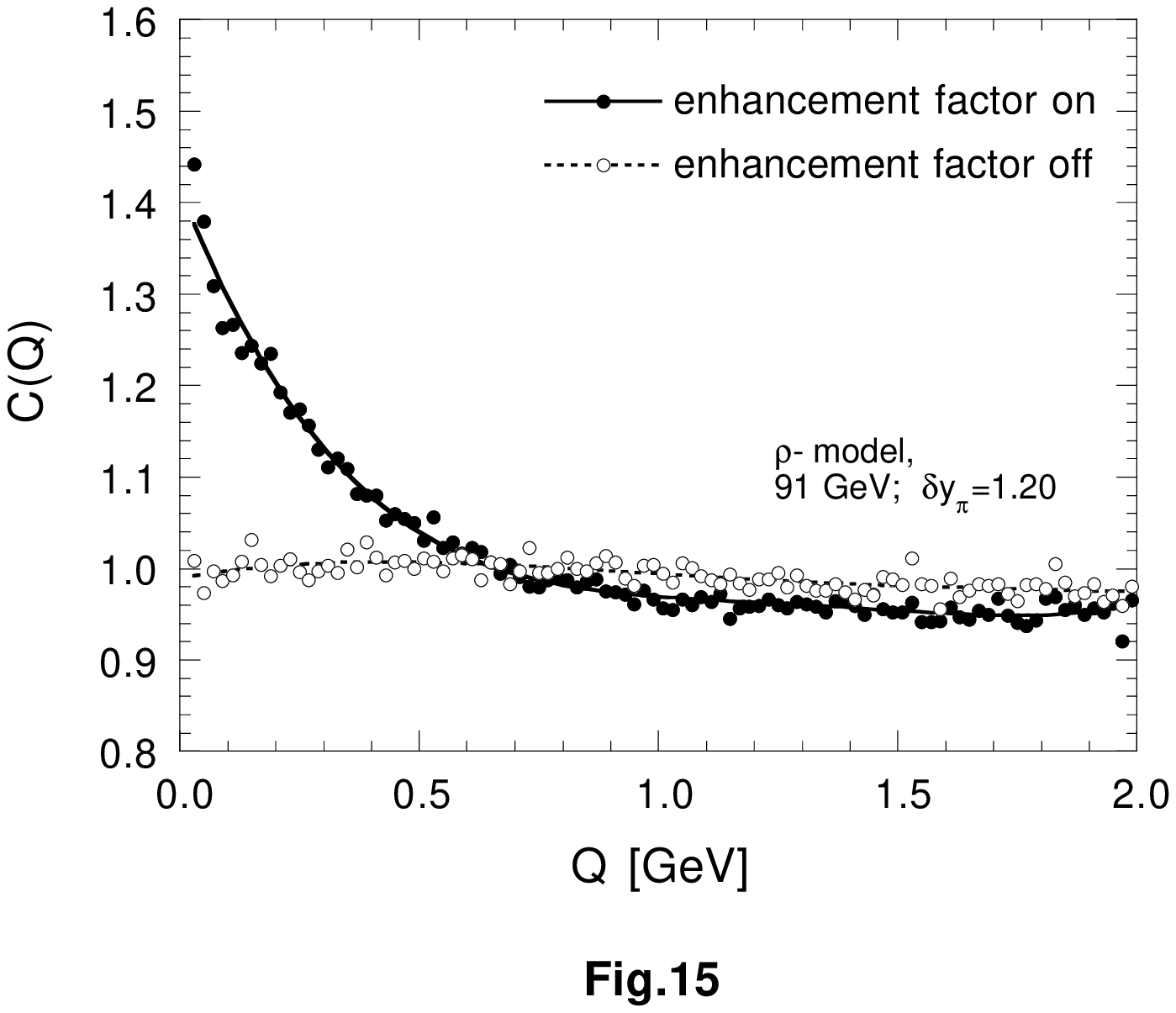}
\end{center}
\end{figure}
\begin{figure}
\begin{center}\leavevmode
\epsfysize=10.0cm
\epsfbox{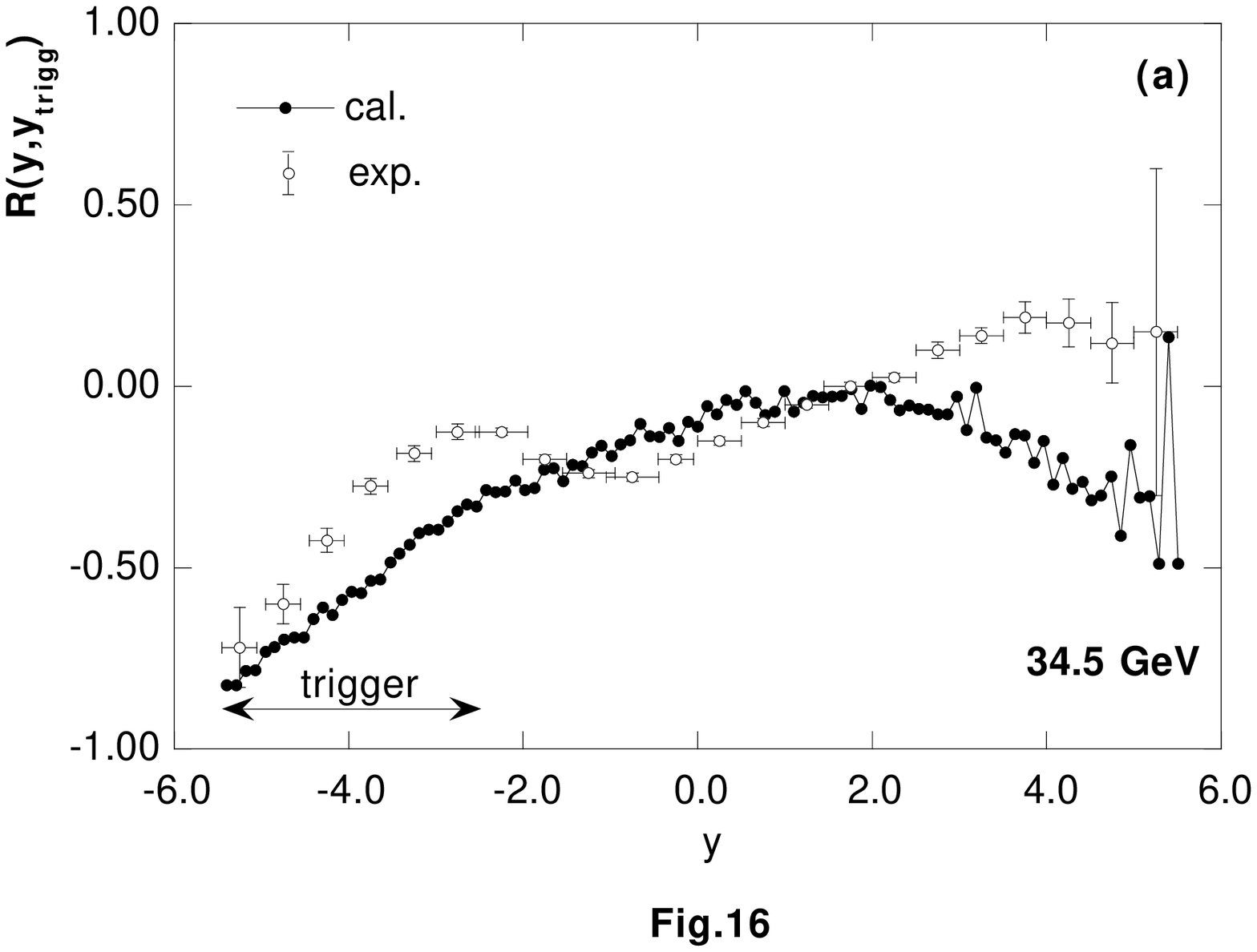}
\end{center}
\end{figure}
\begin{figure}
\begin{center}\leavevmode
\epsfysize=10.0cm
\epsfbox{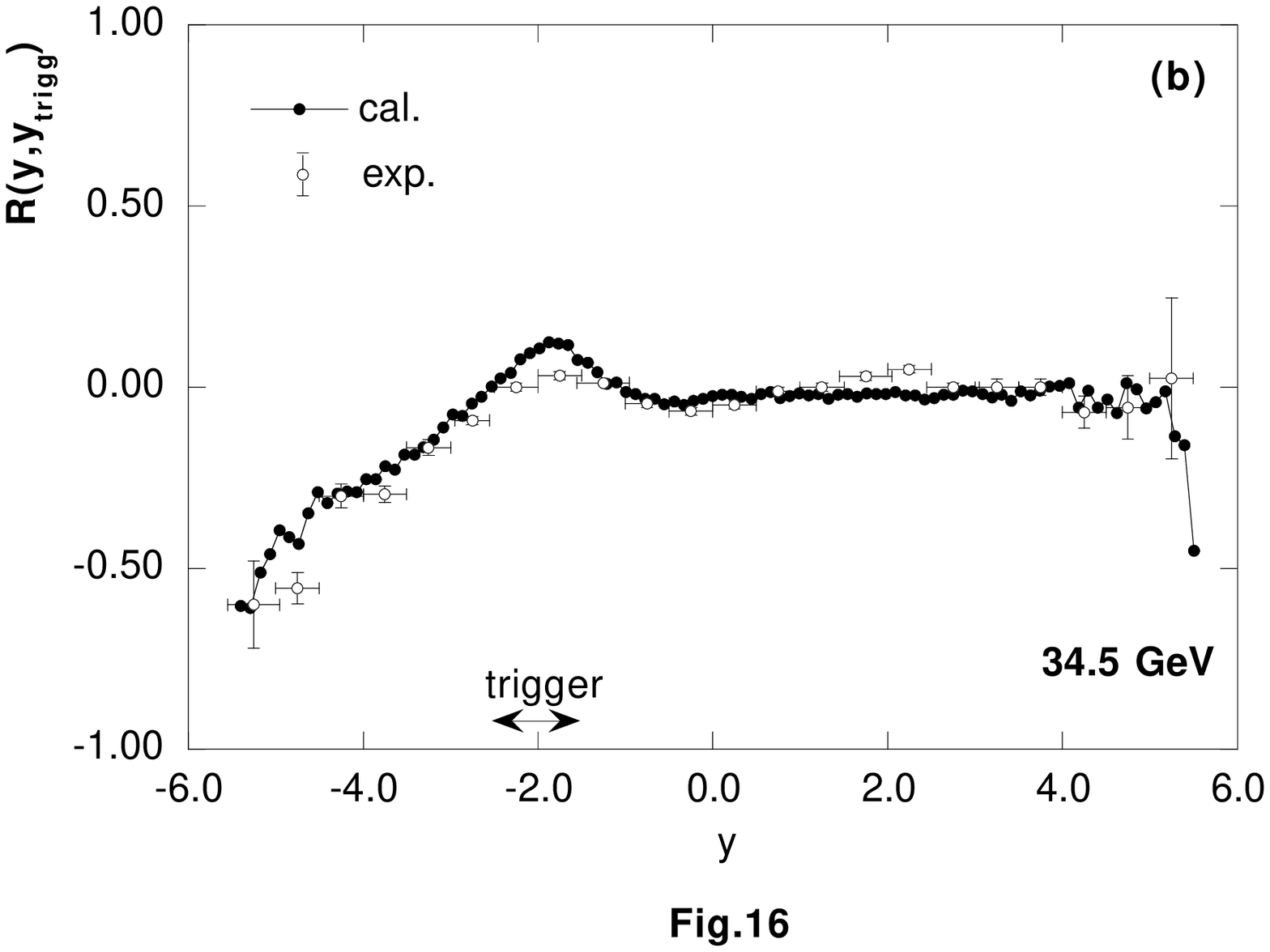}
\end{center}
\end{figure}
\begin{figure}
\begin{center}\leavevmode
\epsfysize=10.0cm
\epsfbox{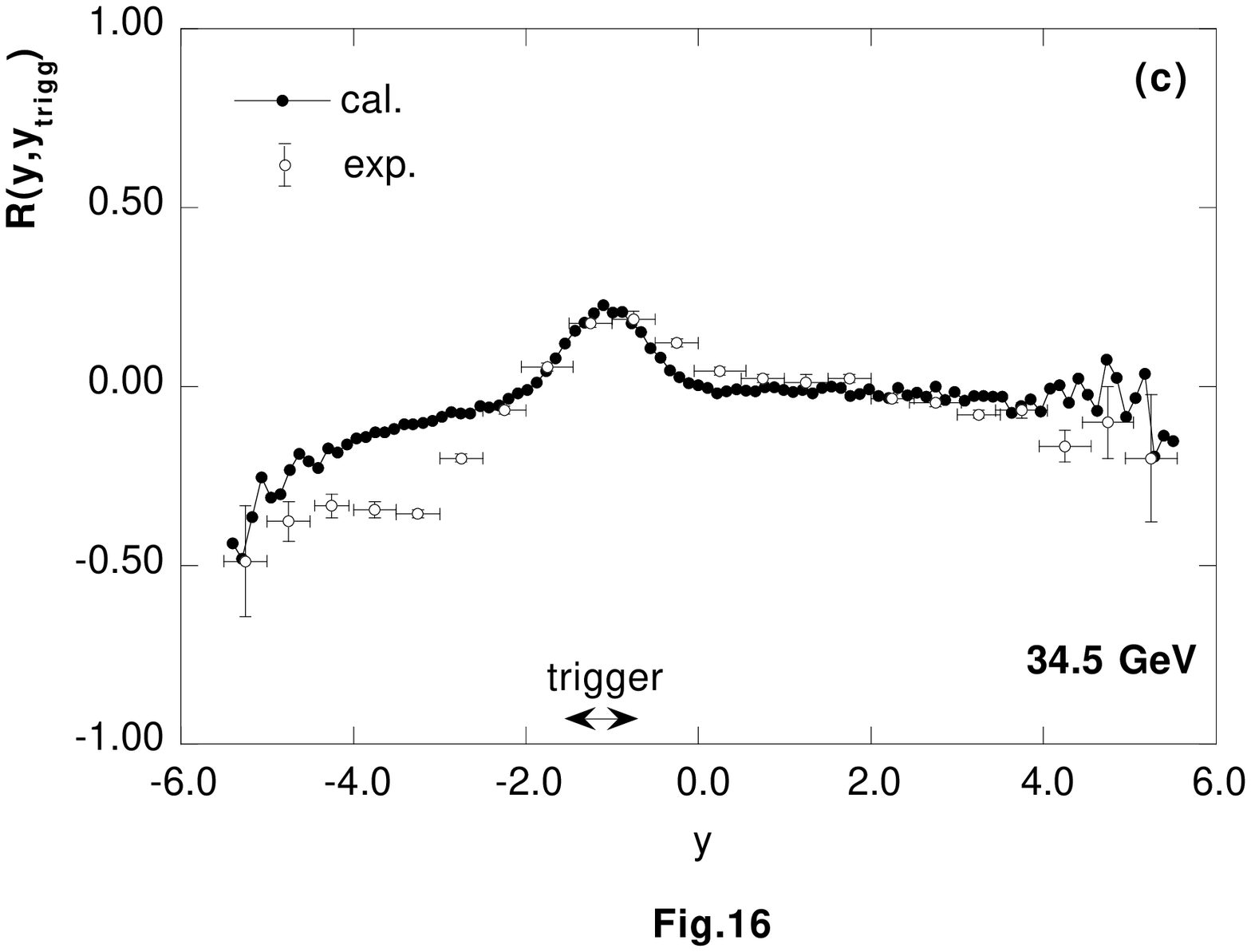}
\end{center}
\end{figure}
\begin{figure}
\begin{center}\leavevmode
\epsfysize=10.0cm
\epsfbox{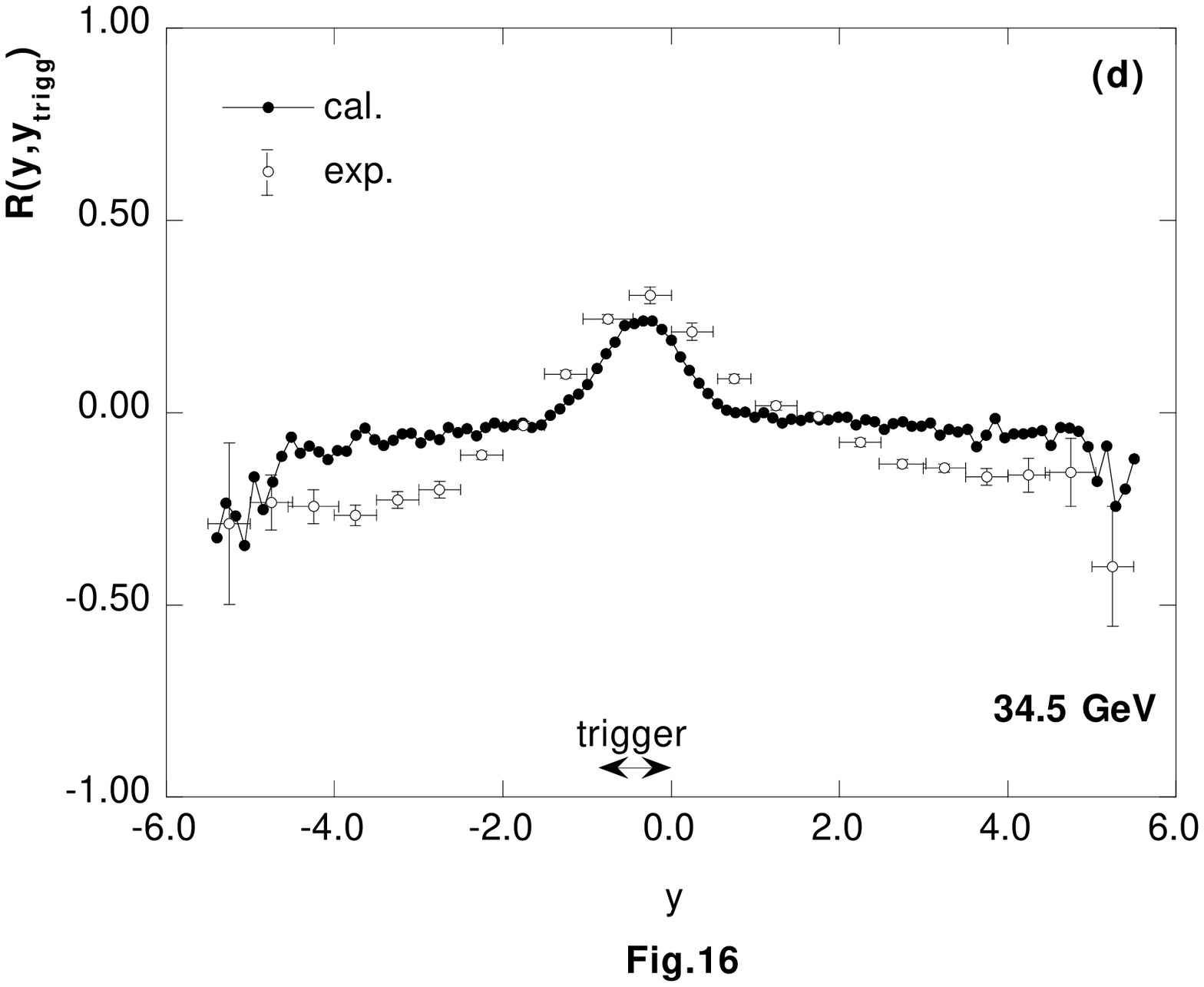}
\end{center}
\end{figure}
\begin{figure}
\begin{center}\leavevmode
\epsfysize=10.0cm
\epsfbox{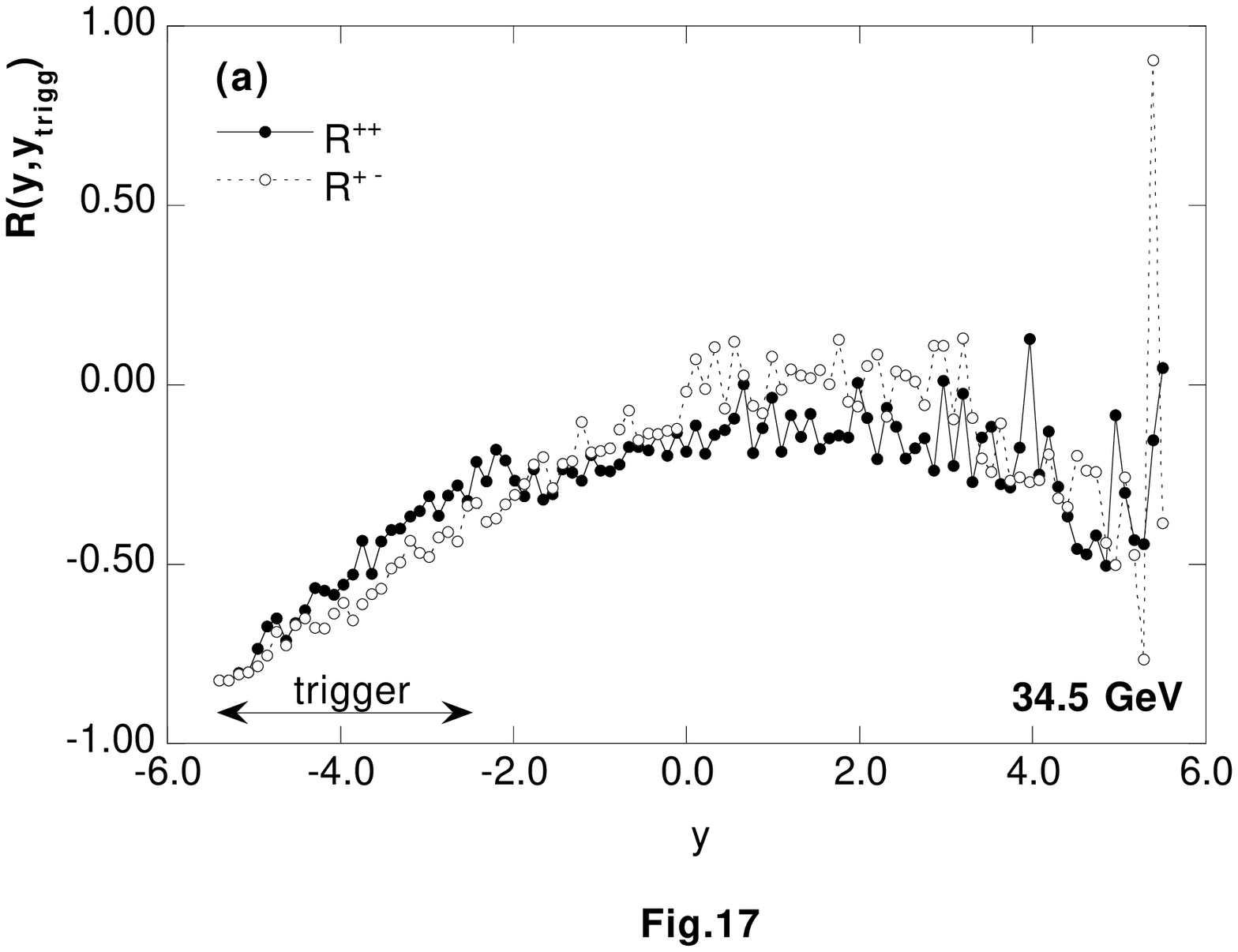}
\end{center}
\end{figure}
\begin{figure}
\begin{center}\leavevmode
\epsfysize=10.0cm
\epsfbox{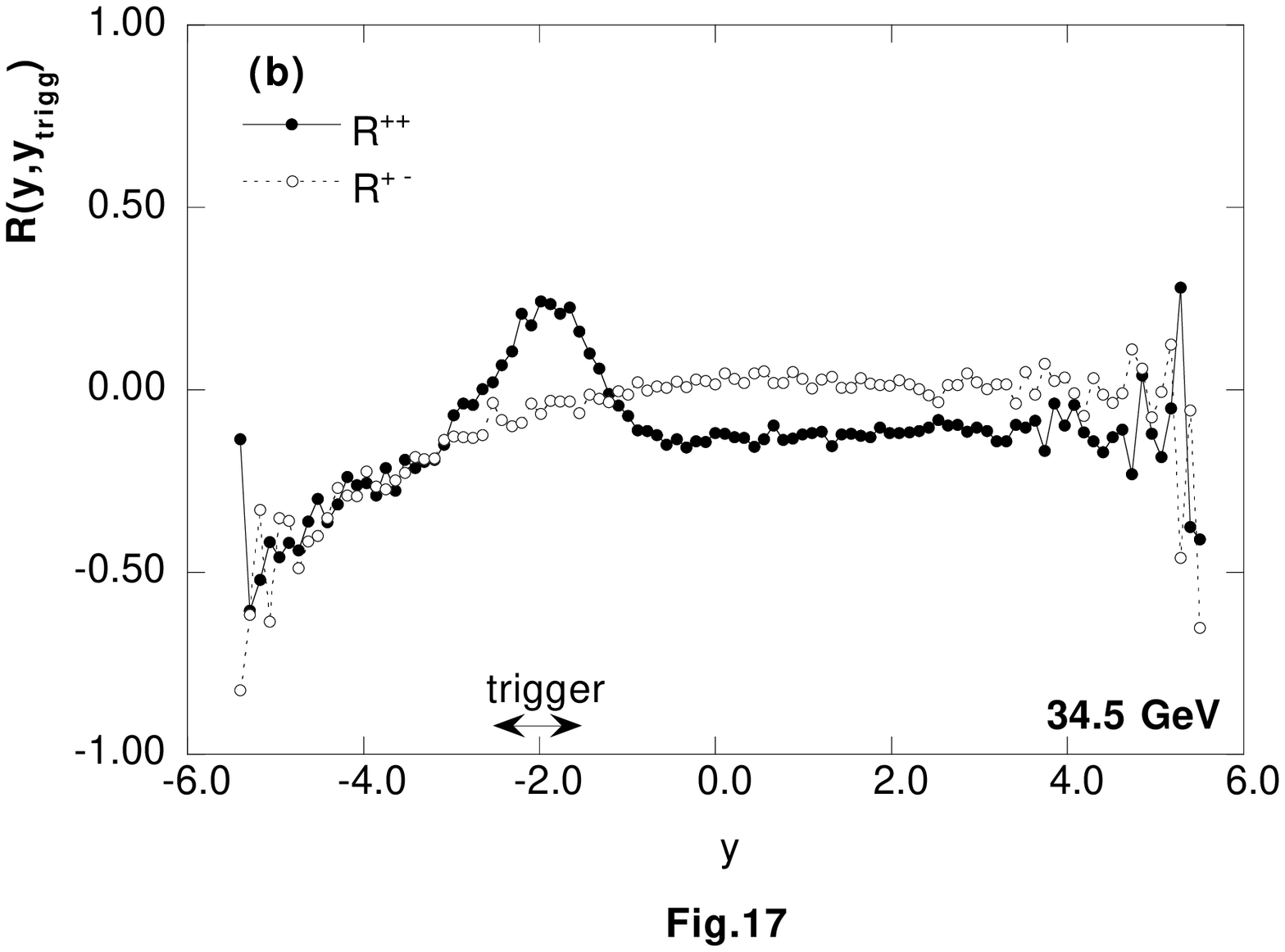}
\end{center}
\end{figure}
\begin{figure}
\begin{center}\leavevmode
\epsfysize=10.0cm
\epsfbox{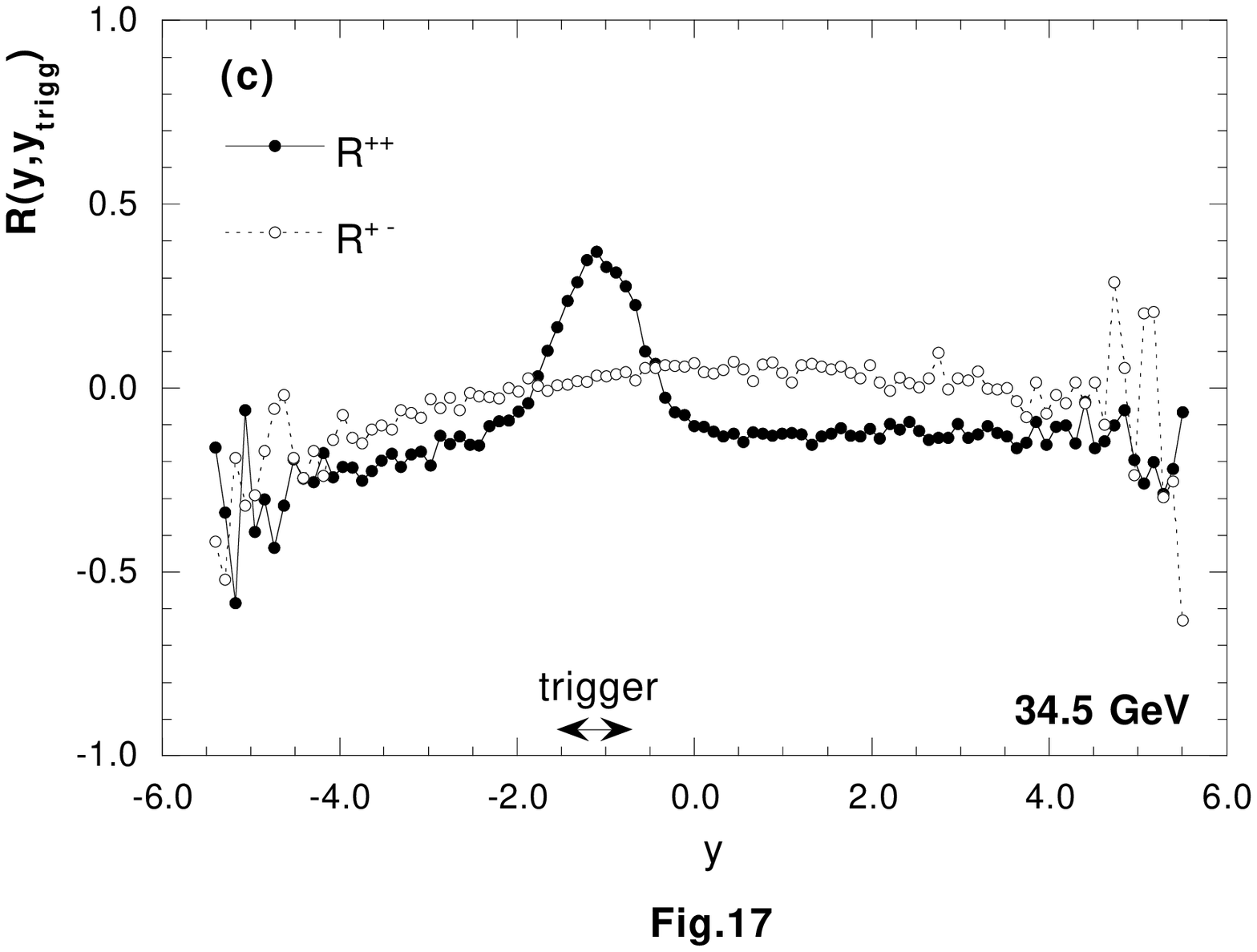}
\end{center}
\end{figure}
\begin{figure}
\begin{center}\leavevmode
\epsfysize=10.0cm
\epsfbox{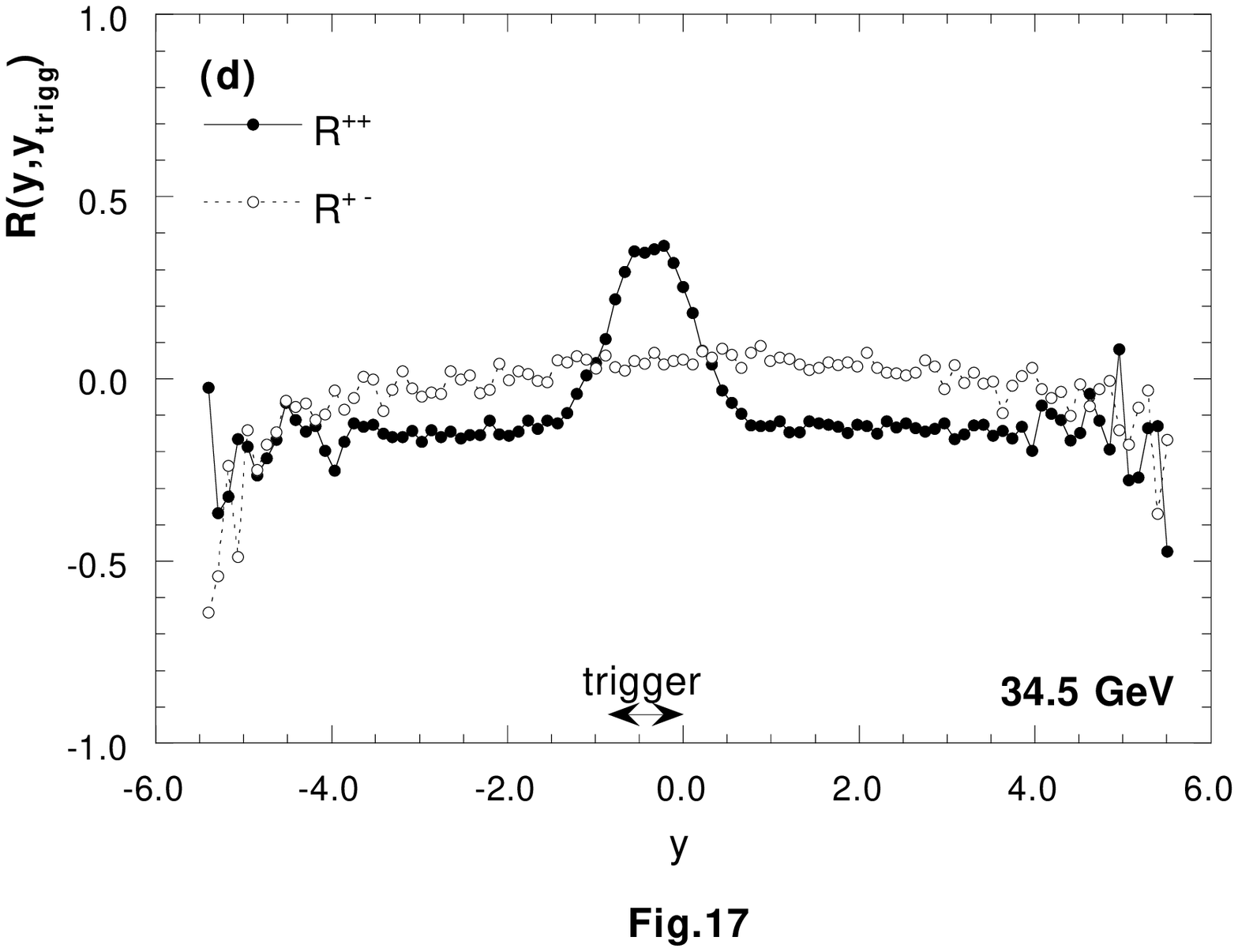}
\end{center}
\end{figure}
\clearpage 
\begin{figure}
\begin{center}\leavevmode
\epsfysize=10.0cm
\epsfbox{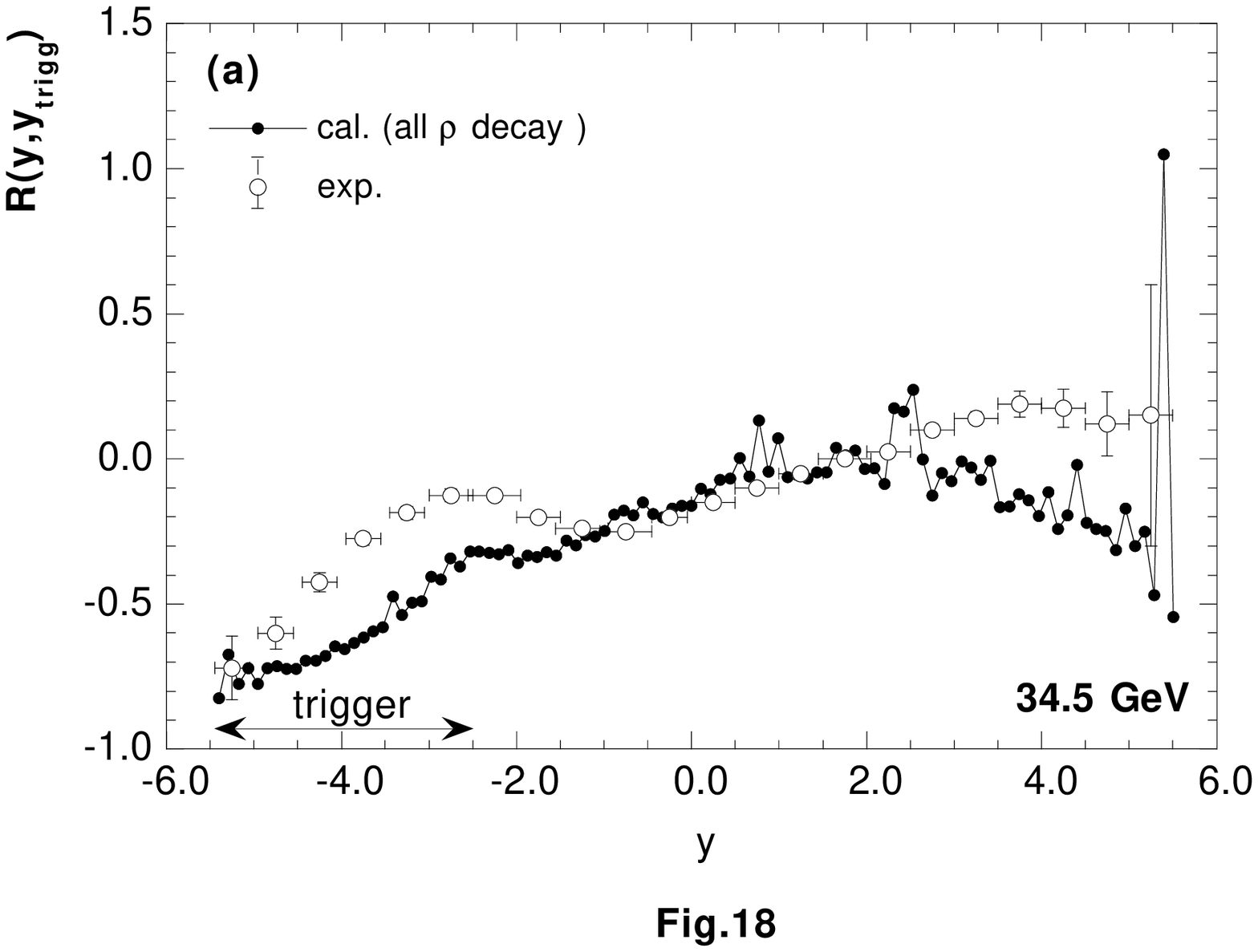}
\end{center}
\end{figure}
\begin{figure}
\begin{center}\leavevmode
\epsfysize=10.0cm
\epsfbox{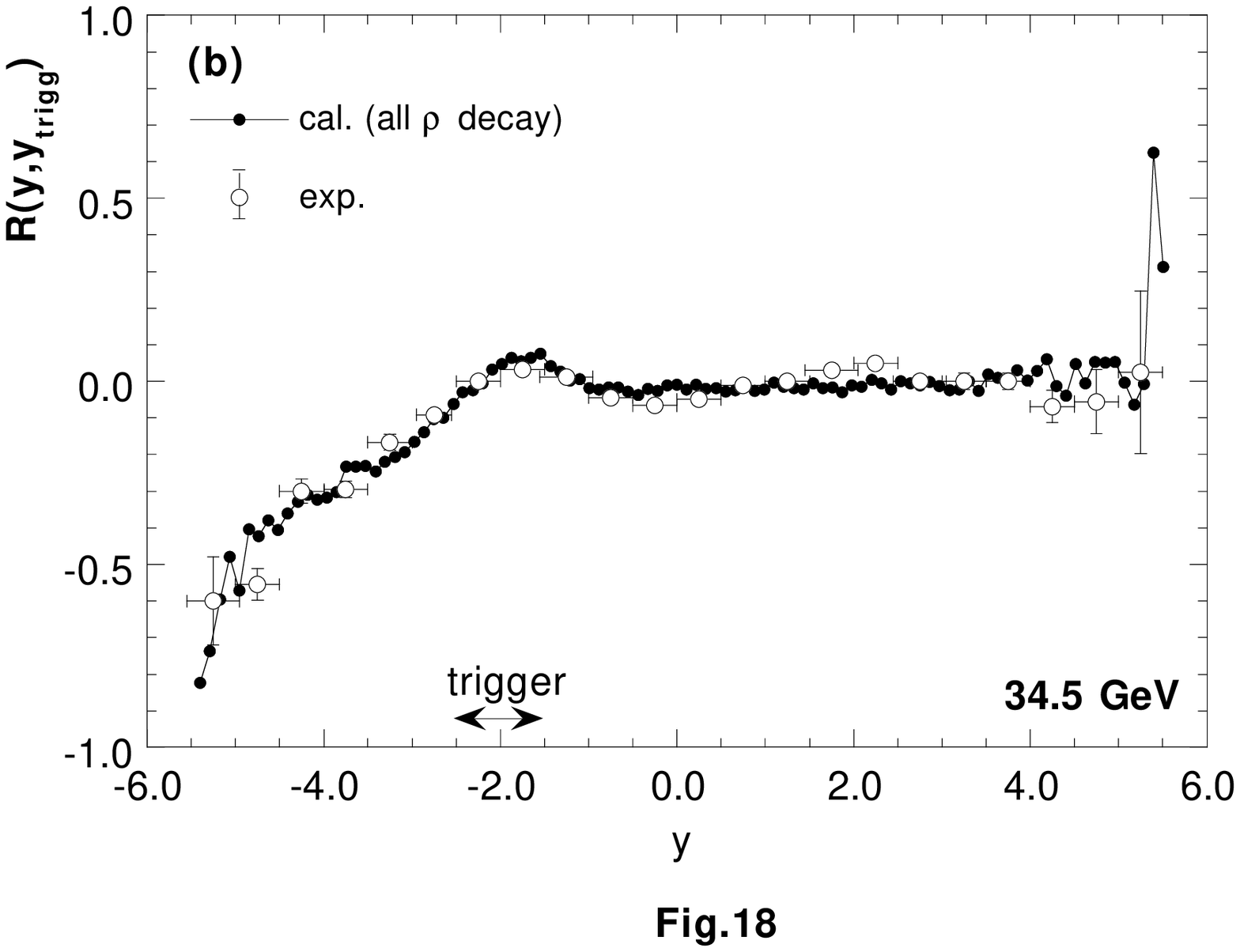}
\end{center}
\end{figure}
\begin{figure}
\begin{center}\leavevmode
\epsfysize=10.0cm
\epsfbox{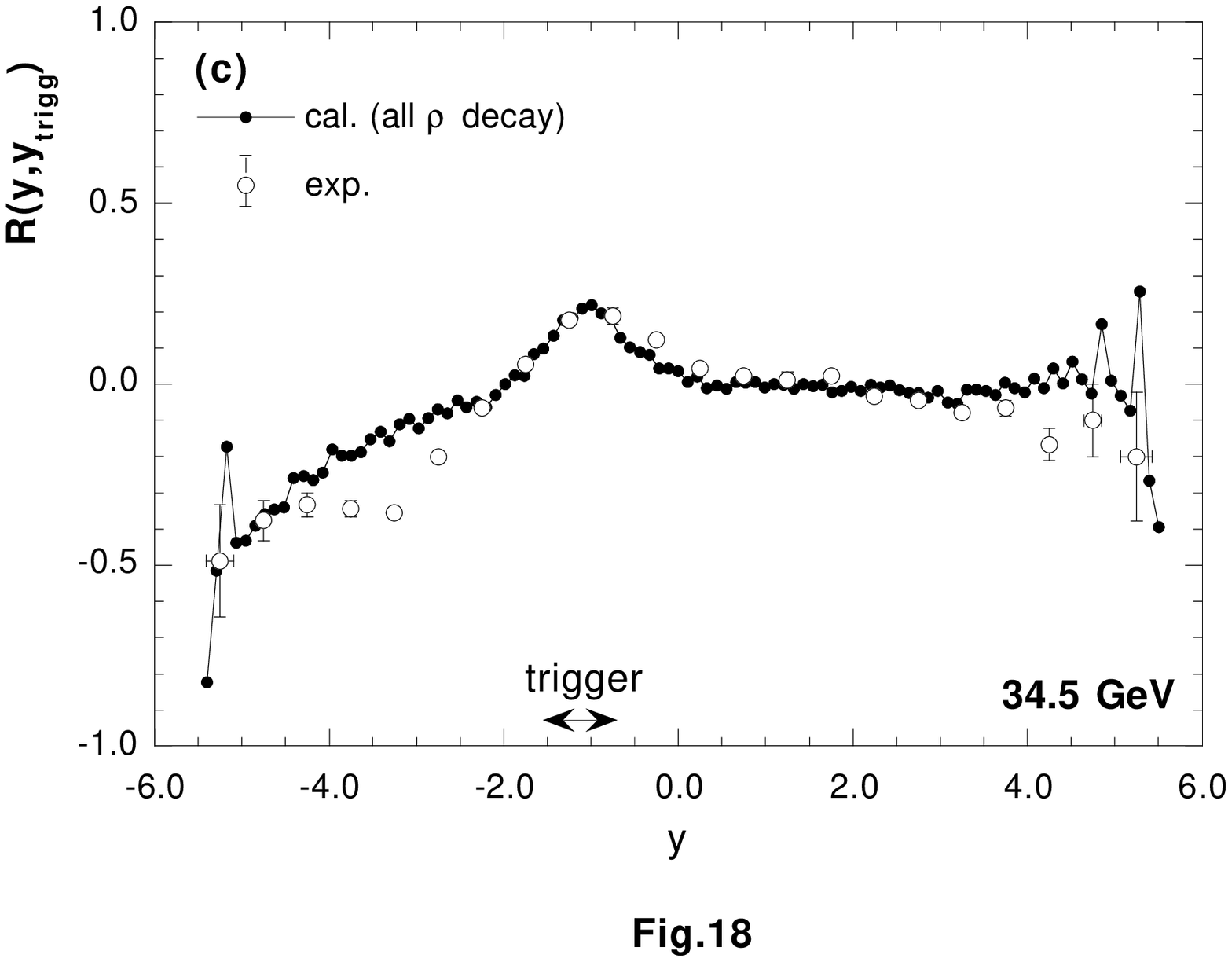}
\end{center}
\end{figure}
\begin{figure}
\begin{center}\leavevmode
\epsfysize=10.0cm
\epsfbox{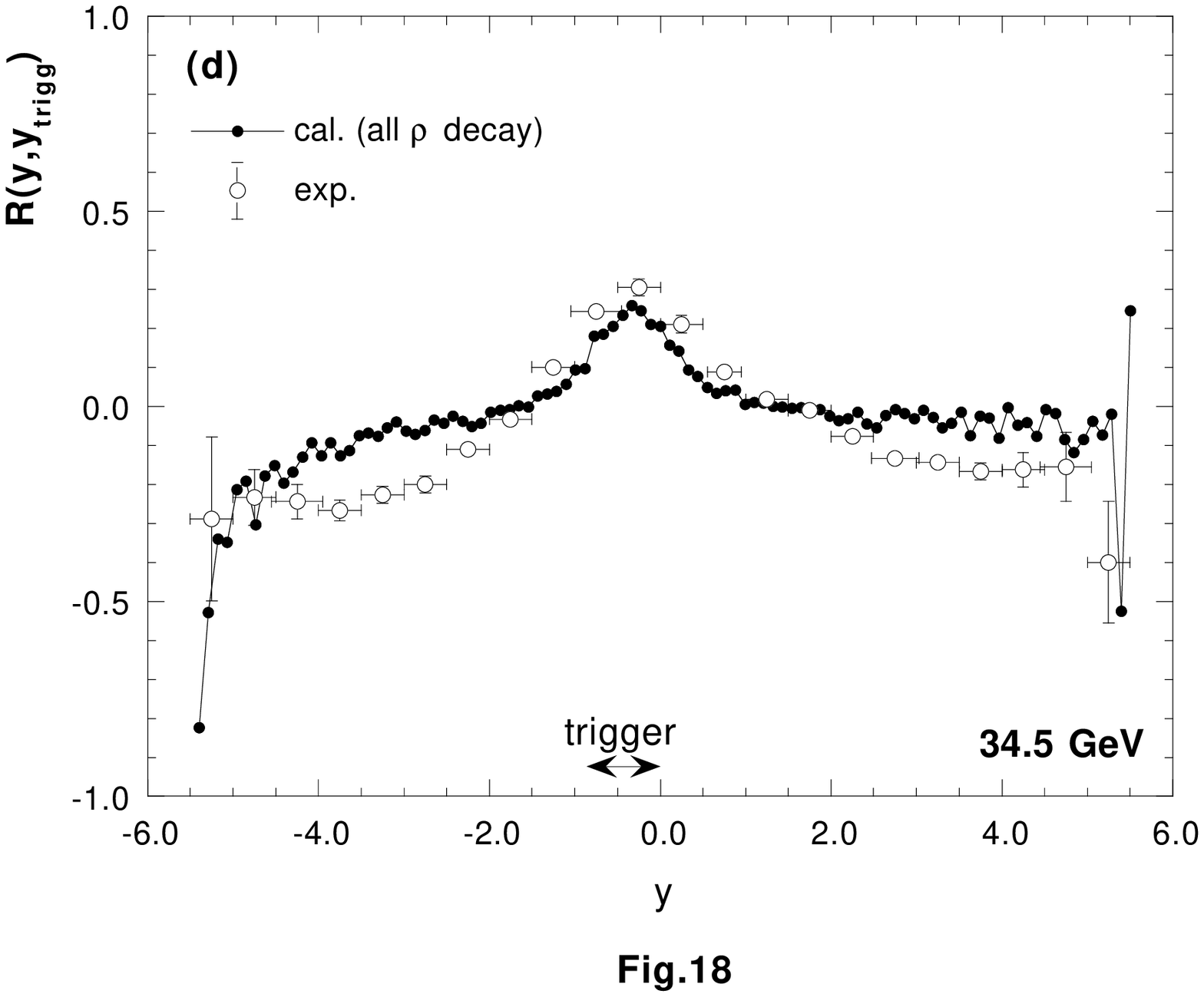}
\end{center}
\end{figure}
\begin{figure}
\begin{center}\leavevmode
\epsfysize=10.0cm
\epsfbox{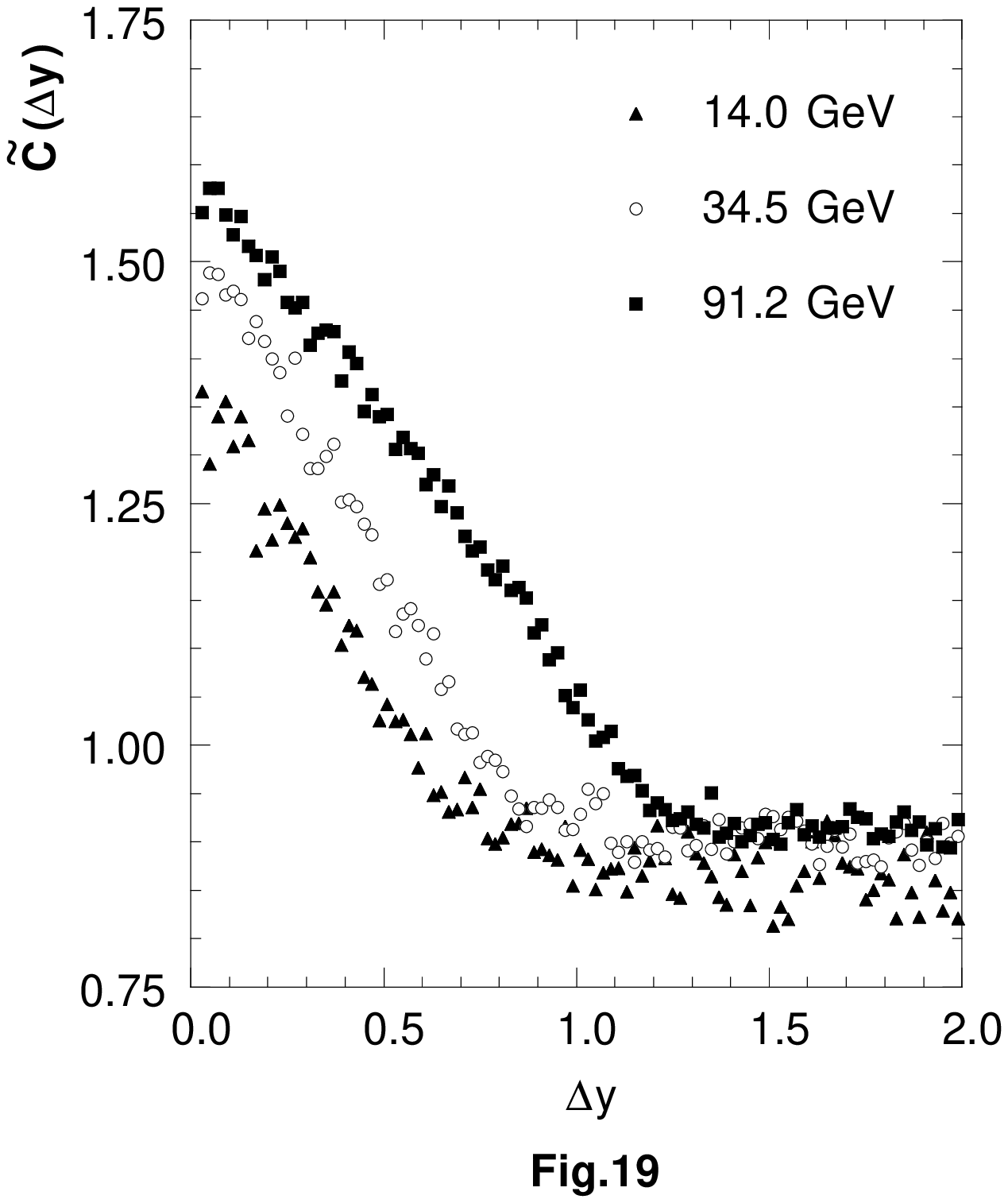}
\end{center}
\end{figure}
\begin{figure}
\begin{center}\leavevmode
\epsfysize=10.0cm
\epsfbox{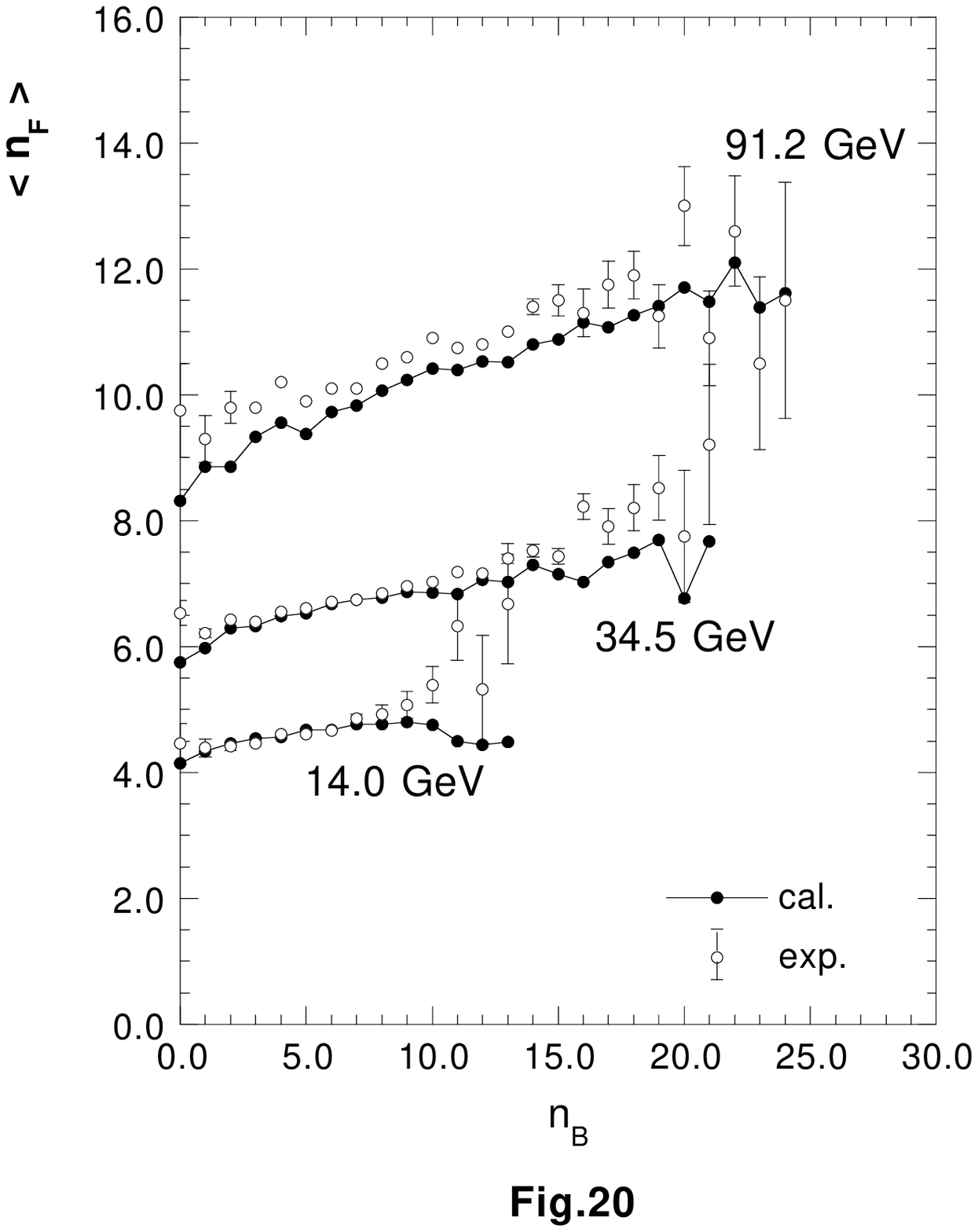}
\end{center}
\end{figure}
\end{document}